\documentclass[showpacs,floatfix]{revtex4}
\usepackage{amsmath}
\usepackage{color}
\usepackage{amssymb}

\usepackage{graphicx}

\begin{document}

\title{Superconductivity in Uranium Ferromagnets}

\author{V.P.Mineev$^{1,2}$}
\affiliation{$^1$
 Commissariat a l'Energie Atomique, UGA, INAC-FELIQS, 38000 Grenoble, France;\\
$^2~$Landau Institute for Theoretical Physics, 119334, Moscow, Russia}

\begin{abstract}
The  theoretical description and the survey of physical properties of superconducting states in the uranium ferromagnetic materials are presented.
On the basis of microscopic theory is shown that the coupling between the electrons in these ferromagnetic metals by means of magnetization fluctuations
 gives rise  the triplet pairing superconducting state and 
the general form of the order parameter dictated by the symmetry is established. 
 The theory allows to explain
some  specific  observations including 
peculiar phenomenon of reentrant superconductivity in URhGe 
in magnetic field perpendicular to the direction of spontaneous magnetization.

In addition we describe several particular topics relating to uranium superconducting ferromagnets:
(i) critical magnetic relaxation in dual localized-itinerant ferromagnets,
(ii)  phase transition to ferromagnetic state in Fermi liquid and UGe$_2$,
(iii) superconducting ordering  in ferromagnetic metals without inversion symmetry.

\end{abstract}

\date{\today}
\maketitle

{\bf CONTENTS}

\smallskip

I. Introduction

II. Order parameters, symmetry of states and quasiparticle spectrum

$~~~$A. Symmetry of superconducting states in orthorhombic ferromagnets

$~~~$B. Superconducting states in UCoGe

$~~~$C. Quasiparticles spectrum in a ferromagnet superconductor with triplet pairing

III.	Superconducting states in microscopic weak coupling theory

$~~~$A.	Triplet pairing by spin-fluctuations exchange

$~~~$B. Magnetic susceptibility of an orthorhombic ferromagnet

$~~~$C. Pairing amplitudes

$~~~$D. Critical temperature of phase transition to the paramagnetic superconducting state in UCoGe

$~~~$E. Phase transition from the paramagnetic to ferromagnetic superconducting state in UCoGe

$~~~$F. Superconducting states in orthorhombic ferromagnets

$~~~$G. Equal-spin-pairing states

$~~~$H. Equal-spin-pairing states near critical temperature

IV. Physical properties

$~~~$A.	Critical temperature

$~~~$B. Upper critical field  parallel to c-axis in UCoGe

$~~~$C. Upper critical field in URhGe

$~~~$D. Zeros in spectrum and specific heat at low temperatures

V. Reentrant superconductivity in URhGe

$~~~$A. Phase transition in an orthorhombic ferromagnet under  magnetic field perpendicular to spontaneous magnetization

$~~~$B. Susceptibilities

$~~~$C. Superconducting state in vicinity of the first order transition

$~~~$D. Concluding remarks

VI. Critical magnetic relaxation in uranium ferromagnets

$~~~$A. Critical magnetic relaxation in ferromagnets

$~~~$B. Magnetic relaxation in dual localized-itinerant ferromagnets

$~~~$C. Concluding remarks

VII.  Anisotropy of nuclear magnetic  relaxation and the upper critical field in UCoGe

$~~~$A. Nuclear magnetic   relaxation rate

$~~~$B. Upper critical field anisotropy

VIII. First order phase transition to ferromagnet state in UGe$_{2}$

$~~~$A. Phase transition to ferromagnetic state in Fermi liquid theory

$~~~$B. Magneto-elastic mechanism of development of the first order type instability

$~~~$C. Specific heat near the Curie temperature

$~~~$D. First-order type transition in UGe$_2$

$~~~$E. Concluding remarks

IX. Superconducting order in UIr

X. Conclusion

Bibiliography

\bigskip

\section{Introduction}
Superconducting and ferromagnetic ordering are usually antagonists each other.  The reason is that the  exchange field exceeds  the  paramagnetic limiting field
 or the field of the Cooper pairs depairing in singlet superconductors. Nevertheless, the singlet superconductivity can coexist with ferromagnetism when the critical temperature of transition to superconducting state surpasses the Curie temperature as it is the case in  so called ternary compounds  actively investigated in nineteen eighties. 
 The coexistence reveals itself
  in a form known as the Anderson-Suhl or crypto-ferromagnetic superconducting state \cite{Maple1983,Maple1995}  characterized by the formation of a periodic domain-like magnetic structure.
The structure period or domain size $\lambda$ is larger than the interatomic distance and smaller than the superconducting coherence length $\xi_0$  what weakens the depairing effect of the exchange field leading to an effective averaging of it to zero.

 The co-existence of superconductivity and ferromagnetism recently discovered \cite{Saxena2000,Aoki2001,Huy2007,Akazawa2004}  in several uranium compounds  UGe$_2$, URhGe, UCoGe, UIr possesses quite different properties.  In the first two compounds the Curie temperatures $T_{Curie}$ is more than the order of magnitude higher than their critical temperatures for superconductivity $T_{sc}$ (Fig.1a,b). In UCoGe the ratio $T_{Curie}/T_{sc}$ at ambient pressure  is about four (Fig.1c).
 This fact and also  that the   upper critical field at low temperatures  strongly exceeds the 
 paramagnetic limiting field in the first three compounds (see reviews \cite{Aoki11,Aoki12,Aoki14})
 indicate that here we deal with the Cooper pairing in the triplet state.
   In  UIr the upper critical field is smaller than the paramagnetic  limiting field \cite{,Akazawa2004}. This, however,
 can be due to the low specimen quality  caused by impurities,  inhomogeneities etc, inasmuch  the  impurities   in unconventional superconductors   strongly suppress the upper critical field.

Ferromagnetism does not suppress the superconductivity with triplet pairing, hence, there is no reason for the formation of a cryptomagnetic state.  Indeed, no traces of a space modulation of magnetic moments directions
on the scale smaller than the coherence length has been revealed \cite{Aoki2001,Aso2005,Kotegawa2005,Visser2009}. 
On the other hand, the neutron depolarization measurements on UGe$_2$ down to 4.2 K  (that is in the ferromagnet but not superconducting region) establish, that the magnetic moment strictly aligned along the $a$-axis, with a typical domain size in the $bc$-plane of the order $4.4\times10^{-4}$  cm \cite{Sakarya2005}
that is about two orders of magnitude larger than the largest superconducting coherence length in the $b$-direction $\xi_b\approx 7\times10^{-6}$ cm. Similar size of domains  has been  recently measured in UCoGe.\cite{Hykel2014}

 So, it is  natural to consider these ferromagnetic superconductors as triplet superconductors similar to superfluid phases of He$^3$. It must be kept in mind, however, that unlike   to the liquid helium which is  completely isotropic neutral Fermi liquid  here we deal with superconductivity developing in strongly anisotropic ferromagnetic metals.  Namely, UGe$_2$, URhGe, UCoGe have an orthorhombic structure with magnetic moment oriented along the $a$-axis in the first of these compounds and along the $c$-axis in the last two of them (Fig.2).  UIr  has monoclinic  PbBi-type structure (space group
P21) without inversion symmetry with magnetic moment oriented along $[10\bar 1]$ direction.\cite{KobayashiHori}

 The magnetic moments in UGe$_2$ \cite{Kernavanois2001}, in URhGe \cite{Prokes2003} and in UCoGe \cite{Prokes2010}
 are mostly concentrated around uranium ions.  At $T=0$ they are  $1.4\mu_B,~0.4\mu_b,~0.07\mu_B$, correspondingly.
 Although these values are significantly smaller than the moment per uranium atom deduced from the susceptibility above $T_{Curie}$: $2.8\mu_B,~ 1.8 \mu_B, ~1.5\mu_B$, correspondingly, this is still not sufficient to treat the uranium  compounds as completely itinerant ferromagnets. They are rather  dual localized-itinerant ferromagnets. The most weakly delocalized material is UGe$_2$\cite{Kernavanois2001,Huxley2001} and the most itinerant is UCoGe\cite{Taupin2015, Butchers2015}.

The local nature of magnetism in  uranium ferromagnetic compounds
put forward as  the most plausible pairing mechanism  the interaction between the conduction electrons by means of spin waves in the system of localized moments. The first such type model has been applied to the superconducting antiferromagnet UPd$_2$Al$_3$ \cite{McHale} and then to the reentrant  superconductivity  in ferromagnetic URhGe  \cite{Hattori}.

The general form  of the order parameters of superconducting states  in orthorhombic ferromagnets dictated by the symmetry was found in the papers \cite{Mineev2002,Champel,Mineev2004}.  Then, the corresponding microscopic description of triplet superconductivity 
based on the pairing interaction due to exchange of magnetization fluctuations in the orthorhombic ferromagnets with strong magnetic anisotropy has been developed \cite{Mineev2011,Mineev2014}.  
This approach  has 
allowed   to explain an interplay of the pressure dependence of the Curie temperature and the critical temperature of
the superconducting transition and
the magnetic field dependence of pairing interaction for   field orientations parallel or perpendicular to the direction of spontaneous magnetization. The latter, in its turn, allows  to explain 
peculiar phenomenon of the reentrant superconductivity in URhGe \cite{Mineev2015}
in magnetic field along  the $b$ direction.

Here we present the survey of the theory and the physical properties  of  the superconducting uranium compounds. Starting from the symmetry   of superconducting states valid for a multi-band 
orthorhombic ferromagnet we limit ourselves by the description of a simplest  two-band spin-up, spin-down superconducting ferromagnet. The structure of the superconducting order parameters and the quasiparticle spectrum  are derived.
Then we expose corresponding weak coupling microscopic theory of pairing due to exchange by magnetic fluctuations in strongly anisotropic media with orthorhombic symmetry. 
 The theory reproduces  the superconducting order parameter dictated by the symmetry. The assumptions made in the previous treatments  are demonstrated explicitly.
Then we discuss  low  temperature  specific heat, upper critical field and other concrete properties of  orthorhombic ferromagnet superconducting materials. 

Special  attention is paid to the peculiar phenomenon of reentrant superconductivity in URhGe \cite{Levy2005}. In frame of the Landau phenomenological theory it is demonstrated that a magnetic field perpendicular to the direction of easy magnetization decreases the Curie temperature  and at strong enough field  the phase transition between anisotropic ferromagnetic and paramagnetic states is changed from the second to the first-order. 
 The pairing interaction strongly increases  in  vicinity of the  transition stimulating the reentrance to the  superconducting state
 suppressed by the orbital mechanism. 
 
 We show that the magnetic field along the direction of spontaneous magnetization suppresses the longitudinal fluctuations of magnetization. This  allows to explain the peculiar phenomena of field direction dependence of nuclear magnetic resonance relaxation \cite{Hattory2012} and sharp anisotropy of the upper critical field  \cite{Aoki2009,Aoki14}  in UCoGe.
 
In addition we describe  several particular topics related to the superconductivity in uranium ferromagnets.

First, we discuss the origin of non-Landau damping of critical magnetic fluctutions in ferromagnets  with dual localized-itinerant nature of $f$-electrons \cite{Huxley2003,Mineev2013}.

Then, we consider  the phase transition to  ferromagnetic state in UGe$_2$ \cite{Mineev2012} and in  Fermi liquid.

Finally, we expose the  general structure   of  superconducting ordering  in ferromagnetic metals without inversion symmetry as it is  in UIr.

\section{order parameters, symmetry of states and quasiparticle spectrum}

\subsection{Symmetry of superconducting states in orthorhombic ferromagnets}

We shall  consider two band  ferromagnetic metal 
with the electron spectra
\begin{equation}
\varepsilon_{\uparrow}({\bf k})=  \xi_{\uparrow}({\bf k})+\mu, ~~~\varepsilon_{\downarrow}({\bf k})=  \xi_{\downarrow}({\bf k})+\mu
\end{equation} 
 for the spin-up and the spin-down bands ( see Fig.3), where  $\xi_{\uparrow},~\xi_{\downarrow}$ are the energies  counted from the chemical potential $\mu$.
 
The spin-triplet   superconducting state arising in  a ferromagnetic metal consists of 
the spin-up, spin-down and spin-zero Cooper pairs 
 described by the matrix order parameter \cite{Book}
\begin{eqnarray}
\Delta_{\alpha \beta}({\bf k},{\bf r})&=& \left(
\begin{array}{cc}
\Delta^{\uparrow}¥({\bf k},{\bf r}) & \Delta^{0}¥({\bf k},{\bf r}) \\
\Delta^{0}¥({\bf k},{\bf r}) &\Delta^{\downarrow}¥({\bf k},{\bf r})
\end{array}\right)=\Delta^{\uparrow}¥({\bf k},{\bf r})|\uparrow\uparrow\rangle +\Delta^{0}¥({\bf k},{\bf r})(|\uparrow\downarrow\rangle+|\downarrow\uparrow\rangle)
+\Delta^{\downarrow}¥({\bf k},{\bf r})|\downarrow\downarrow\rangle\nonumber\\
&=&\left({\bf d}({\bf k},{\bf r})\mbox{\boldmath$\sigma$}\right)i\sigma_{y}
=\left(
\begin{array}{cc}
-d_{x}({\bf k},{\bf r})+i d_{y}({\bf k},{\bf r})&
d_{z}({\bf k},{\bf r}) \\
d_{z}({\bf k},{\bf r}) & d_{x}({\bf k},{\bf r})+
i d_{y}({\bf k},{\bf r}),
\end{array}
\right),
\label{1}
\end{eqnarray}
where $\Delta_{\uparrow}¥({\bf k},{\bf r})$,  $\Delta_{\downarrow}¥({\bf r},{\bf k},{\bf r})$, $ \Delta_{0}¥({\bf k},{\bf r}) $ are the amplitudes of spin-up, spin-down and zero-spin  of superconducting order parameter depending on the Cooper pair centre of gravity coordinate ${\bf r}$ and the momentum ${\bf k }$ of pairing electrons. $\mbox{\boldmath$\sigma$}=(\sigma_{x},\sigma_{y},\sigma_{z})$ are the 
Pauli matrices.
Equivalently, the order parameter can be written as the complex vector 
\begin{equation}
{\bf d}({\bf k},{\bf r})=\frac{1}{2}
\left[-\Delta^{\uparrow}¥({\bf k},{\bf r})(\hat{x}+i\hat{y})+
\Delta^{\downarrow}¥({\bf k},{\bf r})(\hat{x}-i\hat{y})\right]+\Delta^{0}¥({\bf k},{\bf r})\hat z.
\label{e14}
\end{equation}
Here and in what follows $\hat x,\hat y,\hat z$  are the unit vectors along the corresponding coordinate axes.

We consider  a ferromagnetic orthorhombic crystal with strong spin-orbital coupling fixing the
spontaneous magnetization along one of the symmetry axis of the second
order chosen as the $z$-direction. Its point symmetry group or black-white group consists from the rotation on angle $\pi$ around $z$ axis
and the rotations on angle $\pi$ around $x$ and $y$ directions combined with the operation of time inversion $R$ which changes
the  direction of spontaneous magnetization to the opposite one
\begin{equation}
D_{2}(C_{2}^{z})=
(E, C_{2}^{z}, RC_{2}^{x}, RC_{2}^{y}).
\label{e1}
\end{equation}
Being not interested of possible translation invariance breaking at the transition into superconducting state
we do not consider the full space group of the normal state.
Besides the point operations the symmetry group of the normal state includes the group of gauge transformations $U(1)$
\begin{equation}
G_{FM}=U(1)\times D_{2}(C_{2}^{z})=
U(1)\times(E, C_{2}^{z}, RC_{2}^{x}, RC_{2}^{y}).
\label{e2}
\end{equation}

The superconducting states with {\bf different critical temperatures} are described by  the basis functions of {\bf different irreducible
co-representations} of  the symmetry group of the normal state. There are only two different co-representations $A$ and $B$ of the group $G_{FM}$ \cite{Mineev2002,Mineev2004}. The  vector order parameters Eq.(\ref{e14})
$$
{\bf d}_A({\bf k},{\bf r}),~~~~{\bf d}_B({\bf k},{\bf r})
$$
of these states are determined by the amplitudes
\begin{eqnarray}
&\Delta_A^\uparrow({\bf k},{\bf r})=\hat k_x\eta_x^\uparrow({\bf r})+i\hat k_y\eta_y^\uparrow({\bf r}),\nonumber\\
&\Delta_A^\downarrow({\bf k},{\bf r})=\hat k_x\eta_x^\downarrow({\bf r})+i\hat k_y\eta_y^\downarrow({\bf r}),\label{A'}\\
&\Delta_A^0({\bf k},{\bf r})=\hat k_z\eta_z^0({\bf r}),\nonumber
\end{eqnarray}
\begin{eqnarray}
&\Delta_B^\uparrow({\bf k},{\bf r})=\hat k_z\zeta_z^\uparrow({\bf r}),\nonumber\\
&\Delta_B^\downarrow({\bf k},{\bf r})=\hat k_z\zeta_z^\downarrow({\bf r}),\label{B'}\\
&\Delta_B^0({\bf k},{\bf r})=\hat k_x\zeta_x^0({\bf r})+i\hat k_y\zeta_y^{0}({\bf r}).\nonumber
\end{eqnarray}
Here and in what follows $\hat k_x,\hat k_y,\hat k_z$ are the components of the unit vector of momentum $\hat{\bf k}={\bf k}/|{\bf k}|$.
For the state A the pair of scalar complex order parameter amplitudes for spin-up pairing have the common phase 
$(\eta_x^\uparrow, \eta_y^\uparrow)=(|\eta_x^\uparrow|, |\eta_y^\uparrow|)e^{i\varphi^\uparrow} $.
The spin-down pair 
also have the common phase
$(\eta_x^\downarrow, \eta_y^\downarrow)=(|\eta_x^\downarrow|, |\eta_y^\downarrow|)e^{i\varphi^\downarrow} $. And the zero-spin amplitude has its own phase  $\eta_z^0=|\eta_z^0|e^{i\varphi^0}$. We assume that three phases
$\varphi^\uparrow, \varphi^\downarrow, \varphi^0$  either coincide among themselves $\varphi^\uparrow=\varphi^\downarrow= \varphi^0=\varphi$  , or differ on $\pm\pi$. The same property takes place for the spin-up, spin-down and zero-spin order parameter components of the B state.

 One can check that the order parameter ${\bf d}_A({\bf k},{\bf r})$ is invariant in respect to all the transformations of the group $G_A$
 which is isomorphic to the 
 black-white group of the normal state $D_{2}(C_{2}^{z})$ but contains the combined elements of time inversion and gauge transformations
\begin{equation}
G_A=
(E, C_{2}^{z}, e^{2i\varphi}RC_{2}^{x}, e^{2i\varphi}RC_{2}^{y}).
\label{FMA}
\end{equation}
For instance, the element  $RC_{2}^{x}$,  combining transformations $\hat y\to -\hat y,~\hat z\to -\hat z,~\hat k_y\to -\hat k_y,~\hat k_z\to -\hat k_z$ and complex conjugation of the order parameter, transforms it to itself to within the phase factor $e^{-i\varphi}$
such that:
\begin{equation}
e^{2i\varphi}RC_{2}^{x}{\bf d}_A({\bf k},{\bf r})={\bf d}_A({\bf k},{\bf r}).
\nonumber
\end{equation}

The order parameter ${\bf d}_B({\bf k},{\bf r})$ is invariant in respect to all transformations of the group
\begin{equation}
G_B=D_{2}(E)= (E, C_{2}^{z}e^{i\pi}, e^{2i\varphi}RC_{2}^{x}e^{i\pi},
e^{2i\varphi}RC_{2}^{y}).
\label{e4}
\end{equation}

Important to note that the five-component $\eta_x^\uparrow,~ \eta_y^\uparrow,~
\eta_x^\downarrow,~
\eta_y^\downarrow,~
\eta_z^0$ order parameter of the A-state and the four-component $\zeta_z^\uparrow,~
 \zeta_z^\downarrow, \zeta_x^0,~ \zeta_y^{0}$ order parameter of the B-state found from the pure symmetry considerations  include
the zero-spin components. In other words they are not {\bf equal-spin-pairing states}, consisting of the Cooper pairs with the opposite spins.
This fact will be explained below in the frame of microscopic approach.

Writing Eqs.(\ref{A'}) and(\ref{B'}) we were limited by the simplest form of the superconducting states  order parameters.   In general, one must take into account the following complications.

(i) Each term in  Eqs.(\ref{A'}) and(\ref{B'}) 
 can contain $k^2_x, k_y^2, k_z^2$ dependent factors invariant in respect all rotations of the orthorhombic group\cite{Mineev2002}.
  
(ii)  Eqs.(\ref{A'}) and(\ref{B'}) 
 are written as they should be for a two-band spin-up and spin-down ferromagnet. For a multi-band ferromagnet spin-up, spin-down and spin-zero parts of the order parameter should consist of several terms relating to different bands. 

(iii) Also, if necessary, the higher order harmonics (higher powers of $\hat k_x^l\hat k_y^m\hat k_z^n$)
of the same symmetry as the linear in respect of components $\hat{\bf k}$ in the  Eqs.(\ref{A'}) and(\ref{B'})  can be  taken into account \cite{Mineev2002}. 

\subsection{Superconducting states in UCoGe }

Unlike to  URhGe and UGe$_2$, where the superconducting state arises only in the ferromagnetic state,
the phase diagram of UCoGe in Fig.1c includes 
 ferromagnetic  (FM+SC) and paramagnetic (SC) superconducting states \cite{fn4}.
The symmetries of all the states  shown in Fig.1c  obey to the usual for a second order phase transition  subordination rules \cite{MineevAIP}. Namely, the group of symmetry of the ferromagnetic superconducting state A Eq.(\ref{FMA})
\begin{equation}
G_{FM+SC}=
(E, C_{2}^{z}, e^{2i\varphi}RC_{2}^{x}, e^{2i\varphi}RC_{2}^{y})
\end{equation} 
 is the subgroup of the group of symmetry of the ferromagnetic state Eq.(\ref{e2})
 \begin{equation}
G_{FM}=U(1)\times D_{2}(C_{2}^{z})=U(1)\times(E, C_{2}^{z}, RC_{2}^{x}, RC_{2}^{y}),
\end{equation} 
 and the group of symmetry of the paramagnetic superconducting state 
 \begin{equation}
G_{SC}=(E, C_{2}^{z}, C_{2}^{x}, C_{2}^{y})+e^{2i\varphi}R\times(E, C_{2}^{z}, C_{2}^{x}, C_{2}^{y}).
\end{equation} 
In its turn both of these groups are the subgroups of the group of the  paramagnetic normal state 
\begin{equation}
G_{N}=U(1)\times\left\{(E, C_{2}^{z}, C_{2}^{x}, C_{2}^{y})+R\times(E, C_{2}^{z}, C_{2}^{x}, C_{2}^{y})   \right\}.
\end{equation}

The order parameter of the paramagnetic superconducting state looks like the order parameter of superfluid $^3$He-B phase \cite{Book}
\begin{equation}
{\bf d}({\bf k})=k_x\eta_x\hat x+k_y\eta_y\hat y +k_z\eta_z\hat z.
\label{unitary}
\end{equation}
At phase transition to the superconducting ferromagnet state the exchange field lifts  the Kramers degeneracy between the spin-up and the spin-down electron states. Hence, the unitary order  parameter (\ref{unitary}) transforms into the nonunitary order parameter of the type A superconducting ferromagnet state (\ref{A'}):
\begin{eqnarray}
&k_x\eta_x\hat x+k_y\eta_y\hat y +k_z\eta_z\hat z\nonumber\\
&=\frac{1}{2}\left[(k_x\eta_x-ik_y\eta_y)(\hat x+i\hat y)+(k_x\eta_x+ik_y\eta_y)(\hat x-i\hat y)\right ]  +k_z\eta_z\hat z=
\frac{1}{2}
\left[-\Delta^{\uparrow}¥({\bf k})(\hat{x}+i\hat{y})+
\Delta^{\downarrow}¥({\bf k})(\hat{x}-i\hat{y})\right]+\Delta^{0}¥({\bf k})\hat z\nonumber\\
&~{\color{blue}\Longrightarrow}~~\frac{1}{2}(k_x\eta_x^\uparrow({\bf r})-ik_y\eta_y^\uparrow({\bf r}))(\hat x+i\hat y),~~ \frac{1}{2}(k_x\eta_x^\downarrow({\bf r})
+ik_y\eta_y^\downarrow({\bf r})
)(\hat x-i\hat y),~k_z\eta_z^0({\bf r})\hat z.
\label{nu}
\end{eqnarray}

Similar considerations of the phases symmetry and the order parameter transformation can be made for the ferromagnetic superconducting state B.

Experimentally a phase transition  from the paramagnetic superconducting state to the ferromagnetic superconducting state in UCoGe has never been revealed.

\subsection{Quasiparticles spectrum in a ferromagnet superconductor with triplet pairing}

 We consider two band  ferromagnetic metal 
with the electron spectra $ \xi_{\uparrow}({\bf k}),~ \xi_{\downarrow}({\bf k})$ for the spin-up and the spin-down bands.

Even in the absence  of external field in a ferromagnet  there is an internal field  $H_{int}$ acting on the electron charges.
The internal magnetic field in all uranium superconducting ferromagnets is larger than the lower critical field $H_{c1}$ (see, for instance,  the paper \cite{Hykel2014}). Hence, the Meissner state is absent and the superconducting state is always the 
 Abrikosov mixed state with space inhomogeneous distributions of the order parameter and the internal magnetic field.
 In this case to find the elementary excitation energies one has to solve the coupled systems of the Gor'kov and the Maxwell  differential
 equations.
 Some simplifications appear at low temperatures. Here, owing to $H_{int}<<H_{c2}$
 one can work in the London approximation such that the internal magnetic field ${\bf H}_{int}({\bf r})=rot{\bf A}({\bf r})$ is a slow function of coordinates. 
 In the inter-vortex space  the order parameter is constant, and
 the electron (hole) momenta acquire a Doppler  shift ${\bf k}\pm m{\bf v}_s({\bf r})$ 
  due to non-zero velocity of the superfluid component
  $$
 {\bf v}_s({\bf r})=\frac{\hbar}{2m}\left(\nabla\varphi+ \frac{2e}{\hbar c} {\bf A}({\bf r})\right).
  $$
Then, the  Gor'kov equations are 
\begin{equation}
\left( \begin{array}{c}i\omega_n -\frac{1}{2}(\xi_{\uparrow+}+\xi_{\downarrow+})\sigma_0-\frac{1}{2}(\xi_{\uparrow+}-\xi_{\downarrow+})\sigma_z~~~~~~~~~
-i ({\bf d}\mbox{\boldmath$\sigma$})\sigma_{y}\\ 
i\sigma_{y}({\bf d}^*\mbox{\boldmath$\sigma$})~~~~~~~~~i\omega_n +\frac{1}{2}(\xi_{\uparrow-}+\xi_{\downarrow-})\sigma_0+\frac{1}{2}(\xi_{\uparrow-}-\xi_{\downarrow-})\sigma_z
\end{array}\right )
\left( \begin{array}{cc}\hat G& -\hat F\\ 
-\hat F^{\dagger}& -\tilde G_{-k,-\omega}\end{array}\right )=\left( \begin{array}{cc}\sigma_0& 0\\ 
0& \sigma_0,
\label{Gor'kov}
\end{array}\right ),
\end{equation}
where
\begin{equation}
\xi_{\uparrow,\downarrow\pm}({\bf k})=\xi_{\uparrow,\downarrow}({\bf k}\pm m{\bf v}_s)\approx\xi_{\uparrow,\downarrow}({\bf k})\pm{\bf k}{\bf v}_s.
\end{equation}

At $i\omega_n\to E$ the equality of the determinant of this system  to zero gives the energy of elementary excitations
\begin{equation}
E={\bf k}{\bf v}_s+\sqrt{\frac{1}{2}(\xi_{\uparrow}^2+\xi_\downarrow^2)+({\bf d}{\bf d}^*)\pm\sqrt{\frac{1}{4}\left [\xi_{\uparrow}^2-\xi_\downarrow^2+2i({\bf d}\times{\bf d}^*)_z\right ]^2
-(i({\bf d}\times{\bf d}^*)_z)^2+(i({\bf d}\times{\bf d}^*)^2)}}.
\label{E}
\end{equation}   
The excitations energies acquire the simplest form in the so called equal-spin-pairing state, when $d_0=0$
\begin{eqnarray}
E_\uparrow={\bf k}{\bf v}_s+\sqrt{\xi_{\uparrow}^2+{\bf d}{\bf d}^*+i({\bf d}\times{\bf d}^*)_z}={\bf k}{\bf v}_s+\sqrt{\xi_{\uparrow}^2+\Delta_\uparrow^2
},
\label{uparrow}\\
E_\downarrow={\bf k}{\bf v}_s+\sqrt{\xi_{\downarrow}^2+{\bf d}{\bf d}^*-i({\bf d}\times{\bf d}^*)_z}={\bf k}{\bf v}_s+\sqrt{\xi_{\downarrow}^2+\Delta_\downarrow^2}.
\label{downarrow}
\end{eqnarray}  
It is instructive also to write the energy of excitations in nonunitary superconducting state \cite{Book} 
\begin{eqnarray}
E={\bf k}{\bf v}_s+\sqrt{\xi^2+{\bf d}{\bf d}^*\pm |i({\bf d}\times{\bf d}^*|}.
\end{eqnarray}   
arising from  a paramagnetic normal state where $\xi_\uparrow=\xi_\downarrow=\xi$.
In  all the cases the Kramers degeneracy is lifted.

Let us look now what kind of pairing interaction gives rise the A and B   superconducting states in  ferromagnets
with orthorhombic symmetry.

\section{Superconducting states in microscopic weak coupling theory}

\subsection{Triplet pairing by spin-fluctuations exchange}

The interactions between two electrons is assumed to be due to the attraction of one electron by the magnetic polarization cloud of the other.
Unlike to superfluid He$^3$, where the pairing of atoms originates from the magnetic polarization in 
 the isotropic Fermi liquid, pairing of electrons in a ferromagnetic metal  occurs in an anisotropic media due to polarization of the
 electron liquid and the localized moments as well.
 
So, we consider the pairing originating from  the attraction
\begin{equation}
H_{elm}=-\frac{1}{2}\mu_B^2I^2\int d^3{\bf r}d^3{\bf r}^\prime S_i({\bf r})\chi_{ij}({\bf r-{\bf r}^\prime})S_j({\bf r}^\prime)
\label{int}
\end{equation}
between  the electrons with magnetic moments $\mu_B$ by means the electron-magnon interaction in ferromagnetic media with orthorhombic symmetry.
Here, 
$$
{\bf S}({\bf r})=\psi^\dagger_\alpha({\bf r})\mbox{\boldmath$\sigma$}_{\alpha\beta}\psi_\beta({\bf r})
$$ 
is the operator of the electron spin density, $\chi_{ij}({\bf r}) $ is the media susceptibility, $I
$ is an exchange constant.

Transforming the interaction Hamiltonian  in the momentum representation and leaving only the odd in  respect of ${\bf k}$
and ${\bf k}'$ parity terms  we obtain  from Eq.(\ref{int}) after some straightforward algebra \cite{Samokhin} the BCS Hamiltonian for triplet pairing
\begin{equation}
H_{pairing}=\frac{1}{2}\sum_{{\bf k}{\bf k}^\prime}
V_{\alpha\beta\gamma\delta}({\bf k},{\bf k}^\prime)a^\dagger_\alpha({\bf k}) a^\dagger_\beta(-{\bf k}) a_\gamma(-{\bf k}^\prime) a_\delta({\bf k}^\prime),
\end{equation}
where
\begin{equation}
V_{\alpha\beta\gamma\delta}({\bf k},{\bf k}')=V_{ij}({\bf k},{\bf k}')
(i\sigma_i\sigma_y)_{\alpha\beta}(i\sigma_j\sigma_y)^\dagger_{\gamma\delta}
\end{equation}
and
\begin{equation}
V_{ij}({\bf k},{\bf k}')=-\mu_B^2I^2\left(\frac{1}{2}Tr\hat \chi^u({\bf k},{\bf k}')\delta_{ij}-\chi^u_{ij}({\bf k},{\bf k}')\right)
\label{3}
\end{equation}
is expressed   through  the odd part of the media static susceptibility 
$$
\hat \chi^u({\bf k},{\bf k}')=\chi^u_{ij}({\bf k},{\bf k}')=
\frac{1}{2}[\chi_{ij}({\bf k}-{\bf k}')-\chi_{ij}({\bf k}+{\bf k}')].
$$

The critical temperature or the upper critical field are determined as the eigenvalue of the linear equation for the order parameter \cite{Book}
\begin{equation}
\Delta_{\alpha\beta}({\bf k},{\bf q})
=-T
\sum_{n}
\sum_{{\bf k}' }
V_{\beta\alpha\lambda\mu}({\bf k},{\bf k}')
G_{\lambda\gamma}({\bf k}',\omega_n)
G_{\mu\delta}(-{\bf k}'+{\bf q},-\omega_n)\Delta_{\gamma\delta}({\bf k}',{\bf q}).
\label{GLG}
\end{equation}
Here the matrix of order parameter in the momentum representation is
\begin{equation}
\hat \Delta({\bf k},{\bf q})=\int\hat \Delta({\bf k},{\bf r})e^{i{\bf q}{\bf r}} d^3r=\left( \begin{array}{cc}\Delta^{\uparrow}({\bf k},{\bf q})& \Delta^{0}({\bf k},{\bf q})\\ 
\Delta^{0}({\bf k},{\bf q})& \Delta^{\downarrow}({\bf k},{\bf q})
\end{array}\right ),
\end{equation}
$G_{\lambda\gamma}({\bf k}',\omega_n)$
is the matrix of normal metal Green function. 
{\bf In  absence of external field  or when the external magnetic field is strictly parallel to the spontaneous magnetization} it is  diagonal 
\begin{equation}
\hat G_n=\left( \begin{array}{cc}G^{{\uparrow}}& 0\\ 
0 & G^{\downarrow}
\label{matrix}
\end{array}\right ),
\end{equation}
where
\begin{equation}
G^{\uparrow,\downarrow}=\frac{1}{i\omega_n-\xi^{\uparrow,\downarrow}_{{\bf k}}
}.
\label{Gr}
\end{equation}

Matrix equation (\ref{GLG}) is the system of coupled linear equations for the order parameter components
\begin{eqnarray}
\Delta^{\uparrow}({\bf k},{\bf q})
=-T
\sum_{n}
\sum_{{\bf k}' }
\left\{
V^{\uparrow\uparrow}({\bf k},{\bf k}')G^\uparrow G^\uparrow\Delta^{\uparrow}({\bf k}',{\bf q})+
V^{\uparrow\downarrow}({\bf k},{\bf k}')G^\downarrow G^\downarrow\Delta^{\downarrow}({\bf k}',{\bf q})
+V^{\uparrow 0}({\bf k},{\bf k}')\left [G^\downarrow G^\uparrow+G^\uparrow G^\downarrow\right ]
\Delta^{0}({\bf k}',{\bf q})
\right\},~~~~~
\label{up}\\
\Delta^{\downarrow}({\bf k},{\bf q})
=-T\sum_{n}\sum_{{\bf k}' }
\left\{
V^{\downarrow\uparrow}({\bf k},{\bf k}')G^\uparrow G^\uparrow \Delta^{\uparrow}({\bf k}',{\bf q})+
V^{\downarrow\downarrow}({\bf k},{\bf k}')G^\downarrow G^\downarrow \Delta^{\downarrow}({\bf k}',{\bf q})
+V^{\downarrow 0}({\bf k},{\bf k}')\left [G^\downarrow G^\uparrow +G^\uparrow G^\downarrow
\right ]
\Delta^{0}({\bf k}',{\bf q})
\right\},~~~~~
\label{down}\\
\Delta^{0}({\bf k},{\bf q})
=-T\sum_{n}\sum_{{\bf k}' }
\left\{
V^{0\uparrow}({\bf k},{\bf k}')G^\uparrow G^\uparrow \Delta^{\uparrow}({\bf k}',{\bf q})+
V^{0\downarrow}({\bf k},{\bf k}')G^\downarrow G^\downarrow \Delta^{\downarrow}({\bf k}',{\bf q})
+V^{00}({\bf k},{\bf k}')\left [G^\downarrow G^\uparrow +G^\uparrow G^\downarrow
\right ]
\Delta^{0}({\bf k}',{\bf q})
\right\}.~~~~~
\label{0}
\end{eqnarray}
Here the arguments in the Green functions products are as in the matrix equation (\ref{GLG}). For instance 
$$
G^\uparrow G^\uparrow=G^\uparrow({\bf k}',\omega_n)
G^\uparrow(-{\bf k}'+{\bf q},-\omega_n).
$$
The pairing amplitudes found from Eqs. (\ref{3}) are
\begin{eqnarray}
&V^{\uparrow\uparrow}({\bf k},{\bf k}')=V_{xx}+V_{yy}+iV_{xy}-iV_{yx}=-\mu_B^2I^2\chi_{zz}^u,
\label{111}\\
&V^{\downarrow\downarrow}({\bf k},{\bf k}')=V_{xx}+V_{yy}-iV_{xy}+iV_{yx}=-\mu_B^2I^2\chi_{zz}^u
,\label{112}\\
&V^{\uparrow\downarrow}({\bf k},{\bf k}')=-V_{xx}+V_{yy}+iV_{xy}+iV_{yx}=
-\mu_B^2I^2
(\chi^u_{xx}-\chi^u_{yy}-2i\chi^u_{xy})
,
\label{113}\\
&V^{\downarrow\uparrow}({\bf k},{\bf k}')=-V_{xx}+V_{yy}-iV_{xy}-iV_{yx}
=
-\mu_B^2I^2
(\chi^u_{xx}-\chi^u_{yy}+2i\chi^u_{xy}),\label{114}
\\
&V^{00}({\bf k},{\bf k}')=V_{zz}=-\mu_B^2I^2(\chi_{xx}^u+\chi_{yy}^u-\chi_{zz}^u)/2
,\label{115}\\
&V^{\uparrow 0}({\bf k},{\bf k}')=(V^{0\uparrow }({\bf k},{\bf k}'))^*=-V_{xz}+iV_{yz}=-\mu_B^2I^2(\chi_{xz}^u-i\chi_{yz}^u)
,\label{116}\\
&V^{\downarrow 0}({\bf k},{\bf k}')=(V^{0\downarrow}({\bf k},{\bf k}'))^*=V_{xz}+iV_{yz}=-\mu_B^2I^2(-\chi_{xz}^u-i\chi_{yz}^u).
\label{117}
\end{eqnarray}
One can see that the equations for $\Delta^{\uparrow}$,  $\Delta^{\downarrow}$ and $\Delta^{0}$ are entangled. Moreover, the entanglement still exists even in the case of strong spin-up and spin-down bands splitting which allows to omit all the terms including combinations $G^\downarrow G^\uparrow +G^\uparrow G^\downarrow$  corresponding to the interband pairing. Ignoring the inter-band pairing \cite{footnote}, we find 
\begin{eqnarray}
&\Delta^{\uparrow}({\bf k},{\bf q})
=-T
\sum_{n}
\sum_{{\bf k}' }
\left\{
V^{\uparrow\uparrow}({\bf k},{\bf k}')G^\uparrow G^\uparrow\Delta^{\uparrow}({\bf k}',{\bf q})+
V^{\uparrow\downarrow}({\bf k},{\bf k}')G^\downarrow G^\downarrow\Delta^{\downarrow}({\bf k}',{\bf q})
\right\},
\label{up1}\\
&\Delta^{\downarrow}({\bf k},{\bf q})
=-T\sum_{n}\sum_{{\bf k}' }
\left\{
V^{\downarrow\uparrow}({\bf k},{\bf k}')G^\uparrow G^\uparrow \Delta^{\uparrow}({\bf k}',{\bf q})+
V^{\downarrow\downarrow}({\bf k},{\bf k}')G^\downarrow G^\downarrow \Delta^{\downarrow}({\bf k}',{\bf q})
\right\},
\label{down1}\\
\nonumber\\
&\Delta^{0}({\bf k},{\bf q})
=-T\sum_{n}\sum_{{\bf k}' }\left\{
V^{0\uparrow}({\bf k},{\bf k}')G^\uparrow G^\uparrow \Delta^{\uparrow}({\bf k}',{\bf q})+
V^{0\downarrow}({\bf k},{\bf k}')G^\downarrow G^\downarrow \Delta^{\downarrow}({\bf k}',{\bf q})\right\}.
\label{01}
\end{eqnarray}
We see,  that the equation determining the order parameter component $\Delta^{0}$ corresponding to the pairing of particles with the opposite spins still exists. According to Eq. (\ref{01}) the pairing with the opposite spins is induced by the pairing terms with parallel spins. 
{\bf Thus, in general, a superconducting state in a ferromagnetic metal is not equal-spin-pairing state.}
This property originates from  a spin-orbit coupling. Indeed, in what follows,  we shall see  that the pairing amplitudes $V^{0\uparrow },~V^{0\downarrow }$ arise due to the spin-orbital terms in the ferromagnet  gradient energy which are presumably small. So, with good accuracy one can work with Eqs. (\ref{up1}) and  (\ref{down1}) corresponding to equal-spin-pairing superconductivity neglecting  by the amplitude $\Delta^{0}$ induced by the pairing of electrons with spin up-up and down-down spins. 
This case, we deal with two-band superconducting state similar to the $A_2$ state of superfluid $^3$He \cite{Mermin}. This property is supported by the recent low temperature thermal conductivity measurements under magnetic field\cite {Taupin2014}. 

Both the spin up-up $V^{\uparrow\uparrow}$ and the spin down-down  $V^{\downarrow\downarrow}$ pairing amplitudes are determined by the susceptibility component parallel to the direction of spontaneous magnetization $ \chi_{zz}^u$ which strongly exceeds the susceptibility along the other crystallographic directions. On the other hand, the spin-up and the spin-down  Cooper pairs  interact each other due to the susceptibility anisotropy $\chi_{xx}\ne\chi_{yy}$  which gives rise  common  for the spin-up and the spin-down bands phase transition to the
A$_2$-type state.  The susceptibility anisotropy does not exist in the exchange approximation and completely determined by the spin-orbit coupling.
Even in  isotropic liquid   $^3$He the spin-orbit coupling leads to the entanglement between the spin-up and the spin-down order parameters \cite{Yamaguchi2006}.
However,  in view of smallness  of spin-orbit interaction,  the entanglement between the spin-up and the spin-down components is practically absent, what results  in  two subsequent phase transitions under magnetic field: first to the spin-up $A_1$ state and then to the mixed spin-up  and spin-down $A_2$ superfluid state \cite{Mermin}.

Let us now find the susceptibility.

\subsection{Magnetic susceptibility of  an orthorhombic ferromagnet}

We search for the static magnetic susceptibility following phenomenological approach.
The starting point is
the Landau free energy of an orthorhombic ferromagnet 
in
magnetic field ${\bf H}({\bf r})$
\begin{equation}
{\cal F}=\int d V(F_M+F_\nabla),
\label{FE}
\end{equation}
where in expressions for condensation energy density
\begin{equation}
F_M=\alpha_{z}M_{z}¥^{2}+\alpha_{y}M_{y}^{2}+\alpha_{x}M_{x}¥^{2}
+\beta_{z}¥M_{z}¥^{4} 
+\beta_{xy}¥M_{x}¥^{2}¥M_{y}¥^{2}+\beta_{yz}¥M_{z}¥^{2}¥M_{y}¥^{2}¥+\beta_{xz}¥M_{z}¥^{2}¥M_{x}¥^{2}-{\bf M}{\bf  H},
\label{F}
\end{equation}
and gradient energy density
\begin{equation}
F_\nabla=\gamma^{\alpha\beta}_{ij}\frac{\partial M_{\alpha }}{\partial x_i}\frac{\partial  M_\beta }{\partial x_j}
\label{nabla}
\end{equation}
we bear in mind the orthorhombic anisotropy. The $z$ direction is always chosen along the spontaneous magnetization, hence in URhGe  and in UCoGe
 $x, y, z$ coordinates are directed along the $a, b, c$ crystallographic axis, and in UGe$_2$ along  $b,c,a$ axis.
\begin{equation}
\alpha_{z}=\alpha_{z0}¥(T-T_{c}), ~~~\alpha_x>0, ~~~\alpha_y>0,  
\end{equation}
$T_c$ is the Curie temperature, 
and 
\begin{equation}
~~~~~~~\gamma_{ij}^{\alpha=\beta}= \left(\begin{array}{ccc} \gamma^\alpha_{xx} & 0 & 0\\
0 & \gamma^\alpha_{yy} & 0 \\
0 & 0 & \gamma^\alpha_{zz}
\end{array} \right),~~~\alpha=x,y,z
\end{equation}
and
\begin{equation}
\gamma_{ij}^{\alpha\ne\beta} =\gamma_{ij}^{ij} = \left(\begin{array}{ccc} 
0 & \gamma_{xy} & \gamma_{xz}\\
\gamma_{xy} & 0& \gamma_{yz} \\
\gamma_{xz} & \gamma_{yz} &0
\end{array} \right).
\end{equation}
The corresponding energy density in the exchange approximation is
\begin{equation}
F^{exchange}_M+F^{exchange}_\nabla=\alpha {\bf M}^{2}
+\beta {\bf M}^{4}
-{\bf M}{\bf  H}+\gamma_{ij}\frac{\partial {\bf M}}{\partial x_i}\frac{\partial {\bf M}}{\partial x_j}.
\label{sum}
\end{equation}
where the matrix $\gamma_{ij}$ 
is 
\begin{equation}
\gamma_{ij} = \left(\begin{array}{ccc} \gamma_{xx} & 0 & 0\\
0 & \gamma_{yy} & 0 \\
0 & 0 & \gamma_{zz}
\end{array} \right)
\end{equation}
such that the gradient energy is determined only by 3 constants instead 12 constants involved in general case taking into account the relativistically small interactions.

Let us take the magnetic field as the sum of the constant part in $z$ direction and the coordinate dependent small addition 
\begin{equation}
{\bf H}({\bf r})=\delta H_x({\bf r})\hat x+\delta H_y({\bf r})\hat y+(H_z+\delta H_z({\bf r}))\hat z.
\end{equation}
By variation of functional (\ref{FE}) in respect to the components of magnetization we arrive to the equations
\begin{eqnarray}
2\alpha_xM_x+2\beta_{xy}M^2_yM_x+2\beta_{xz}M^2_zM_x-2\gamma^x_{ij}\frac{\partial^2 M_x}{\partial x_i\partial x_j}
-\gamma_{xy}\frac{\partial^2 M_y}{\partial x\partial y}-\gamma_{xz}\frac{\partial^2 M_z}{\partial x\partial z}
=\delta H_x,\nonumber\label{va}
\\
 2\alpha_yM_y+2\beta_{xy}M^2_xM_y+2\beta_{yz}M^2_zM_y-2\gamma^y_{ij}\frac{\partial^2 M_y}{\partial x_i\partial x_j}
-\gamma_{xy}\frac{\partial^2 M_x}{\partial x\partial y}-\gamma_{yz}\frac{\partial^2 M_z}{\partial y\partial z}=\delta H_y,
\\
2\alpha_zM_z+4\beta_zM_z^3+2\beta_{xz}M^2_xM_z+2\beta_{yz}M^2_yM_z-2\gamma^z_{ij}\frac{\partial^2 M_z}{\partial x_i\partial x_j}
-\gamma_{xz}\frac{\partial^2 M_x}{\partial x\partial z}-\gamma_{yz}\frac{\partial^2 M_y}{\partial y\partial z}
=H_z+\delta H_z.\nonumber
\end{eqnarray}

The equilibrium magnetization projections are determined by the equations:
\begin{eqnarray}
&M_x=0,~~~~~~~~M_y=0,
\label{xy}\\
&M_z^2=-\frac{\alpha_z}{2\beta_z}
+\frac{H_z}{4\beta_zM_z}.
\label{z}
\end{eqnarray}
The first and the last term in this expression correspond to the spontaneous and the induced part of magnetization along $z$ direction.

Taking magnetization as the sum of the constant part and the coordinate dependent small addition 
\begin{equation}
{\bf M}({\bf r})=M_z\hat z+\delta M_x({\bf r})+\delta M_y({\bf r})+\delta M_z({\bf r}),
\end{equation}
we obtain from the Eqs.(\ref{va})  the linear equations for the Fourier components  of  $\delta {\bf M}({\bf k})$:
\begin{eqnarray}
2(\alpha_x+\beta_{xz}M_z^2+\gamma_{ij}^xk_ik_j)\delta M_x({\bf k})+\gamma_{xy}k_xk_y\delta M_y({\bf k})+\gamma_{xz}k_xk_z\delta M_z({\bf k})=\delta H_x({\bf k}),\nonumber\\
\gamma_{xy}k_xk_y\delta M_x({\bf k})+2(\alpha_y+\beta_{yz}M_z^2+\gamma_{ij}^yk_ik_j)\delta  M_y({\bf k})+\gamma_{yz}k_yk_z\delta M_z({\bf k})=\delta H_y({\bf k}),\label{A}
\\
\gamma_{xz}k_xk_z\delta M_x({\bf k})+\gamma_{yz}k_yk_z\delta M_y({\bf k})+2(\alpha_z+6\beta_zM_z^2+\gamma_{ij}^zk_ik_j)\delta M_z({\bf k})=\delta H_x({\bf k}).\nonumber
\end{eqnarray}
The coupling between the magnetization components in Eqs.(\ref{A}) are due to the anisotropy terms originating from the small relativistic interactions.
Hence, solving  Eqs.(\ref{A}) one can neglect 
by  all the products of the terms  like 
$\gamma_{xy}k_xk_y\gamma_{yz}k_yk_z$ etc and  obtain:
\begin{eqnarray}
&\chi_{xx}=\frac{\delta M_x}{\delta H_{x}}\approx\frac{1}{2(\alpha_x+\beta_{xz}M_z^2+\gamma_{ij}^xk_ik_j)},\\
&\chi_{yy}=\frac{\delta M_y}{\delta H_{y}}\approx\frac{1}{2(\alpha_y+\beta_{yz}M_z^2+\gamma_{ij}^yk_ik_j)},\\
&\chi_{zz}=\frac{\delta M_z}{\delta H_{z}}\approx\frac{1}{2(\alpha_z+6\beta_zM_z^2+\gamma_{ij}^zk_ik_j)},\label{59}\\
&\chi_{xy}=\frac{\delta M_x}{\delta H_{y}}=\frac{\delta M_y}{\delta H_{x}}\approx-\frac{\gamma_{xy}k_xk_y}{4(\alpha_x+\beta_{xz}M_z^2+\gamma_{ij}^xk_ik_j)(\alpha_y+\beta_{yz}M_z^2+\gamma_{ij}^yk_ik_j)},\\
&\chi_{xz}=\frac{\delta M_x}{\delta H_{z}}=\frac{\delta M_z}{\delta H_{x}}\approx-\frac{\gamma_{xz}k_xk_z}{4(\alpha_x+\beta_{xz}M_z^2+\gamma_{ij}^xk_ik_j)(\alpha_z+6\beta_zM_z^2+\gamma_{ij}^zk_ik_j)},\\
&\chi_{yz}=\frac{\delta M_y}{\delta H_{z}}=\frac{\delta M_z}{\delta H_{y}}\approx-\frac{\gamma_{yz}k_yk_z}{4(\alpha_y+\beta_{yz}M_z^2+\gamma_{ij}^yk_ik_j)(\alpha_z+6\beta_zM_z^2+\gamma_{ij}^zk_ik_j)}.
\end{eqnarray}

The expressions for the susceptibility components depend from  the wave vector through the combinations like $\gamma k^2$. They are found at wave vectors much smaller than the inverse interatomic 
distance $a^{-1}$. 
The  corresponding wave vector dependent  terms in the susceptibility
components found from an appropriate 
microscopic
model  will be given by the combinations of trigonometric functions like  $\gamma\sin^2ka/a^2$. 
 These combinations
at small $k$   reproduce our  phenomenological expressions and at   $k\cong k_F$ they are of the order $\gamma/a^2$ as the combinations 
 in the phenomenological theory. 
This means that our formulas for susceptibilities
are still qualitatively valid  at large wave vectors transfer $k\cong k_F$ determining pairing interaction. The Fermi momenta $k_F$ at different Fermi surface points are different, hence, $k_F$ is a function of direction in the reciprocal space possessing full orthorhombic symmetry.

The odd part 
of $z$-component of susceptibility 
is found as 
\begin{eqnarray}
\chi^u_{zz}({\bf k},{\bf k}')=\frac{1}{2}[\chi_{zz}({\bf k}-{\bf k}')-\chi_{zz}({\bf k}+{\bf k}')]\nonumber\\
=\frac{\gamma_{ij}k_ik_j^\prime}{
(\alpha_z+6\beta_zM_z^2
+\gamma_{ij}(k_ik_j+k_i^{\prime}k_j^{\prime}))^2-(2\gamma_{ij}k_ik_j^{\prime})^2}.
\label{ch}
\end{eqnarray}
According to Eqs.(\ref{111}) and (\ref{112}) the pairing interaction is  mostly determined by this formula.  The situation is similar to the case of weak coupling singlet pairing, where the zero-frequency limit of phonon propagator plays the role of the potential for the phonon-mediated attraction between the electrons.
We  are interested in the pairing interaction inside the ferromagnetic state where $\alpha_z+6\beta_zM_z^2$ has a finite positive
 magnitude.   
 At the Curie temperature this  combination is equal to zero  and $\chi^u_{zz}({\bf k},{\bf k}')$ diverges at the coincident arguments corresponding the Cooper pairing. This is inevitable property of  a model with static interaction. To avoid this pairing interaction divergency 
 D. Fay and J .Appel \cite{Fay1980} 
in their  theory of $p$-wave superconductivity in an itinerant ferromagnet  have  introduced  a cutoff depending from the distance from the ferromagnetic phase transition. As result  the critical temperature of phase transition to the superconducting state having a finite value  both in the ferromagnetic and the paramagnetic state proved to be equal to zero  at transition between them.  This misleading property does not take place in a model taking into account a retardation effect in the pairing interaction.

At finite $\alpha_z+6\beta_zM_z^2$ 
we can  keep only the  angular dependence of interaction in the numerator of Eq.(\ref{ch}) neglecting  by the angular dependence of $k_F$ and the orthorhombic symmetry  terms in denominator $\gamma^z_{ij}(k_ik_j+k_i^{\prime}k_j^{\prime})\approx2\gamma^z k_F^2$
as well as by all the higher angular harmonics of interaction determined by the last term in denominator \cite{footnote}. 
The calculations without these simplifications are much more cumbersome but do not give rise the  qualitatively different results.
Hence, we obtain
\begin{equation}
\chi^u_{zz}({\bf k},{\bf k}')\cong
\frac{\gamma^z_{ij}k_F^2}{
a_z^2}\hat k_i \hat k_j^\prime,
\label{Chi}
\end{equation}
where
\begin{equation}
a_z=\alpha_z+6\beta_zM_z^2+2\gamma^z k_F^2=2\beta_z(3M_z^2-M_{z0}^2)+2\gamma^z k_F^2,
\label{az}
\end{equation}
where $M_z$ is the solution of  Eq.(\ref{z}) and $M_{z0}=M_{z}(H_z=0)=(-\alpha_z/2\beta_z)^{1/2}$.
At temperatures noteceably smaller than the Curie temperature one can use experimental values for the field dependent
magnetization $M_z(H_z)$ and its almost temperature independent spontaneous part $M_{z0}=M_{z}(H_z=0)$.

Found in similar manner the odd part 
of the susceptibility $x$ and $y$ -components are
\begin{eqnarray}
\chi^u_{xx}({\bf k},{\bf k}')\cong
\frac{\gamma^x_{ij}k_F^2}{a_x^2}\hat k_i \hat k_j^\prime,
~~~~~~
\chi^u_{yy}({\bf k},{\bf k}')\cong
\frac{\gamma_{ij}^yk_F^2}{
a_y^2}\hat k_i\hat k_j^\prime,
\label{Chi4}
\end{eqnarray}
where
\begin{equation}
a_x=\alpha_x+\beta_{xz}M_z^2+2\gamma^x k_F^2,~~~~~
a_y=\alpha_y+\beta_{yz}M_z^2+2\gamma^y k_F^2.
\label{ax}
\end{equation}

All off-diagonal components of susceptibility are linear in respect of anisotropy terms determined by the spin-orbital coupling: 
\begin{eqnarray}
\chi^u_{xy}({\bf k},{\bf k}')\cong\frac{\gamma_{xy}k_F^2}
{4\tilde a_x\tilde a_y}(\hat k_x\hat k_y^\prime+\hat k_x^\prime \hat k_y),
\label{xy1}\\
\chi^u_{xz}({\bf k},{\bf k}')\cong\frac{\gamma_{xz}k_F^2}
{4\tilde a_x\tilde a_z
}(\hat k_x\hat k_z^\prime+\hat k_x^\prime\hat  k_z),
\label{xz1}\\
\chi^u_{yz}({\bf k},{\bf k}')\cong\frac{\gamma_{yz}k_F^2}
{4\tilde a_y\tilde a_z
}(\hat k_y\hat k_z^\prime+\hat k_y^\prime\hat  k_z),
\label{yz1}
\end{eqnarray}
\begin{equation}
\tilde a_x=\alpha_x+\beta_{xz}M_z^2,~~~~~
\tilde a_y=\alpha_y+\beta_{yz}M_z^2,~~~~~\tilde a_z=\alpha_z+6\beta_zM_z^2=4\beta_zM_z^2+\frac{H_z}{2M_z}.
\label{tilde}
\end{equation}
Here we have completely neglected by the quartic terms in respect of  the wave vector components. They have the same symmetry as Eqs. (\ref{xy1})-(\ref{yz1}) but strongly complicate the corresponding
expressions.

\subsection{Pairing amplitudes}

The equations ({\ref{111})-(\ref{117}) express the pairing amplitudes through the susceptibility components in a ferromagnetic metal  
with arbitrary symmetry.
The explicit formula for the susceptibility component in an orthorhombic ferromagnet  are found in the previous section. 
So, the pairing amplitudes in this particular case are
\begin{eqnarray}
&V^{\uparrow\uparrow}({\bf k},{\bf k}')=V^{\downarrow\downarrow}({\bf k},{\bf k}')=-\mu_B^2I^2\chi_{zz}^u
=
-\frac{\mu_B^2I^2k_F^2\gamma_{ij}^z\hat k_i\hat k_j^\prime}{4\left[\beta_z(3M_z^2-M_{z0}^2)+\gamma^z k_F^2\right]^2}
=-V_{1ij}\hat k_i\hat k_j^\prime
,\\
\label{211}
&V^{\uparrow\downarrow}({\bf k},{\bf k}')=-V_{2ij}\hat k_i \hat k_j^\prime+iV_3(\hat k_x\hat k_y^\prime+\hat k_y\hat k_x^\prime)
,\\
\label{22}
&V^{\downarrow\uparrow}({\bf k},{\bf k}')=(V^{\uparrow\downarrow}({\bf k},{\bf k}'))^*
,\\
\label{23}
&V^{00}({\bf k},{\bf k}')=-W_{1ij}\hat k_i \hat k_j^\prime
,\\
\label{24}
&V^{\uparrow0}({\bf k},{\bf k}')=(V^{0\uparrow}({\bf k},{\bf k}'))^*=-W_2(\hat k_x\hat k_z^\prime+\hat k_z\hat k_x^\prime)+iW_3(\hat k_y\hat k_z^\prime+\hat k_z\hat k_y^\prime),\\
\label{25}
&V^{\downarrow0}({\bf k},{\bf k}')=(V^{0\downarrow}({\bf k},{\bf k}'))^*=
-(V^{\uparrow0}({\bf k},{\bf k}'))^*
\label{26}
\end{eqnarray}
Here, the constants are
\begin{eqnarray}
&V_{1ij}=\mu_B^2I^2k_F^2\frac{\gamma_{ij}^z}{a_z^2}=
\frac{\mu_B^2I^2k_F^2\gamma_{ij}^z}{4\left[\beta_z(3M_z^2-M_{z0}^2)+\gamma^z k_F^2\right]^2},
\label{V1}\\
&V_{2ij}=\mu_B^2I^2k_F^2\left (\frac{\gamma_{ij}^x}{a_x^2}-\frac{\gamma_{ij}^y}{a_y^2}  \right ),~~~
V_3=\frac{\mu_B^2I^2k_F^2\gamma_{xy}}{4\tilde a_x\tilde a_y},\\
&W_{1ij}=\frac{\mu_B^2I^2k_F^2}{2}\left (\frac{\gamma_{ij}^x}{a_x^2}+ \frac{\gamma_{ij}^y}{a_y^2}-\frac{\gamma_{ij}^z}{a_z^2}   \right ),~~~
W_2=\frac{\mu_B^2I^2k_F^2\gamma_{xz}}{4\tilde a_x\tilde a_z}
,~~~
W_3=\frac{\mu_B^2I^2k_F^2\gamma_{yz}}{4\tilde a_y\tilde a_z}.
\end{eqnarray}

The pairing  interaction  between the particles in the same  spin-up or spin-down band plays the most important role. The corresponding amplitude originates from the 
 odd part of the magnetic susceptibility component $\chi_{zz}^u$ which is the largest and the 
temperature and the magnetic field dependent component of susceptibility.

The amplitudes $V_{1ij}$ and $W_{1ij}$ are determined mainly by the exchange interaction. 
The amplitude $V_{2ij}$ is equal to zero in the exchange approximation. It has, however,  
non-negligible magnitude corresponding to  the strong enough orthorhombic anisotropy of susceptibility $\chi_{xx}
\ne
\chi_{yy}$.  The amplitudes $V_3, ~W_2,~W_3$ are determined by the spin-orbit  terms in the gradient energy of an orthorhombic ferromagnet, and we will treat them as the smallest amplitudes.

\subsection{Critical temperature of phase transition to the paramagnetic superconducting state in UCoGe}

The equations (\ref{up})-(\ref{0}) are also applicable to determination of the critical temperature of the phase transition from the  normal to the paramagnetic superconducting state taking place in UCoGe at high pressures  (see Fig.1c). This case an internal magnetic field is absent, hence, the normal state Green functions for the spin-up and the spin-down electrons are equal $G^\uparrow=G^\downarrow=G$ and the  order parameter is homogeneous in space. So, the equations take the form
\begin{eqnarray}
\Delta^{\uparrow}({\bf k})
=-T
\sum_{n}
\sum_{{\bf k}' }
\left\{
V^{\uparrow\uparrow}({\bf k},{\bf k}')
\Delta^{\downarrow}({\bf k}')+
V^{\uparrow\downarrow}({\bf k},{\bf k}')
\Delta^{\downarrow}({\bf k}')
+2V^{\uparrow 0}({\bf k},{\bf k}')
\Delta^{0}({\bf k}')
\right\}G({\bf k}',\omega_n)
G(-{\bf k}',-\omega_n),~~~~~
\label{upp}\\
\Delta^{\downarrow}({\bf k})
=-T\sum_{n}\sum_{{\bf k}' }
\left\{
V^{\downarrow\uparrow}({\bf k},{\bf k}')
\Delta^{\uparrow}({\bf k}')+
V^{\downarrow\downarrow}({\bf k},{\bf k}')
 \Delta^{\downarrow}({\bf k}')
+2V^{\downarrow 0}({\bf k},{\bf k}')
\Delta^{0}({\bf k}')
\right\}G({\bf k}',\omega_n)
G(-{\bf k}',-\omega_n),~~~~~
\label{downn}\\
\Delta^{0}({\bf k})
=-T\sum_{n}\sum_{{\bf k}' }
\left\{
2V^{0\uparrow}({\bf k},{\bf k}')\Delta^{\uparrow}({\bf k}')+
V^{0\downarrow}({\bf k},{\bf k}') \Delta^{\downarrow}({\bf k}')
+V^{00}({\bf k},{\bf k}')
\Delta^{0}({\bf k}')
\right\}G({\bf k}',\omega_n)
G(-{\bf k}',-\omega_n).~~~~~
\label{00}
\end{eqnarray}
 Substitution  the paramagnetic state order parameter components  (see  Chapter IIB)
 \begin{equation}
 \Delta^\uparrow=-k_x\eta_x+ik_y\eta_y,~~~\Delta^\downarrow=k_x\eta_x+ik_y\eta_y,~~~  \Delta^0=k_z\eta_z\hat z
 \end{equation}
to these equations gives rise 5 equations for 3 amplitudes $\eta_x,\eta_y,\eta_z$. Two of these equations coincide with two others,
so, the 3 independent equations are 
\begin{eqnarray}
(\lambda^{-1}-g_{1x}+g_{2x})\eta_x+g_{3y}\eta_y +2w_{2z}\eta_z=0,\nonumber\\
g_{3x}\eta_x+(\lambda^{-1}-g_{1y}-g_{2y})\eta_y+2w_{3z}\eta_z=0,
\label{syst}\\
2w_{2x}\eta_x+2w_{3y}\eta_y+(\lambda^{-1}-w_{1z})\eta_z=0.\nonumber
\end{eqnarray}
Here
\begin{eqnarray}
&g_{1x}=V_{1xx}\langle\hat k_x^2N_0({\bf k}) \rangle,~g_{2x}=V_{2xx}\langle\hat k_x^2N_0( {\bf k}) \rangle, ~g_{1y}=V_{1yy}\langle\hat k_y^2N_0({\bf k}) \rangle,~g_{2y}=V_{2yy}\langle\hat k_y^2N_0( {\bf k}) \rangle,~
 w_{1z}=W_{1zz}\langle\hat k_z^2N_0({\bf k}) \rangle,\nonumber\\
&g_{3x}=V_3\langle \hat k_x^2N_0( {\bf k}) \rangle,~g_{3y}=V_3\langle \hat k_y^2N_0( {\bf k}) \rangle,\nonumber\\
&w_{2x}=W_{2}\langle\hat k_x^2N_0( {\bf k}) \rangle,~w_{2z}=W_{2}\langle\hat k_z^2N_0( {\bf k}) \rangle,~
w_{3z}=W_3\langle \hat k_z^2N_0( {\bf k}) \rangle,~w_{3y}=W_3\langle \hat k_y^2N_0( {\bf k}) \rangle\nonumber
\end{eqnarray}
are  the constants of pairing interaction, the angular brackets mean the averaging over the Fermi surface, $N_0( {\bf k})$ is the angular dependent density of electronic states at the Fermi surface. 
The function 
\begin{equation}
\lambda(T)=2\pi T\sum_{n\geq 0}\frac{1}{\omega_n}=\ln\frac{\varepsilon}{T},\nonumber
\end{equation} 
where $\varepsilon=\frac{2\gamma\varepsilon_0}{\pi}$, $\ln\gamma=0.577$ is the Euler constant, and $\varepsilon_0$  is an energy cutoff for the pairing interaction.
The critical temperature of phase transition to the paramagnetic superconducting state is
\begin{equation}
T_{sc}=\varepsilon \exp\left (-\frac{1}{g}\right),
\end{equation}
where  the constant $g$ is the maximal eigen value of the system of equations (\ref{syst}).

\subsection{Phase transition from the paramagnetic to ferromagnetic superconducting state in UCoGe}

The  equations (\ref{upp})-(\ref{00}) are also applicable for the determination of the critical temperature 
of phase transition which has to separate
the paramagnetic superconducting and the ferromagnetic superconducting states in UCoGe (see Fig.1c)
but has not been revealed experimentally. This case, the Green function in this system is the Green function of the paramagnetic superconducting state
\begin{equation}
G({\bf k},\omega_n)=-\frac{i\omega_n+\xi_{\bf k}}{\omega_n^2+\xi_{\bf k}^2+\eta_x^2\hat k_x^2+\eta_y^2\hat k_y^2+\eta_z^2\hat k_z^2}.
\end{equation}
Substitution  the ferromagnetic state order parameter components  (see  Chapter IIB)
 \begin{equation}
 \Delta^\uparrow=-k_x\eta^\uparrow_x+ik_y\eta^\uparrow_y,~~~\Delta^\downarrow=k_x\eta^\downarrow_x+ik_y\eta^\downarrow_y,~~~  \Delta^0=k_z\eta_z^0\hat z
 \end{equation}
to the equations (\ref{upp})-(\ref{00}) gives rise 5 equations for 5 amplitudes $\eta_x^\uparrow,\eta_y^\uparrow,\eta_x^\downarrow,\eta_y^\downarrow,\eta^0_z$. The maximum eigen value of this system determines the critical temperature of phase transition from the paramagnetic  to the ferromagnetic superconducting state. 

This phase transition  occurs in the itinerant electron subsystem. 
Mathematically it is described   by the smooth  development  of an inequality  in the spin-up and the spin-down amplitudes of the order parameter
that is by the development of spontaneous magnetic moment of pure superconducting nature. Simultaneously  the magnetization 
not related to the itinerant electron subsystem but to the subsystem of localized  moments will appear. Its emergence induced by the superconducting electrons magnetic moment 
 is similar to the crossover between paramagnetic and ferromagnetic  normal states under an external magnetic field.  One can expect that, due to the magnetization smallness,  below the transition line the superconductor still remains in the Meissner state.

There is an other possible scenario of phase transition from the paramagnetic to ferromagnetic superconducting state. It is realized when 
the transition driving force is the ordering 
in the subsystem of localized moments. This case the subsystem of superconducting electrons  is  tuned up  to ferromagnetic superconducting state 
due to  the emergency of spontaneous magnetization.

The proper  theory of phase transition from the paramagnetic to the ferromagnetic superconducting state must include 
the effect of  emergency of  supercurrents  and the field dependence of magnetization which can  be important in view of divergency of magnetic susceptibility near the Curie temperature.
A satisfactory treatment of this phenomenon
is at the moment absent \cite{fn5}.

\subsection{Superconducting states in orthorhombic ferromagnets}

Let us find now what kind of superconducting state emerges at phase transition from the normal ferromagnetic to the superconducting ferromagnetic state.
Performing the Taylor expansion of Eqs.(\ref{up})-(\ref{0}) in powers of ${\bf q}$ up to the second order and  and then transforming them  to the coordinate representation, that means simple substitution
\begin{equation}
{\bf q}\to {\bf D}=-i\nabla_{\bf r}+2e{\bf A}({\bf r}),
\end{equation}
we obtain  equations 
\begin{eqnarray}
&\Delta^{\uparrow}({\bf k},{\bf r})
=T\sum_n\int\frac{d^3{\bf k}^\prime}{(2\pi)^3} V_{1ij}\hat k_i\hat k_j^\prime\left(G^\uparrow
G^\uparrow
+\frac{1}{2}G^\uparrow
\frac{\partial^2G^\uparrow
}{\partial k^\prime_l\partial k^\prime_m}D_lD_m
\right)\Delta^{\uparrow}({\bf k}^\prime,{\bf r})
\label{upe1}\\
&+T\sum_{n}\int\frac{d^3{\bf k}^\prime}{(2\pi)^3}[V_{2ij}\hat k_i\hat k_j^\prime-iV_{3}
(\hat k_x\hat k_y^\prime+\hat k_x^\prime \hat k_y)]
\left(G^\downarrow
G^\downarrow
+\frac{1}{2}G^\downarrow
\frac{\partial^2G^\downarrow
}{\partial k^\prime_l\partial k^\prime_m}D_lD_m
\right)\Delta^{\downarrow}({\bf k}^\prime,{\bf r})
\nonumber\\
&+T\sum_n\int\frac{d^3{\bf k}^\prime}{(2\pi)^3}[W_{2}(\hat k_x\hat k_z^\prime+\hat k_z\hat k_x^\prime)-iW_{3}(\hat k_y\hat k_z^\prime+\hat k_z\hat k_y^\prime)]\left(G^\uparrow
G^\downarrow
+\frac{1}{2}G^\uparrow
\frac{\partial^2G^\downarrow
}{\partial k^\prime_l\partial k^\prime_m}D_lD_m
+
G^\downarrow
G^\uparrow
+\frac{1}{2}G^\downarrow
\frac{\partial^2G^\uparrow
}{\partial k^\prime_l\partial k^\prime_m}D_lD_m
\right)\Delta^{0}({\bf k}^\prime,{\bf r}),\nonumber
\end{eqnarray}
\begin{eqnarray}
&\Delta^{\downarrow}({\bf k},{\bf r})=
T\sum_{n}\int\frac{d^3{\bf k}^\prime}{(2\pi)^3}(V_{2ij}\hat k_i\hat k_j^\prime+iV_{3}(\hat k_x\hat k_y^\prime+\hat k_x^\prime \hat k_y))
\left(G^\uparrow
G^\uparrow
+\frac{1}{2}G^\uparrow
\frac{\partial^2G^\uparrow
}{\partial k^\prime_l\partial k^\prime_m}D_lD_m
\right)\Delta^{\uparrow}({\bf k}^\prime,{\bf r})
\label{downe1}\\
&+T\sum_{n}\int\frac{d^3{\bf k}^\prime}{(2\pi)^3} V_{1ij}\hat k_i\hat k_j^\prime\left(G^\downarrow
G^\downarrow
+\frac{1}{2}G^\downarrow
\frac{\partial^2G^\downarrow(
}{\partial k^\prime_l\partial k^\prime_m}D_lD_m
\right)\Delta^{\downarrow}({\bf k}^\prime,{\bf r})
\nonumber
\\
&+T\sum_n\int\frac{d^3{\bf k}^\prime}{(2\pi)^3}
[-W_{2}(\hat k_x\hat k_z^\prime+\hat k_z\hat k_x^\prime)-iW_{3}(\hat k_y\hat k_z^\prime+\hat k_z\hat k_y^\prime)]\left(G^\uparrow
G^\downarrow
+\frac{1}{2}G^\uparrow
\frac{\partial^2G^\downarrow
}{\partial k^\prime_l\partial k^\prime_m}D_lD_m
+G^\downarrow
G^\uparrow
+\frac{1}{2}G^\downarrow
\frac{\partial^2G^\uparrow
}{\partial k^\prime_l\partial k^\prime_m}D_lD_m
\right)\Delta^{0}({\bf k}^\prime,{\bf r})
,\nonumber
\end{eqnarray}
\begin{eqnarray}
&\Delta^{0}({\bf k}^\prime,{\bf r})=T\sum_{n}\int\frac{d^3{\bf k}^\prime}{(2\pi)^3}[W_{2}(\hat k_x\hat k_z^\prime+\hat k_z\hat k_x^\prime)+iW_{3}(\hat k_y\hat k_z^\prime+\hat k_z\hat k_y^\prime)]\left(G^\uparrow
G^\uparrow
+\frac{1}{2}G^\uparrow
\frac{\partial^2G^\uparrow
}{\partial k^\prime_l\partial k^\prime_m}D_lD_m
\right)\Delta^{\uparrow}({\bf k}^\prime,{\bf r})\nonumber\\
&+T\sum_{n}\int\frac{d^3{\bf k}^\prime}{(2\pi)^3}[-W_{2}(\hat k_x\hat k_z^\prime+\hat k_z\hat k_x^\prime)+iW_{3}(\hat k_y\hat k_z^\prime+\hat k_z\hat k_y^\prime)]
\left(G^\downarrow
G^\downarrow
+\frac{1}{2}G^\downarrow
\frac{\partial^2G^\downarrow
}{\partial k^\prime_l\partial k^\prime_m}D_lD_m
\right)\Delta^{\downarrow}({\bf k}^\prime,{\bf r})
\label{0e1}\\
&+T\sum_n\int\frac{d^3{\bf k}^\prime}{(2\pi)^3} W_{1ij}\hat k_i\hat k_j^\prime\left(G^\uparrow
G^\downarrow
+\frac{1}{2}G^\uparrow
\frac{\partial^2G^\downarrow
}{\partial k^\prime_l\partial k^\prime_m}D_lD_m
+G^\downarrow
G^\uparrow
+\frac{1}{2}G^\downarrow\frac{\partial^2G^\uparrow
}{\partial k^\prime_l\partial k^\prime_m}D_lD_m
\right)\Delta^{0}({\bf k}^\prime,{\bf r}).\nonumber
\end{eqnarray}
Here, as before,  the arguments in the Green functions products are
$$
G^\uparrow G^\uparrow=G^\uparrow({\bf k}',\omega_n)
G^\uparrow(-{\bf k}',-\omega_n),~~~G^\uparrow
\frac{\partial^2G^\uparrow
}{\partial k^\prime_l\partial k^\prime_m}=G^\uparrow({\bf k}',\omega_n)
\frac{\partial^2G^\uparrow(-{\bf k}',-\omega_n)
}{\partial k^\prime_l\partial k^\prime_m},~~~...
$$

In the single domain approximation in the absence of external field $H=0$ or at the external field directed along the axis of spontaneous magnetization $\hat z$
the order parameter components are the $z$-coordinate independent and the long derivatives are
\begin{equation}
D_x=-i\frac{\partial}{\partial x},
~~
D_y=-i\frac{\partial}{\partial y}+\frac{2e}{c}(H+ H_{int})x.
\end{equation}
Here, we have  introduced the internal electromagnetic field corresponding  to the spontaneous magnetization $H_{int}=4\pi M$ and ignore  the difference between the external field and the magnetic field induced inside the media by the external field.

Taking into account the wave vector dependence of pairing interaction in the equations (\ref{upe1})-(\ref{0e1}) we can choose the superconducting order parameter as the following linear combinations of the 
momentum  direction projections on the coordinate axis
\begin{eqnarray}
&\Delta^\uparrow({\bf k},{\bf r})=\hat k_x\eta_x^\uparrow({\bf r})+i\hat k_y\eta_y^\uparrow({\bf r})+\hat k_z\zeta_z^\uparrow({\bf r}),\nonumber\\
&\Delta^\downarrow({\bf k},{\bf r})=\hat k_x\eta_x^\downarrow({\bf r})+i\hat k_y\eta_y^\downarrow({\bf r})+\hat k_z\zeta_z^{\downarrow}({\bf r}),\nonumber\\
&\Delta^0{\bf k},{\bf r})=\hat k_x\zeta_x^0({\bf r})+i\hat k_y\zeta_y^0({\bf r})+\hat k_z\eta_z^0({\bf r}).\nonumber
\end{eqnarray}
One can check that substitution of these expressions to Eqs. (\ref{upe1})-(\ref{0e1}) leads  to  two independent systems of the differential equations 
\begin{equation}
\eta_\alpha({\bf r})=A_{\alpha\beta}\eta_\beta({\bf r}),~~~~\zeta_\alpha({\bf r})=B_{\alpha\beta}\zeta_\beta({\bf r})
\label{AB}
\end{equation}
for the components of vectors 
\begin{equation}
\eta_\alpha({\bf r})=(\eta_x^\uparrow({\bf r}),\eta_x^\downarrow({\bf r}),\eta_y^\uparrow({\bf r}),\eta_y^\downarrow({\bf r}),\eta_z^0({\bf r}))
\end{equation}
and
\begin{equation}
\zeta_\alpha({\bf r})=(\zeta_z^\uparrow({\bf r}),\zeta_z^\downarrow({\bf r}),\zeta_x^0({\bf r}),\zeta_y^0({\bf r}))
\end{equation}
corresponding to {\bf two different superconducting states with different critical temperatures  relating to co-representations A and B. Thereby, the derived microscopic equations confirm the conclusions made in Chapter II  from the pure symmetry considerations.}

\subsection{Equal-spin-pairing states}

In what follows we will work with Eqs. (\ref{up1}) and  (\ref{down1}) corresponding to the equal-spin-pairing superconductivity.
This case, the four component state A is 
\begin{eqnarray}
&\Delta^\uparrow({\bf k},{\bf r})=\hat k_x\eta_x^\uparrow({\bf r})+i\hat k_y\eta_y^\uparrow({\bf r})
,
\label {Aup}\\
&\Delta^\downarrow({\bf k},{\bf r})=\hat k_x\eta_x^\downarrow({\bf r})+i\hat k_y\eta_y^\downarrow({\bf r})
\label{Adown}
\end{eqnarray}
and the   two component state B is 
\begin{eqnarray}
&\Delta^\uparrow({\bf k},{\bf r})=
\hat k_z\zeta_z^\uparrow({\bf r}),
\label{Bup}\\
&\Delta^\downarrow({\bf k},{\bf r})=
\hat k_z\zeta_z^{\downarrow}({\bf r}).
\label{Bdown}
\end{eqnarray}
The corresponding equations  (\ref{AB}) for  the critical temperatures of A and B states are determined by the following 4x4 and 2x2 matrices
\begin{widetext}
\begin{eqnarray}
A_{\alpha\beta} =
 \left(\begin{array}{cccc}
 \medskip
g^\uparrow_{1x}\lambda+L_{1x}^\uparrow &
g_{2x}^\downarrow\lambda+L_{2x}^\downarrow+iL_{3yx}^\downarrow & 
iL_{1xy}^\uparrow & 
-g_{3y}^\downarrow\lambda+iL_{2xy}^\downarrow-L_{3y}^\downarrow
\\
\medskip
g_{2x}^\uparrow\lambda+L_{2x}^\uparrow-iL_{3yx}^\uparrow & 
g^\downarrow_{1x}\lambda+L_{1x}^\downarrow  & 
g_{3y}^\uparrow\lambda+iL_{2xy}^\uparrow +L_{3y}^\uparrow& 
iL_{1xy}^\downarrow
 \\
\medskip
-iL_{1yx}^\uparrow & 
g_{3x}^\downarrow\lambda-iL_{2yx}^\downarrow+L_{3x}^\downarrow &  
g_{1y}^\uparrow \lambda+L_{1y}^\uparrow& 
g_{2y}^\downarrow\lambda+L_{2y}^\downarrow+iL_{3xy}^\downarrow 
 \\
\medskip
-g_{3x}^\uparrow\lambda-iL_{2yx}^\uparrow-L_{3x}^\uparrow& 
-iL_{1yx}^\downarrow & 
g_{2y}^\uparrow\lambda+L_{2y}^\uparrow-iL_{3xy}^\uparrow & 
g_{1y}^{\downarrow}\lambda+L_{1y}^\downarrow 
\end{array} \right),
\label{matrixA}
\end{eqnarray}
\begin{eqnarray}
B_{\alpha\beta} =
 \left(\begin{array}{cc}
 g_{1z}^\uparrow\lambda+L_{1z}^\uparrow & 
 g_{2z}^\downarrow\lambda+L_{2z}^\downarrow 
\\
\medskip
 g_{2z}^\uparrow\lambda+L_{2z}^\uparrow & 
 g_{1z}^\downarrow\lambda+L_{1z}^\downarrow 
 \end{array} \right).
 \label{matrixB}
\end{eqnarray}
\end{widetext}

Here, 
\begin{equation}
g_{1x}^\uparrow=V_{1xx}\langle\hat k_x^2N_0^\uparrow({\bf k}) \rangle=\frac{\mu_B^2I^2k_F^2\gamma_{xx}^z\langle\hat k_x^2N_0^\uparrow({\bf k}) \rangle}{4
\left[\beta_z(3M_z^2-M_{z0}^2)+\gamma^z k_F^2\right]^2}
\label{g1}
\end{equation}
is one of  the constants of pairing interaction, the angular brackets mean the averaging over the Fermi surface, $N_0^\uparrow( {\bf k})$ is the angular dependent density of electronic states at the Fermi surface of the band $\uparrow$. Correspondingly
\begin{equation}
g_{2x}^{\downarrow}=V_{2xx}\langle\hat k_x^2N_0^\downarrow( {\bf k}) \rangle,~~~~~~
g_{3x}^{\downarrow}=V_3\langle \hat k_x^2N_0^\downarrow( {\bf k}) \rangle.
\end{equation}
All the other constants of pairing interaction  are obtained by the obvious substitutions $x\leftrightarrow y$ and $\uparrow \leftrightarrow\downarrow$ or $x\rightarrow z$. 

The function 
\begin{equation}
\lambda(T)=2\pi T\sum_{n\geq 0}\frac{1}{\omega_n}=\ln\frac{\epsilon}{T},
\end{equation} where $\epsilon=\frac{2\gamma\varepsilon_0}{\pi}$, $\ln\gamma=0.577$ is the Euler constant, and $\varepsilon_0$  is an energy cutoff for pairing interaction. We assume here that it has the same value for both bands. 

The first type of differential operators is defined as follows
\begin{equation}
L_{1x}^\uparrow=\frac{1}{2}V_{1xx}T\sum_n\int\frac{d^3{\bf k}}{(2\pi)^3}  \hat k_x^2
G^\uparrow({\bf k},\omega_n){\cal D}^{\uparrow},
\end{equation}
and $L_{2y}^\downarrow$ and the other operators with  same structure are obtained by obvious substitutions $(x\rightarrow y,z)$, $(1\rightarrow2)$ and $(\uparrow\rightarrow\downarrow)$, 
but  similar operator with index 3 is 
\begin{equation}
L_{3x}^\uparrow=\frac{1}{2}V_3T\sum_n\int\frac{d^3{\bf k}}{(2\pi)^3} \hat k_y^2
G^\uparrow({\bf k},\omega_n){\cal D}^{\uparrow},
\end{equation}
here,
\begin{equation}
{\cal D}^{\uparrow}=
\frac{\partial^2G^\uparrow(-{\bf k},-\omega_n)}{\partial k_x^2 }D_x^2+\frac{\partial^2G^\uparrow(-{\bf k},-\omega_n)}{\partial k_y^2 }D_y^2
.
\end{equation}
The second type of operators is
\begin{equation}
L_{1xy}^\uparrow=\frac{1}{2}V_{1xx}T\sum_n\int\frac{d^3{\bf k}}{(2\pi)^3} \hat k_x\hat k_y
G^\uparrow({\bf k},\omega_n)\frac{\partial^2G^\uparrow(-{\bf k},-\omega_n)}{\partial k_x\partial k_y }(D_xD_y+D_yD_x),
\end{equation}
and $L_{2yx}^\uparrow$ and the others operators of this type are obtained by obvious substitutions $(x\rightarrow y)$, $(1\rightarrow 2)$, $(\uparrow\rightarrow\downarrow)$. Similar  operators with index 3 is defined  as
\begin{equation}
L_{3xy}^\uparrow=\frac{1}{2}V_3T\sum_n\int\frac{d^3{\bf k}}{(2\pi)^3}\hat  k_x \hat k_y
G^\uparrow({\bf k},\omega_n)\frac{\partial^2G^\uparrow(-{\bf k},-\omega_n)}{\partial k_x\partial k_y }(D_xD_y+D_yD_x),
\end{equation}

\subsection{Equal-spin-pairing states near critical temperature}

As we already mentioned the internal field acting on the electron charges in the uranium ferromagnets is much smaller than the upper critical field at zero temperature. In this case the gradient terms  produce only small of order of 
$ O(\frac{H_{int}}{H_{c2}(T=0)})$ correction  to the eigenvalues of linear differential equations (\ref{AB}) for the order parameter components.
 Then, these equations
 are transformed into the algebraic equations:
\begin{eqnarray}
\eta_x^\uparrow=(g_{1x}^\uparrow\eta_x^\uparrow+g_{2x}^{\downarrow}\eta_x^\downarrow+g_{3y}^{\downarrow}\eta_y^\downarrow)\lambda,
\nonumber\\
\eta_x^\downarrow=(g_{2x}^{\uparrow}\eta_x^\uparrow+g_{1x}^\downarrow\eta_x^\downarrow-g_{3y}^{\uparrow}\eta_y^\uparrow
)\lambda,
\\
\eta_y^\uparrow=(g_{1y}^\uparrow\eta_y^\uparrow+g_{2y}^{\downarrow}\eta_y^\downarrow-g_{3x}^{\downarrow}\eta_x^\downarrow)\lambda,
\nonumber\\
\eta_y^\downarrow=(g_{2y}^{\uparrow}\eta_y^\uparrow+g_{1y}^\downarrow\eta_y^\downarrow+g_{3x}^{\uparrow}\eta_x^\uparrow
)\lambda.
\nonumber
\end{eqnarray} 
for the A state and 
\begin{eqnarray}
\zeta_z^\uparrow=(g_{1z}^\uparrow\zeta_z^\uparrow+g_{2z}^{\downarrow}\zeta_z^\downarrow)\lambda,
\\
\zeta_z^\downarrow=(g_{2z}^{\uparrow}\zeta_z^\uparrow+g_{1z}^\downarrow\zeta_z^\downarrow
)\lambda.
\nonumber
\end{eqnarray} 
for the B state. 
 Taking into account that 
 the constants of interaction with indices 1,2 strongly exceed 
the constants with index 3  originating from the spin-orbit terms in the gradient energy
$$
g_1,~g_2~>>~g_3,
$$
we come to three independent systems of equations for the  x, y and z components of the order parameter in two band superconductor
\begin{eqnarray}
\eta_x^\uparrow=(g_{1x}^\uparrow\eta_x^\uparrow+g_{2x}^{\downarrow}\eta_x^\downarrow)\lambda,
\nonumber\\
\eta_x^\downarrow=(g_{2x}^{\uparrow}\eta_x^\uparrow+g_{1x}^\downarrow\eta_x^\downarrow
)\lambda,
\label{Ax}
\end{eqnarray}
\begin{eqnarray}
\eta_y^\uparrow=(g_{1y}^\uparrow\eta_y^\uparrow+g_{2y}^{\downarrow}\eta_y^\downarrow)\lambda,
\nonumber\\
\eta_y^\downarrow=(g_{2y}^{\uparrow}\eta_y^\uparrow+g_{1y}^\downarrow\eta_y^\downarrow
)\lambda,
\label{Ay}
\end{eqnarray}
and
\begin{eqnarray}
\zeta_z^\uparrow=(g_{1z}^\uparrow\zeta_z^\uparrow+g_{2z}^{\downarrow}\zeta_z^\downarrow)\lambda,
\nonumber\\
\zeta_z^\downarrow=(g_{2z}^{\uparrow}\zeta_z^\uparrow+g_{1z}^\downarrow\zeta_z^\downarrow
)\lambda.
\label{Bz}
\end{eqnarray}
Thus, in the exchange approximation for the energy of magnetic inhomogeneity we have three different superconducting states $(\hat k_x\eta_x^\uparrow,~\hat k_x\eta_x^\downarrow),$  $(\hat k_y\eta_y^\uparrow,~\hat k_y\eta_y^\downarrow)$ and $(\hat k_z\zeta_z^\uparrow,~\hat k_z\zeta_z^\downarrow)$ with different critical temperatures defined by the determinants of Eqs.  (\ref{Ax}), (\ref{Ay}) and (\ref{Bz}). 

\section{physical properties}

\subsection{Critical temperature}

Assuming that the largest critical temperature corresponds to the $(\hat k_x\eta_x^\uparrow,~\hat k_x\eta_x^\downarrow)$ superconducting state the zero of determinant of the system (\ref{Ax}) yields the BCS-type formula
\begin{equation}
T=\varepsilon~exp\left (-\frac{1}{g}  \right ),
\label{21}
\end{equation}
where the constant of interaction
\begin{equation}
g=\frac{g_{1x}^\uparrow+g_{1x}^\downarrow}{2}+\sqrt{\frac{(g_{1x}^\uparrow-g_{1x}^\downarrow)^2}{4}+g_{2x}^{\uparrow}g_{2x}^{\downarrow}}
\end{equation}
is the function of temperature and  magnetic field. Thereby the formula (\ref{21}) is, in fact, an equation for the determination of the critical temperature of the transition to the superconducting state. Let us look on it in the easiest case of  a single-band (say spin-up) superconducting state when $g=g_{1x}^\uparrow$.

In URhGe the transition to the superconducting state occurs at temperature much lower than the Curie temperature. Hence, one can
neglect  by the temperature dependence of the constant of interaction. 
Then  the critical temperature   is determined by 
\begin{equation}
\ln\frac{\varepsilon}{T_{sc}}\cong\frac{1}{g_{1x}^\uparrow}\propto\frac{\left(\alpha_{0z}T_{c}+\gamma_zk_F^2\right )^2}{\mu_B^2I^2\gamma_{xx}^zk_F^2\langle\hat k_x^2N_0^\uparrow({\bf k})\rangle},
\label{22}
\end{equation}
where we have used Eqs. (\ref{g1}), (\ref{az})  for $g_{1x}^\uparrow$ in the absence of magnetic field.
The Curie temperature  $T_c$  in URhGe is an increasing function of pressure (see Fig.1b). Pressure dependence of all other quantities in the right hand side in this equation is unknown. In assumption that the right hand side  as whole is also increased with pressure we see
that this should be accompanied by the slow decrease of the temperature of transition to the superconducting state. And vice versa, when the right hand side decreases with pressure, the $T_{sc}({P})$  is grow up. The first obviously corresponds to the observed pressure dependences  $T_c({P})$ and $T_{sc}({P})$ in URhGe, the second one to the situation in UCoGe( see Fig.1c). In the latter case, of course, this argumentation is applicable to the pressure dependences in the region where $T_{sc}$ is significantly smaller than $T_c$.

We do not consider here UGe$_2$ where the superconducting state   arises in the phase diagram region below the line of the first-order transition from the paramagnet to the ferromagnet state.

\subsection{Upper critical field parallel to c-axis in UCoGe}

The upper critical field $H_{c2}(T)$ parallel to the axis of spontaneous magnetization in UCoGe \cite{Taupin} possesses  the definite upward curvature in the slope
near the critical temperature (Fig.4). The natural explanation of this phenomenon is that the critical temperature itself is a function of external field.
Indeed, near the critical temperature $T_{sc}$  the upper critical field is 
\begin{equation}
H_{c2}=AT_{sc}(T_{sc}-T),
\label{H}
\end{equation}
where $A\approx\frac{\phi_0}{v_F^2}$ is a constant  and the critical temperature is the function of pairing amplitude    $T_{sc}=\epsilon\exp(-\frac{1}{g})$.  Again, assuming that the largest critical temperature corresponds to the $(\hat k_x\eta_x^\uparrow,~\hat k_x\eta_x^\downarrow)$ superconducting state,   we have in the single band approximation
\begin{equation}
\ln \frac{\varepsilon}{T_{sc}}=\frac{1}{g_{1x}^\uparrow}\propto\left [\beta_z(3M_z^2-M_{z0}^2)+\gamma^z k_F^2\right]^2.
\label{H}
\end{equation}
At temperatures well below the Curie temperature the magnetization is almost temperature independent. 
On the other hand the UCoGe  magnetic moment under the field in $c$ direction  quite rapidly increases \cite{Huy2008}. At fields  about 1 Tesla $M_z=M_z(H)$ is about twice larger than $M_{z0}=M_z(H=0)$. Hence, in accordance with (\ref{H}),
the magnetic  field increase decreases the constant of interaction $g_{1x}^\uparrow$ and the critical temperature $T_{sc}(g_{1x}^\uparrow)$.

The temperature dependence of the upper critical field (\ref{H}) can be rewritten as the field dependence of the temperature of phase transition to the superconducting state
 $T_{sc}^{orb}$
determined  by the orbital effect and  by the  field dependence of pairing interaction $g_{1x}^\uparrow=g_{1x}^\uparrow(H)$:
\begin{equation}
T_{sc}^{orb}=T_{sc}(g_{1x}^\uparrow)-\frac{H}{AT_{sc}(g_{1x}^\uparrow)}.
\end{equation}
Obviously, the  field dependence $T_{sc}(g_{1x}^\uparrow(H))$ not only shifts down  the linear field dependence of $T_{sc}^{orb}(H)$  but also creates  an upward curvature in accordance with the experimental data shown in Fig.4.

In  URhGe the temperature dependence of the upper critical field parallel to spontaneous magnetization (see Fig.5) does not reveal an upward curvature \cite{Hardy2005}.  Unlike  to UCoGe  in this material the change of magnetic moment in the field $H_z$ 
 smaller than 1 Tesla is negligibly small \cite{Hardy2012}.
Hence, the field dependence of the pairing constant  plays no role.

\subsection{Upper critical field in URhGe}

The superconducting critical temperature in all uranium ferromagnets increases with the sample quality  as it should be  in unconventional superconducting states 
where the $T_{sc}(l)$ dependence  from electron mean free path  at $l>\xi_0$ is described by \cite{Book}
\begin{equation}
T_{sc}\approx T_{sc0}-\frac{\pi v_F}{8l}
\end{equation} 
and the zero temperature upper critical field increases with the sample purity as square of the critical temperature
\begin{equation}
H_{c2}\approx\frac{\phi_0}{\pi\xi_0^2}\propto T_{sc}^2.
\end{equation}
The latter relation has been demonstrated by measurements of upper critical field in URhGe on the samples of different quality \cite{Hardy2005} (Fig.5).
 
An other peculiar property revealed by Hardy and Huxley \cite{Hardy2005} is the temperature dependence of the upper critical field anisotropy.
The ratio of $H_{c2}(T)$ along the $c$ axis to that along the $b$
axis is independent of temperature. However, the ratio of
$H_{c2}(T)$ parallel to the $a$ axis divided by the value along the $b$
axis (or $ c$ axis) increases linearly by approximately 20\%  as
the temperature is decreased from $T_c$ to zero (Fig.6). 
This behavior is
consistent with a choice of an equal-spin-paired gap having
a line node in the $bc$ plane.
Namely, working in the single-band approximation and taking the order parameter amplitude at $H=0$ as 
$$ \Delta^\uparrow({\bf k}, {\bf r})=\eta_x^\uparrow k_x
,$$
one can show \cite{Scharnberg1985} that  the solutions of the linear Gor'kov equations corresponding to the maximal upper critical field for different field directions are
\begin{eqnarray}
&H\parallel b&~~~~~~\Delta^\uparrow({\bf k},{\bf r})\sim A(H,T)(k_x+ik_z)\psi_0(x,z)+B(H,T)(k_x-ik_z)\psi_2(x,z),\\
&H\parallel c&~~~~~~\Delta^\uparrow({\bf k},{\bf r})\sim A(H,T)(k_x+ik_y)\psi_0(x,y)+B(H,T)(k_x-ik_y)\psi_2(x,y),\\
&H\parallel a&~~~~~~\Delta^\uparrow({\bf k},{\bf r})\sim k_x\psi_0(y,z),
\end{eqnarray}
where $\psi_n(x,y)$ are  the Landau functions of particle with charge $2e$ in a magnetic field, $n$ is the Landau level number, and the coefficients $A(H,T)$ and $B(H,T)$ are the functions of magnetic field and temperature.
We see, that the solutions for the field along  the $c$ and the $b$ axes  have the same structure and differ from the solution for the field along the $a$ direction,
what naturally  explains the observed temperature dependence of the upper critical field anisotropy.  

This property is still valid in a multi-band superconductor with  equal-spin-pairing  if we  assume (as we did in the single band case) that our superconducting state is the particular A-state such that
the order parameter spin-up and spin-down 
amplitudes at zero field in different bands  have the form:
\begin{equation}
 \Delta^\uparrow({\bf k}, {\bf r})=\eta_x^\uparrow k_x,~~~~ \Delta^\downarrow({\bf k}, {\bf r})=\eta_x^\downarrow k_x.
 \label{D}
\end{equation}
 Thus, the observed behavior of the temperature dependence of upper critical field anisotropy strongly
points  on the  preferable order parameter structure in URhGe.

\subsection{Zeros in spectrum and specific heat at low temperatures}

As we already pointed out  even in the absence  of an external field in a ferromagnetic superconductor there is an internal field  $H_{int}$ acting on the electron charges.
The internal magnetic field in all uranium ferromagnets is larger than the lower critical field $H_{c1}$. Hence, the Meissner state is absent and the superconducting state is always the
 Abrikosov mixed state with space inhomogeneous distributions of the order parameter and the internal magnetic field. At low temperatures, when due to $H_{int}<<H_{c2}$ the distance between vortices 
 is much larger than the core radius,  one can separate the  specific heat inputs arising from the vortex cores and the inter-vortex space.
 
 It is usually accepted to operate with the ratio $C/T=\gamma$ which is  in a normal metal directly proportional to the electron density of states.  The cores contribution to the specific heat is  due to the almost gapless excitations localized in the vortex cores.
 Hence, due to the vortex cores  $\gamma$  keeps a finite value in the superconducting state at low temperatures
 \begin{equation}
 \gamma_v\approx\frac{H_{int}}{H_{c2}}\gamma_N,
 \end{equation}
where $\gamma_N$ is  the normal state value of $\gamma$.

Another contribution to the density of states originates from so called the Volovik effect \cite{Volovik1993} taking place in the inter-vortex space.  This case, the energy of excitations is 
 given by Eqs. (\ref{uparrow}) and (\ref{downarrow}).
In the absence an additional phase transition inside of  superconducting state
the order parameter of the superconducting state belonging  to A co-representation is given by  Eqs. (\ref{Aup}), (\ref{Adown}), or to B co-representation is  given by Eqs.(\ref{Bup}), (\ref{Bdown}).
The A-state order parameter is equal to zero in isolated points $k_x=k_y=0$, hence the inter-vortex space  contribution to the density of states is given by \cite{Volovik1993}
 \begin{equation}
 \gamma^A_{iv}\approx\frac{H_{int}}{H_{c2}}\ln\left (\frac{H_{c2}}{H_{int}}\right )\gamma_N.
 \label{A4}
 \end{equation}
The B-state order parameter is equal to zero at line $k_z=0$, hence the inter-vortex space  contribution to the density of states is given by \cite{Volovik1993}
 \begin{equation}
 \gamma^B_{iv}\approx\sqrt{\frac{H_{int}}{H_{c2}}}\gamma_N.
 \label{B444}
 \end{equation}
 As we pointed out in the previous Chapter the mixing of $x$ and $y$ component of the order parameter in A-state is in fact quite small, owing to the smallness of $V_3$ amplitude of pairing. So, the gap in A state spectrum is almost equal to zero either at line $k_x=0$, or at line $k_y=0$. Due to this reason the inter-vortex contribution 
 to the density of states in the A-state can be given by the same square-root formula as for the B state.
 
The Eqs.(\ref{A4}), (\ref{B444}) are applicable to the defect free superconducting crystals.  In presence of inhomogeneities created by impurities,  dislocations, domain walls  the gap in the quasiparticle spectrum is suppressed in finite vicinity of the order parameter zeros \cite{Book} as well as in the finite vicinity of the inter-domain walls.  As result, the zero energy density of states acquires a  field independent contribution.   At high enough impurity concentration the square root field dependence   \label{B4}  can  also be modified \cite{Barash}.

Qualitatively the total low temperature $\gamma_0$ value for superconducting ferromagnets at moderate impurities amount is described by
\begin{equation}
\gamma_0=\gamma_{dw}+ \gamma_{iv}+\gamma_v\approx\left (a +\sqrt{\frac{H_{int}}{H_{c2}}}
+\frac{H_{int}}{H_{c2}} \right )\gamma_N,
 \label{Cc}
 \end{equation}
where constant $a<<1$.

One can estimate magnitude of internal field  as
\begin{equation}
H_{int}=const\frac{\mu_u}{a_{uu}^3},
\label{Hint}
\end{equation}
where $\mu_u$ is the magnetic moment per uranium atom, $a_{uu}$ is the distance between nearest neighbor uranium atoms.  The inter-uranium distances in UCoGe, URhGe and UGe$_2$ 
are close to each other. On the other hand, the corresponding zero temperature magnetic moments 0.05$\mu_B$, 0.4$\mu_B$ and $\mu_B$ are quite different  what determines the difference in $H_{int}$ in these materials. The indeterminacy is introduced by  the unknown pre-factors in Eq. (\ref{Hint}). 
Another  way to determine the internal field is just to take it  equal to the
external field  along the direction of spontaneous magnetization that suppresses  ferromagnet many domain structure. 

The internal field  estimated in review \cite{Aoki14}  is about 100G  for UCoGe,  800G for URhGe and 
2800 G for UGe$_2$ according to higher value of magnetic moment in this material.  The  zero temperature upper critical field directed along spontaneous magnetization in UCoGe  is 1.2 T ,  in UGe$_2$ it is approximately 2.2 T.  Known value of $H_{c2}$ for URhGe 
is 0.6 T has been measured, however,  in low RRR=21 single crystal. So, one can expect that the real value of  low temperature upper critical field  in URhGe is roughly the same as in UCoGe. 
Thus, the field depend part of  ratio $\gamma_0/\gamma_N$ found making use Eq. (\ref{Cc})  approximately  is 0.1 for UCoGe, 0.3 for URhGe and 0.5 for UGe$_2$.   Corresponding experimentally established values are presented in Fig.7.

\section{Reentrant superconductivity in UR${\bf h}$G${\bf e}$}

URhGe has   a peculiar property.  At low enough temperature  the magnetic field about 1.3 Tesla 
 directed along the $b$-axis suppresses the superconducting state \cite{Hardy2005} but at much higher field of about 10 Tesla the superconductivity is recreated and exists 
 till the field about 13 Tesla.\cite{Levy2005} The  maximum of the superconducting critical temperature in this field interval is $\approx 0.4~K$. In the same field interval the material transfers from the ferromagnet to the paramagnet state by means of the first-order type transition. The superconducting state exists not only inside of the ferromagnetic state but also in the paramagnetic state separated from the ferromagnetic state by the phase transition of the first order (Fig.8).

The observation of the abrupt collapse  of spontaneous magnetization under a strong enough external field along the $b$ axis has been reported already in the first publication about magnetic field-induced superconductivity in ferromagnet URhGe.\cite{Levy2005} Recently, the first-order character of transition has been confirmed by the direct observation of hysteresis \cite{AokiKnebel2014} in the Hall resistivity in the vicinity of the transition field $H_R\sim 12.5~T$.

In this Chapter  we develop the Landau type phenomenological description of the ferromagnet-paramagnet phase transition under an external magnetic field perpendicular to spontaneous magnetization.
We   find the components of magnetic susceptibility determining the superconducting pairing interaction and show that the magnetic susceptibility corresponding to longitudinal magnetic fluctuations strongly increases  in the vicinity of the first-order transition stimulating the reentrance of the superconducting state.

\subsection{Phase transition in orthorhombic ferromagnet under  magnetic field perpendicular to spontaneous magnetization}

The Landau free energy of an orthorhombic ferromagnet 
in
magnetic field ${\bf H}({\bf r})$ is
\begin{equation}
{\cal F}=\int d V(F_M+F_\nabla),
\label{FE1}
\end{equation}
where in 
\begin{eqnarray}
F_M=\alpha_{z}M_{z}¥^{2}+\beta_{z}¥M_{z}¥^{4}+\delta_zM_z^6~~~~~~~~~~~~~~~~~\nonumber\\
 +\alpha_{y}M_{y}^{2}+\alpha_{x}M_{x}¥^{2}+\beta_{xy}M_x^2M_y^2+\beta_{yz}¥M_{z}¥^{2}¥M_{y}¥^{2}¥+\beta_{xz}¥M_{z}¥^{2}¥M_{x}¥^{2}-{\bf M}{\bf  H},
\label{F11}
\end{eqnarray}
we bear in mind the orthorhombic anisotropy and also the term of the sixth order in powers of $M_z$.
The density of gradient energy is  taken in the exchange approximation,
\begin{equation}
F_\nabla=\gamma_{ij}\frac{\partial {\bf M}}{\partial x_i}\frac{\partial {\bf M}}{\partial x_j}.
\label{nabla1}
\end{equation}
Here,  $x, y, z$ are the coordinates
 pinned to the $a, b, c$
crystallographic directions correspondingly, 
$a, b, c$,
\begin{equation}
\alpha_{z}=\alpha_{z0}(T-T_{c0}), \alpha_x>0,~~ \alpha_y>0 \end{equation}
and 
\begin{equation}
\gamma_{ij} = \left(\begin{array}{ccc} \gamma_{xx} & 0 & 0\\
0 & \gamma_{yy} & 0 \\
0 & 0 & \gamma_{zz}
\end{array} \right).
\end{equation}

In constant magnetic field ${\bf H}=H_y\hat y$
the equilibrium magnetization projections along the $x,y$ directions are obtained by minimization of free energy (\ref{F11}) in respect of $M_x,M_y$
\begin{equation}
M_x=0,~~~~M_y=\frac{H_y}{2(\alpha_y+\beta_{yz}M_z^2)}.
\label{My}
\end{equation}
 Substituting these expressions back to (\ref{F11})
we obtain
\begin{eqnarray}
F_M=\alpha_{z}M_{z}¥^{2}
+\beta_{z}M_{z}^{4}+\delta_zM_z^6-\frac{1}{4}\frac{H_y^2}{\alpha_y+\beta_{yz}M_z^2},
\label{F1}
\end{eqnarray}
that gives after expansion of the denominator in the last term, 
\begin{equation}
F_M=-\frac{H_y^2}{4\alpha_y}+\tilde\alpha_{z}M_{z}¥^{2}
+\tilde\beta_{z}¥M_{z}¥^{4}+\tilde\delta_zM_z^6+\dots,
\label{F2}
\end{equation}
where
\begin{eqnarray}
&\tilde\alpha_{z}=\alpha_{z0}(T-T_{c0})+\frac{\beta_{yz}H_y^2}{4\alpha_y^2},\\
&\tilde\beta_{z}=\beta_z-\frac{\beta_{yz}}{\alpha_y}\frac{\beta_{yz}H_y^2}{4\alpha_y^2}\label{beta}\\
&\tilde\delta_{z}=\delta_z+\frac{\beta_{yz}^2}{\alpha_y^2}\frac{\beta_{yz}H_y^2}{4\alpha_y^2}
\end{eqnarray}

We see that under a magnetic field perpendicular to the direction of spontaneous magnetization  the Curie temperature decreases as
\begin{equation}
T_c=T_c(H_y)=T_{c0}-\frac{\beta_{yz}H_y^2}{4\alpha_y^2\alpha_{z0}}.
\label{Cur}
\end{equation}
The coefficient $\tilde\beta_z$ also decreases with $H_y$ and reaches  zero at
\begin{equation}
H_y^{cr}=\frac{2\alpha_y^{3/2}\beta_z^{1/2}}{\beta_{yz}}.
\end{equation}
At this field under fulfillment the  condition,
\begin{equation}
\frac{\alpha_{z0}\beta_{yz}T_{c0}}{\alpha_y\beta_z}>1
\end{equation}
the Curie temperature (\ref{Cur}) is still positive and at $$H_y>H_y^{cr}$$  phase transition from a paramagnetic to a ferromagnetic state becomes the transition of the first order
(Fig.9). The point $(H_y^{cr},T_c(H_y^{cr}))$ on the line  paramagnet-ferromagnet phase transition is a tricritical point.

The minimization of the free energy Eq. (\ref{F2})  gives the value of the order parameter in the ferromagnetic state,
\begin{equation}
M_z^2=\frac{1}{3\tilde\delta_z}[-\tilde\beta_z+\sqrt{\tilde\beta_z^2-3\tilde\alpha_z\tilde\delta_z}].
\end{equation}
The minimization of the free energy in the paramagnetic state,
\begin{equation}
F_{para}=\alpha_yM_y^2-H_yM_y
\label{p}
\end{equation}
in respect $M_y$ gives the equilibrium value of magnetization projection on axis $y$ in paramagnetic state,
\begin{equation}
M_y=\frac{H_y}{2\alpha_y}.
\end{equation}
Substitution back in Eq. (\ref{p}) yields the equilibrium value of free energy in the paramagnetic state,
\begin{equation}
F_{para}=-\frac{H_y^2}{4\alpha_y}.
\end{equation}
On the line of the phase transition of the first order from the paramagnetic to ferromagnetic state
determined by  the equations \cite{StatPhysI}
\begin{equation}
F_M=F_{para},~~~~~\frac{\partial F_M}{\partial M_z}=0
\end{equation}
the order parameter $M_z$ has the  jump (Fig.10) from
\begin{equation}
 M_z^{\star^2}=-\frac{\tilde\beta_z}{2\tilde\delta_z}.
 \label{jump}
\end{equation}
in the ferromagnetic state to zero in the paramagnetic state.
Its substitution  back in  equation $F_M=F_{para}$ gives the equation of the first-order transition line,
\begin{equation}
4\tilde\alpha_z\tilde\delta_z=\tilde\beta_z^2,
\label{line}
\end{equation}
that is
\begin{equation}
T^\star=T^\star(H_y)=T_{c0}-\frac{\beta_{yz}H_y^2}{4\alpha_y^2\alpha_{z0}}+\frac{\tilde\beta_z^2}{4\alpha_{z0}\tilde\delta_z}.
\end{equation}

The corresponding  jump of $M_y$ (see Fig.10) is given by 
\begin{equation}
M_y^\star=M_y^{ferro}-M_y^{para}=\frac{H_y}{2(\alpha_y+\beta_{yz}M_z^{\star^2})}-\frac{H_y}{2\alpha_y}.
\label{jumpy}
\end{equation}

\subsection{Susceptibilities}

In a perpendicular field
magnetic susceptibilities along all directions are found in the same manner as this has been  performed in the Section IIIB for the case parallel field.
In the ferromagnetic state $T<T^\star$ they are
\begin{eqnarray}
&\chi^f_{xx}({\bf k})\cong \frac{1}{2(\alpha_x+\beta_{xz} M_z^2+\beta_{xy}M_y^2+\gamma_{ij}k_ik_j)},\nonumber\\
&\chi^f_{yy}({\bf k})\cong \frac{1}{2(\alpha_y+\beta_{yz}M_z^2+\gamma_{ij}k_ik_j)},
\label{sf}\nonumber\\
&\chi^f_{zz}({\bf k})\cong \frac{1}{2(\alpha_z+6\beta_zM_z^2+15\delta_zM_z^4+\beta_{yz}M_y^2+\gamma_{ij}k_ik_j)}= 
\frac{1}{2(4\beta_zM_z^2+12\delta_zM_z^4+\gamma_{ij}k_ik_j)}.
\end{eqnarray}
In the paramagnetic state  $T>T^\star$ they are:
\begin{eqnarray}
&\chi^p_{xx}({\bf k})\cong \frac{1}{2(\alpha_x+\beta_{xy}M_y^2+\gamma_{ij}k_ik_j)},\nonumber\\
&\chi^p_{yy}({\bf k})\cong \frac{1}{2(\alpha_y+\gamma_{ij}k_ik_j)},
\label{sp}\nonumber\\
&\chi^p_{zz}({\bf k})\cong \frac{1}{2(\tilde\alpha_z+\gamma_{ij}k_ik_j)}=\frac{1}{2(\alpha_{z0}(T-T_c(H_y))+\gamma_{ij}k_ik_j)}.
\end{eqnarray}

The spin-triplet pairing interaction is expressed through the odd part of the susceptibility components :
\begin{equation}
\chi^u_{ii}({\bf k},{\bf k}')=\frac{1}{2}[\chi_{ii}({\bf k}-{\bf k}')-\chi_{ii}({\bf k}+{\bf k}')],~~~~i=x,y,z.
\end{equation}
Thus  for the ferromagnetic state  $T<T^\star$ we have 
\begin{eqnarray}
\chi^{fu}_{xx}({\bf k},{\bf k}')\cong
\frac{\gamma_{ij}k_F^2}{(a^f_{x})^2}\hat k_i \hat k_j^\prime,~~~\chi^{fu}_{yy}({\bf k},{\bf k}')\cong
\frac{\gamma_{ij}k_F^2}{
(a^f_{y})^2}\hat k_i\hat k_j^\prime,~~~~\chi^{fu}_{zz}({\bf k},{\bf k}')\cong
\frac{\gamma_{ij}k_F^2}{
(a^f_{z})^2}\hat k_i \hat k_j^\prime,
\label{su}
\end{eqnarray}
where
\begin{eqnarray}
&a^f_x=\alpha_x+\beta_{xz} M_z^2+\beta_{xy}M_y^2+2\gamma k_F^2,\nonumber\\
&a^f_y=\alpha_y+\beta_{yz}M_z^2+2\gamma k_F^2,
\label{sf1}\nonumber\\
&a^f_z=
4\beta_zM_z^2+12\delta_zM_z^4+2\gamma k_F^2.
\label{suu}
\end{eqnarray}
Here  $M_z(H_y), M_y(H_y)$ are the equilibrium values of the magnetization components. It is instructive to compare the obtained expressions for the odd parts of the susceptibility components with Eqs.(\ref{Chi})-(\ref{ax})  found  at $\delta_z=0,~H_y=0$ but $H_z\ne 0$.

For the paramagnetic state $T>T^\star$ they are
\begin{eqnarray}
\chi^{pu}_{xx}({\bf k},{\bf k}')\cong
\frac{\gamma_{ij}k_F^2}{(a^p_{x})^2}\hat k_i \hat k_j^\prime,~~~\chi^{pu}_{yy}({\bf k},{\bf k}')\cong
\frac{\gamma_{ij}k_F^2}{
(a^p_{y})^2}\hat k_i\hat k_j^\prime,~~~~\chi^{pu}_{zz}({\bf k},{\bf k}')\cong
\frac{\gamma_{ij}k_F^2}{
(a^p_{z})^2}\hat k_i \hat k_j^\prime,
\label{spu}
\end{eqnarray}
\begin{eqnarray}
&a^p_x=\alpha_x+\beta_{xy}M_y^2+2\gamma k_F^2,\nonumber\\
&a^p_y=\alpha_y+2\gamma k_F^2,
\label{sp1}\nonumber\\
&a^p_z=2(\alpha_{z0}(T-T_c(H_y))+
2\gamma k_F^2.
\end{eqnarray}

 Thus, at the first-order transition from the paramagnetic to the ferromagnetic state the components of susceptibility abruptly change their values.

As we have seen in Chapter IIIA the pairing interaction in the ferromagnetic state  is mostly determined by the odd part of
the $z$ component of susceptibility
\begin{equation}
\chi^{fu}_{zz}({\bf k},{\bf k}')\cong
\frac{\gamma_{ij}k_F^2}{4
(2\beta_zM_z^2+6\delta_zM_z^4+
\gamma k_F^2
)^2}\hat k_i \hat k_j^\prime.
\label{chii}
\end{equation}
  The equilibrium magnetization  $M_z(H_y)$ decreases with magnetic field $H_y$
(see Fig.10). 
One can  expect,  that the jump of   $M_z$ on the first-order transition
 is much smaller than the low temperature magnetization   at zero field $H_y=0$ :
 \begin{equation}
 M_z|_{H_y=0,T=0}>>M_z^\star.
 \end{equation}
Then  according to the equation (\ref{chii}) the $\chi^{fu}_{zz}$ on the line of the first order strongly exceeds  its own initial value at $H_y=0$
what stimulates 
 the reentrance of superconductivity near the first-order transition.

\subsection{Superconducting state in vicinity of the first order transition}

The  suppression of the Curie temperature by the magnetic field perpendicular to spontaneous magnetization leads to effective increase of pairing interaction. This effect can in principle compensate the suppression of superconductivity by the orbital depairing.
In URhGe the Curie temperature is much higher than $T_{sc}$. Hence, the orbital effect  succeeds to suppress the superconducting state  ($H_{c2}^b(T=0)\approx 1.3~T$ see \cite{Hardy2005})
much before the effect of 
decreasing of the Curie temperature and the stimulation of pairing intensity reveals itself.  But at fields higher than 10 Tesla the latter effect starts to overcome the orbital depairing and the superconducting state recreates. The critical temperature of superconducting transition begins grow up and approaches to the line of the first order transition from ferromagnetic to paramagnetic state and intersects it \cite{Levy2005,AokiKnebel2014}. 
Here we have a look what is going on with line of superconducting phase transition at the intersection with the line of ferromagnet-paramagnet first order phase transition $T^\star(H_y)$.

If the external field  orientated along  b-axis, that is perpendicular to the exchange field $h$ (see Fig.11),
 it is natural to choose the spin quantization axis along the  direction of the total magnetic field $
 h\hat z+H_y\hat y$. Then the normal state  matrix Green function is  diagonal 
\begin{equation}
\hat G_n=\left( \begin{array}{cc}G^{{\uparrow}}& 0\\ 
0 & G^{\downarrow}
\end{array}\right ),
\end{equation}
where
\begin{equation}
G^{{\uparrow,\downarrow}}=\frac{1}{i\omega_n-\xi^{{\uparrow,\downarrow}}_{{\bf k}}\pm\mu_B\sqrt{h^2+H_y^2}
}.
\label{nG1}
\end{equation}

All the formula obtained in the Chapter IIIA are  valid. The only modification is that one need to work with the susceptibility tensor written in the new coordinate frame
\begin{equation}
\chi_{ij}\to\tilde\chi_{ij}=R_{il}\chi_{lm}R_{jm},
\end{equation}
where 
\begin{equation}
\hat R=\left( \begin{array}{ccc}1& 0 & 0\\ 
0 &\cos\varphi&-\sin\varphi\\
0&\sin\varphi&  \cos\varphi
\end{array}\right )
\end{equation} 
is the matrix of rotation around $\hat x$ direction on the angle given by 
$$
\tan\varphi=\frac{H_y}{h}.
$$

For simplicity, one can work  with the equal-spin-pairing superconductivity neglecting  by the amplitude $\Delta^{0}$. 
 Also, the gradient  energy of orthorhombic ferromagnet Eq.(\ref{nabla1}) was taken in the exchange approximation such that the  pairing amplitude $V_3=0$. Unlike to the case of parallel field,  {\bf here we neglect by the orbital effects } ignoring the order parameter coordinate dependence.
This case the critical temperature of transition in superconducting state $T_{sc}(H_y)$ is determined from the self-consistency equations  
\begin{widetext}
\begin{eqnarray}
&\Delta^{\uparrow}({\bf k})
=\mu_BI^2T
\sum_{n}
\sum_{{\bf k}' }
\left\{\left[\chi^{fu}_{zz}({\bf k},{\bf k}')\cos^2\varphi+\chi^{fu}_{yy}({\bf k},{\bf k}')\sin^2\varphi\right]
G_{1}^\uparrow
G_{2}^\uparrow
\Delta^{\uparrow}({\bf k}')\right.\nonumber\\
&+\left.\left[
(\chi^{fu}_{xx}({\bf k},{\bf k}')-\chi^{fu}_{yy}({\bf k},{\bf k}'))\cos^2\varphi+(\chi^{fu}_{xx}({\bf k},{\bf k}')-\chi^{fu}_{zz}({\bf k},{\bf k}'))
\sin^2\varphi\right]
G_{1}^\downarrow
G_{2}^\downarrow
\Delta^{\uparrow}({\bf k}')
\right\},
\label{e11}
\end{eqnarray}
\begin{eqnarray}
&\Delta^{\downarrow}({\bf k})
=\mu_BI^2T
\sum_{n}
\sum_{{\bf k}' }
\left\{\left[(\chi^{fu}_{xx}({\bf k},{\bf k}')-\chi^{fu}_{yy}({\bf k},{\bf k}'))\cos^2\varphi+(\chi^{fu}_{xx}({\bf k},{\bf k}')-\chi^{fu}_{zz}({\bf k},{\bf k}'))\sin^2\varphi\right]
G_{1}^\uparrow
G_{2}^\uparrow
\Delta^{\uparrow}({\bf k}',{\bf q})\right.\nonumber\\
&+\left.\left[\chi^{fu}_{zz}({\bf k},{\bf k}')\cos^2\varphi+\chi^{fu}_{yy}({\bf k},{\bf k}')\sin^2\varphi\right]
G_{1}^\downarrow
G_{2}^\downarrow
\Delta^{\downarrow}({\bf k}')
\right\}.
\label{e21}
\end{eqnarray}
\end{widetext}
Here, 
$G_{1}^\uparrow=G^{\uparrow}({\bf k}',\omega_n)$, 
$G_{2}^\uparrow=G^{\uparrow}(-{\bf k}',-\omega_n)$ and similarly for the  $G_{1}^\downarrow$ and 
$G_{2}^\downarrow$ Green functions given by Eq.(\ref{nG1}).
In the ferromagnetic state near the first order transition the angle $\varphi\approx \pi/4$ , and the susceptibilities are  determined by Eqs. (\ref{su}) and(\ref{sf1}).

In the paramagnetic state the susceptibilities are determined by Eqs. (\ref{spu}) and(\ref{sp1}). The angle $\varphi=\pi/2$,  
\begin{equation}
G^{{\uparrow,\downarrow}}_{para}=\frac{1}{i\omega_n-\xi_{{\bf k}}\pm\mu_BH_y}
\end{equation}
and one can work with  the equations for 
$(\Delta^{\uparrow},\Delta^{\downarrow})$  independently  from equation for $\Delta^{0}$ \cite{Mineev2011}
\begin{widetext}
\begin{eqnarray}
&\Delta^{\uparrow}({\bf k},{\bf q})
=\mu_BI^2T
\sum_{n}
\sum_{{\bf k}' }
\left\{\chi^{pu}_{yy}({\bf k},{\bf k}')
G_{1}^\uparrow
G_{2}^\uparrow
\Delta^{\uparrow}({\bf k}')
+
(\chi^{pu}_{xx}({\bf k},{\bf k}')-\chi^{pu}_{zz}({\bf k},{\bf k}'))
G_{1}^\downarrow
G_{2}^\downarrow
\Delta^{\downarrow}({\bf k}',{\bf q})
\right\},
\label{e111}
\end{eqnarray}
\begin{eqnarray}
&\Delta^{\downarrow}({\bf k},{\bf q})
=\mu_BI^2T
\sum_{n}
\sum_{{\bf k}' }
\left\{(\chi^{pu}_{xx}({\bf k},{\bf k}')-\chi^{pu}_{zz}({\bf k},{\bf k}'))
G_{1}^\uparrow
G_{2}^\uparrow
\Delta^{\uparrow}({\bf k}',{\bf q})
+\chi^{pu}_{yy}({\bf k},{\bf k}')
G_{1}^\downarrow
G_{2}^\downarrow
\Delta^{\downarrow}({\bf k}',{\bf q})
\right\}.
\label{e211}
\end{eqnarray}
\end{widetext}

 As we mentioned already the components of susceptibility undergo a finite jump at first order phase transition  from ferromagnetic to paramagnetic state.
The  Fermi surfaces of split spin-up and spin-down electron bands, and  the average density of states on them  also undergo an abrupt changes. Finally, the structure of equations for determination of
the critical temperature of transition in superconducting state $T_{sc}(H_y)$  is quite different on both  sides of the ferromagnet-paramagnet phase transition. 
So,  the line of $T_{sc}(H_y)$  
should  undergo a jump at the intersection of line of the first order phase transition $T^\star(H_y)$. 
The experiment \cite{Levy2005} clearly demonstrates an abrupt fall of the critical temperature at this transition.

\subsection{ Concluding remarks}

Making use the phenomenological description of the phase diagram  in URhGe
  under a magnetic field along $b$ axis perpendicular to the direction of spontaneous magnetization
 we have found that the phase transition between the  ferromagnetic and the paramagnetic states under a strong enough magnetic field perpendicular to the direction of easy magnetization 
 changes  its order  from the second to the first order type. 
 The reentrance of superconductivity is explained by  the strong increase in magnetic susceptibility 
 in  the vicinity of the first-order transition in comparison with its zero-field value.  
 The reentrant superconductivity near the first-order transition line $T^\star(H_y)$ 
 exists both in ferromagnet and paramagnet state. The critical temperature of the transition to the superconducting state undergoes an abrupt fall down at the intersection with the line of the ferromagnet-paramagnet phase transition.  
 
 The suppression of the Curie temperature by a magnetic field perpendicular to the spontaneous magnetization increases the  pairing interaction. This effect  
 compensates the suppression of superconductivity by the orbital depairing.
  In UCoGe, where the Curie temperature  does not strongly exceed the temperature of transition to  superconducting state, this mechanism  stimulates  the upturn of the upper critical field 
for the field along $b$ direction above 
5 Tesla shown in Fig.12. The enhancement of superconductivity in UCoGe is accompanied by an enhancement
of nuclear relaxation rate caused by the increase of magnetic susceptibility in process of approach to the Curie temperature in field  parallel to $b$ axis \cite{Hattory2014}.

The $a$ direction is magnetically much harder
than the $b$ one: $\alpha_x\gg\alpha_y$. Hence, the suppression of the Curie temperature  by magnetic field $H_x$  is much less effective
and practically unobservable in available magnetic fields \cite{Hattory2014}. However, one can expect  the similar development of superconductivity stimulation   at much higher fields   along the $a$ direction.

In the vicinity of the phase transition of the first order caused by  the magnetic field along the $b$-axis $H_y\approx 12~T$
the application of some field along the $a$ axis increasing the magnitude of the  total magnetic field $H=\sqrt{H_y^{\star^2}+H_x^2}$ introduces negligible changes in the pairing interaction that was in the absence of the field along the hard $a$-axis. 
 At the same time the orbital upper critical field in the $a$ direction is one and a half times larger than in the $b$ direction 
 \cite{Hardy2005}.
 This roughly explains the stability of reentrant superconductivity in URhGe 
up to the fields $H=\sqrt{H_y^{\star^2}+H_x^2}\approx 30$ Tesla \cite{Levy2007}.

It is known that   in the presence of an external field along the direction of spontaneous magnetization $\hat z$ the line of the first order transition $T^\star(H_y)$ spreads to two surfaces of the first order transition $T^\star(H_y,\pm H_z)$. At these surfaces the jump in ferromagnet spontaneous magnetization
decreases  with increase of $|H_z|$ and disappears completely on some lines 
 beginning at the tricritical point  $T_c(H_y^{cr},H_z=0)$. 
It was suggested \cite{Levy2007} that these lines
are finished at zero temperature in some quantum critical points on the  $(H_y, H_z)$ plane. 
 The quantum critical magnetic fluctuations have been put forward as a source stimulating superconductivity in the vicinity of the line of the ferromagnet-paramagnet 
 first-order phase transition. This idea looks  plausible. One 
can remark, however, that in general the tricritical line $T^{cr}(H_y,H_z)$  can never reach zero temperature  or simply be dislocated far enough from the superconducting region on the phase diagram. 

Motivated by the idea of critical fluctuations
the recent  measurements  by Y.Tokunaga et al \cite{Tokunaga2015} have demonstrated the enormous increase of NMR relaxation rate  in URh$_{0.9}$Co$_{0.1}$Ge  at $H_y\approx 13$ Tesla.
These experiments
 have been performed at temperature 1.6 Kelvin that is in the region near the second-order phase transition between the  ferromagnetic and the paramagnetic states. The reentrant superconductivity  appears at much lower temperatures  near the line of the first-order transition where the role of critical fluctuations is certainly less important.

Here we have demonstrated that the reentrant superconductivity in URhGe can arise even in the absence of critical fluctuations due to a drastic increase in the longitudinal susceptibility in the  vicinity
 on the first-order transition line from the paramagnet to the ferromagnet state.

 \section{  Critical magnetic relaxation in uranium ferromagnets}
 
 \subsection{Critical relaxation in ferromagnets}

The excitations in magnetic systems are measured by the neutron scattering.
The inelastic magnetic neutron scattering intensity 
\begin{equation}
I({\bf Q},\omega)=A(k_i,k_f)(\delta_{\alpha\beta}-\hat Q_\alpha \hat Q_\beta)|F({\bf Q})|^2S_{\alpha\beta}({\bf q},\omega)
\label{cross}
\end{equation}
is related to the 
 dynamical structure factor 
 $$S_{\alpha\beta}({\bf q},\omega)=\int_{-\infty}^\infty dt e^{i\omega t}\langle M_{\alpha{\bf q}}(t)M_{\beta-{\bf q}}(0)\rangle$$
  which is a wave vector - frequency dependent magnetic moments correlation function \cite{Hove} related by the  fluctuation-dissipation theorem to the imaginary part of susceptibility
 \begin{equation}
 S_{\alpha\beta}({\bf q},\omega)=\frac{2}{1-\exp(-\frac{\omega}{T})} \chi^{\prime\prime}_{\alpha\beta}({\bf q},\omega).
 \end{equation}   
 The total wave-vector transfer is ${\bf Q}={\bf q}+\mbox{\boldmath$\tau$}$, $\mbox{\boldmath$\tau$}$ is a reciprocal lattice vector and ${\bf q}$ lies in the first Brillouin zone. $\hat Q_\alpha$ is the direction cosine of ${\bf Q}$ along coordinate axis $\alpha$. $F({\bf Q})$ is the magnetic form factor measured by the elastic neutron scattering. 
We put the Planck constant $\hbar=1$.  To be compatible with neutron scattering literature we shall use in this chapter the notation ${\bf q}$ 
for the wave vectors. 

 For each crystallographic direction one can fit the imaginary part of susceptibility as
 \begin{equation}
 \frac{\chi^{\prime\prime}({\bf q},\omega)}{\omega}=\frac{A}{\omega^2+\Gamma_{\bf q}^2},
 \end{equation}
 that is to express it through  the experimentally measured  amplitude $A$ and the  width $\Gamma_{\bf q}$ of dynamical form-factor. Then the  Kramers-Kronig relation allows to find the real part of static susceptibility
 \begin{equation}
 \chi({\bf q})=\chi^{\prime}({\bf q},0)=\frac{1}{\pi}\int\frac{\chi^{\prime\prime}({\bf q},\omega)}{\omega}d\omega=\frac{A}{\Gamma_{\bf q}}.
 \end{equation}

In the absence of walls and spin-orbital coupling the magnetization is a conserved quantity, hence, in a Heisenberg ferromagnet above Curie temperature  the only mechanism  leading to the magnetization relaxation is the spin diffusion that results in \cite{Hove,Forster}
\begin{equation}
S({\bf q},\omega)=\frac{2\omega\chi({\bf q})}{1-\exp(-\frac{\omega}{T})}\frac { \Gamma_{\bf q}}{\omega^2+\Gamma_{\bf q}^2},
\label{d}
\end{equation}
such that line width of quasi elastic scattering 
\begin{equation}
\Gamma_{\bf q}=Dq^2
\label{diff}
\end{equation}
is determined by the diffusion coefficient $D$. 
The $q^2$ law dependence was observed in a wide temperature range above $T_c$ in Ni and  Fe ( see Ref.\cite{Shirane} and references therein) reducing at $T=T_c$ to $\Gamma\propto q^{2.5}$ dependence according to predictions of mode-mode coupling theory \cite{Hohenberg}.  

In weak itinerant ferromagnets above Curie temperature another mechanism of dissipationless relaxation  can dominate with structure factor given by the same Eq. (\ref{d}) 
but with 
the linewidth determined by equality \cite{Hertz,Moriya}
\begin{equation}
\chi({\bf q})\Gamma_{\bf q}=\chi_P\omega({\bf q}), 
\label{Landau}
\end{equation}
where $\chi_P$ is the noninteracting Pauli susceptibility, $\omega({\bf q})$  is the Landau damping frequency equal to  $\frac{2}{\pi}qv_F$ for the spherical Fermi surface.
The linear in wave vector line width was observed  in MnSi \cite{Ishikawa}, however, in the other weak itinerant ferromagnets MnP \cite{Yamada} and Ni$_3$Al 
\cite{Semadeni} the linewidth q-dependence is closer to the dynamic scaling theory predictions \cite{Hohenberg}.

\subsection{ Magnetic relaxation in dual localized-itinerant ferromagnets}

  The magnetic susceptibility in uranium ferromagnets along easy axis is much larger than in a direction   perpendicular to it.  In UGe$_2$  the easy axis is along the $a$ crystallographic direction. 
Neutron scattering measurements reported in the paper \cite{Huxley2003} with scattering wave vector ${\bf q}$  parallel to the crystal $a$ axis revealed no extra scattering relative to the background while for the ${\bf q}$ parallel to the c-axis  a strongly temperature dependent contribution was found as it should be according to Eq.(\ref{cross}).
However, unlike both Eqs (\ref{diff}) and (\ref{Landau}) $\Gamma_{\bf q}$ does not vanish as $q\to 0$ for temperatures different from $T_c$. The same result was found in UCoGe \cite{Stock}  for the scattering with ${\bf q}$ parallel to the a-axis because the easy axis in this material  is along $c$-direction \cite{residual}. 
Hence, in uranium ferromagnets there is some mechanism  of the uniform magnetization relaxation 
 to the equilibrium.

The magnetization in an electron gas relaxes due to the spin-flip processes caused by the spin-orbit coupling either between the electrons \cite{Overhauser} 
or  by the spin-orbit  coupling between the  itinerant  Bloch electron spins with  potential of ions in vibrating lattice \cite{Elliott,Feher}. Both mechanisms produce so tiny homogeneous
relaxation rate, that it is  unobservable in  ferromagnetic materials, while the relaxation rate $\Gamma_{{\bf q}=0}$  found in UGe$_2$ \cite{Huxley2003} is of the order of several Kelvin. Hence, the magnetization relaxation in this material is due to  some different mechanism.
 In what follows, for the concreteness, we shall discuss mostly UGe$_2$.

 Magnetic susceptibility of single UGe$_2$ crystals has been measured by several groups \cite{Menovsky,Sakon,Troc}). The easy axis magnetization  at zero temperature was found 1.43 $\mu_B/$f.u. that  in the  case of itinerant ferromagnetism corresponds to completely polarized single electron band.
On the other hand the  neutron scattering measurements of the magnetic form factor \cite{Kernavanois2001} shows that:
(i) the shape of its $q$ dependence  is not distinguishable from the wave vector dependences of the  form factors of free U$^{3+}$ or U$^{4+}$ ions, (ii) its  low temperature value at $q\to 0$ coincides with the magnetization measured by magnetometer with accuracy of the order of 1 percent. 
Thus, practically whole magnetic moment  both in paramagnetic and in ferromagnetic states concentrated at uranium atoms \cite{footnote1} 

The static magnetic properties of UGe$_2$  are well described in \cite{Troc} in terms of crystal field splitting of the  $U^{4+}$ state, which is the $^3H_4$  term of the  $5f^2$ configuration of localized electrons, despite of the presence of the itinerant electrons filling the bands formed by two $7s$, one $6d$ and one $5f$ uranium and also germanium orbitals. So, UGe$_2$  is actually a dual system where local and itinerant states of $f$-electrons coexist.
The $^3H_4$  term of each atom of UGe$_2$ in paramagnetic state mostly consists of superposition of three quasi-doublets and three singlets  arising from the state with a fixed value of total momentum $J=4$ split by the crystal field \cite{Troc}.  The temperature decrease causes the change in probabilities of  populations of crystal field states revealing itself in temperature dependence of  the magnetic moment. The degeneracy removal of the ground state formed by the lower quasi-doublet allows the system to order magnetically with the ordered moment of $\sim1.5\mu_B$  twice smaller than  the Curie-Weiss moment deduced from susceptibility above the Curie temperature. 
  The itinerant electron subsystem formed by $7s$, $6d$ and partly $5f$ electrons is also present  providing about $0.02\mu_B$ long range magnetic correlations
as demonstrated by muon spin relaxation measurements \cite{Yaouanc,Sakarya}. 
All mentioned observations  as well the theoretical treatment \cite{Troc} unequivocally point on the local nature of UGe$_2$ ferromagnetism. 
This means that the quasi-elastic neutron scattering occurs  mostly on the fluctuations of magnetization   in the localized moments subsystem.

The interaction between localized and itinerant electron subsystems   leads to the magnetization relaxation measured by neutron scattering  in paramagnetic and ferromagnetic state of UGe$_2$.
 This type of relaxation can be considered as an analog of {\bf spin-lattice relaxation} well known in physics of nuclear magnetic resonance \cite{Slichter}. In our case {\bf the magnetization created by the local moments of uranium atoms  giving the dominate contribution to the neutron scattering plays the role of "spin" subsystem, whereas the itinerant electrons present the "lattice" degrees of freedom} absorbing and dissolving fluctuations of magnetization. Unlike to 
  the  NMR relaxation determined by the nucleus and the electrons magnetic moments  interaction {\bf the spin-lattice relaxation between the localized and conducting electrons is determined by spin-spin exchange processes} and has no relativistic smallness typical for NMR relaxation. 
 A deviation of magnetization from the equilibrium value  relaxes by transfer   to the itinerant electrons. 
According to this, we shall treat the  magnetization almost completely determined by the local moments of uranium atoms as not conserved quantity  \cite{Mineev2013}.

  The process of the easy axis magnetization relaxation to equilibrium  is described by the Landau-Khalatnikov kinetic equation \cite{Landau} 
\begin{equation}
\frac{\partial M}{\partial t} =-A\frac{\delta {\cal F}}{\delta M}.
\label{kin}
\end{equation} 
Here
\begin{equation}
 {\cal F}=\int dV\left (\alpha_0(T-T_c)(T)M^2+\gamma_{ i j}\frac{\partial M}{\partial x_i}\frac{\partial M}{\partial x_j} -MH\right )
 \end{equation}
is the  energy of  the order parameter fluctuations above the Curie temperature
in a quasi stationary magnetic field along the easy axis.  The gradient energy in orthorhombic crystal  written in the  exchange approximation  is determined by three nonzero constants $\gamma_{xx}, \gamma_{yy},
\gamma_{zz}$.  The $x, y, z$ axes are pinned to the $b, c, a$ directions.
The kinetic equation can be rewritten as
 \begin{equation}
\frac{\partial M}{\partial t} + \nabla_i j_{i}=-\frac{M_\alpha}{\tau}+AH,
\label{kine}
\end{equation}
where
$
\tau^{-1}=2A\alpha_{0}(T-T_c)$,
\begin{equation}
 j_{ i} =-2A\gamma_{ij}
\frac{\partial {M}}{\partial x_j}
\end{equation} 
are the components of the spin diffusion currents.
 
Substituting in Eq.(\ref{kine}) $M=m_{{\bf q}\omega} e^{i({\bf q}{\bf r}- \omega t)}$ and $H=h_{{\bf q}\omega}e^{i({\bf q}{\bf r}- \omega t)}$ we obtain susceptibility
  \begin{equation}
\chi({\bf q},\omega)=\frac{m_{{\bf q}\omega} }{h_{{\bf q}\omega} }=\frac{A}{-i\omega+\Gamma_{\bf q}}.
\label{chiii}
\end{equation}
The width of quasi-elastic scattering   for ${\bf q}\parallel \hat c$  axis is
\begin{equation}
\Gamma_{q}=2A\left [\alpha_0(T-T_c)+\gamma_{yy}q_c^2\right].
\label{Gammapara}
\end{equation} 
  
Below the Curie temperature in the ferromagnetic state the equilibrium magnetization  $M=M_0(T)$  
and energy of fluctuations is  
\begin{equation}
 {\cal F}=\int dV\left (2\alpha_0(T_c-T)(M-M_0)^2+\gamma_{ i j}\frac{\partial M}{\partial x_i}\frac{\partial M}{\partial x_j} -(M-M_0)H\right )
 \end{equation}  
 By the similar derivation we obtain for the susceptibility the same expression as Eq.(\ref{chiii}) with the following width of quasielastic scattering 
\begin{equation}
\Gamma_{q}=2A\left [2\alpha_0(T_c-T)+\gamma_{yy}q_c^2\right].
\label{Gammaferro}
\end{equation}

\subsection{ Concluding remarks}

Experimentally there were determined  two independent values: the width $\Gamma_{\bf q}$ and the amplitude $A=\chi({\bf q})\Gamma_{\bf q}$ of distribution given by Eq. (\ref{d}) (see Fig.13). Here, we have established  the following.
The line width of quasi-elastic neutron scattering near the Curie temperature proves to be a  linear function of $T-T_c$. The  absolute value of the derivative $|d\Gamma_{q_c}/dT|$ in a ferromagnetic region is roughly twice as large as the corresponding derivative in a paramagnetic region. The dependence of the wave vector $q_c$ is parabolic. All of these findings are  in qualitative correspondence with the experimental observations reported in the paper \cite{Huxley2003}  (see Fig.14a,b). 

At the same time there has been found \cite{Huxley2003} (see Fig.14d) that the product $\chi({\bf q})\Gamma_{\bf q}$ is temperature independent above Curie temperature but reveals the fast drop below $T_c$.
Such type behavior means that   below $T_c$ the  decrease of susceptibility $\chi({\bf q})$  with temperature proves to be  much faster than it is in the accordance with the mean field theory. 
This unusual behavior of susceptibility in UGe$_2$ far enough from the critical region deserves a special investigation.
 
    Many experimental observations point to the local nature of magnetism  in uranium ferromagnetic superconductors. The interaction between localized and itinerant electron subsystems gives rise to a specific mechanism  of magnetization relaxation similar to "spin-lattice" relaxation known in physics of nuclear magnetic resonance. This relaxation determined by the exchange spin-spin coupling is much faster than NMR relaxation  supported by much weaker  interaction between electron and nuclei magnetic moments.
The phenomenological description of quasi-elastic magnetic relaxation 
is based on specific for heavy fermionic  ferromagnet uranium compounds property that  magnetization  supported by the moments located at uranium atoms  is not a conserved quantity. As result the line width of quasi elastic neutron scattering at $q\to 0$ acquires non-vanishing value at all temperatures besides the Curie temperature.  

This conclusion has been confirmed by the microscopic analysis \cite{Chubukov2014} where was shown explicitly that near the ferromagnetic instability in a system consisting of localized and itinerant electrons the magnetization concentrating mostly on the localized subsystem is not conserved.

\section{Anisotropy of nuclear magnetic resonance relaxation and the upper critical field in UC${\bf o}$G${\bf e}$}

The nuclear magnetic resonance measurements on $^{59}$Co nucleus  in UCoGe show that the  magnetic field along the $c$-axis strongly suppresses the magnetic fluctuations along this direction
\cite{Ihara2010,Hattory2012}. Also, there was revealed that the upper critical field value in this superconducting  material  drops abruptly
 at small field declinations  from both   the $a$ and the $b$ crystallographic directions towards the direction of the spontaneous magnetization $c$ \cite{Aoki2009,Aoki14}.
So, the component of magnetic field along the $c$ axis very effectively suppresses the superconducting state.
As we have seen, the triplet pairing in uranium ferromagnet superconductors mostly provided by the longitudinal fluctuations of magnetization with the amplitude  proportional to the odd part of the susceptibility $\chi_{zz}^u$. Here we shall demonstrate that both mentioned phenomena 
have the same origin and they are explained by the strong increase of magnetization  \cite{Huy2008,Knafo2012} and the corresponding decrease of differential susceptibility (\ref{59}) in UCoGe in magnetic field along $c$-axis.

\subsection{ Nuclear magnetic resonance  relaxation rate}

Nuclear spin-lattice relaxation rate measured in a field  along the $\alpha$ direction is expressed in terms of the imaginary part of the dynamic susceptibility along 
the $\beta$ and $\gamma$ directions perpendicular to $\alpha $ as
\begin{equation}
\frac{1}{T_1^\alpha}
\propto T\sum_{\bf k}
\left[ |A_{hf}^\beta|^2\frac{\chi^{\prime\prime}_\beta({\bf k},\omega)}{\omega}+  
 |A_{hf}^\gamma|^2\frac{\chi^{\prime\prime}_\gamma({\bf k},\omega)}{\omega}
\right ].
\end{equation}
At low temperatures $1/T_1$ for $H\parallel c$ is more than order magnitude smaller  those measured in the other two field directions \cite{Ihara2010}. So, if we are interested in the relaxation rate in the field tilted on angle  $\theta$ in respect to the  $b$ axis in the  $bc$ plane, such that  $\theta$ is noticeably smaller than $\pi/2$, we can use the following expression
\begin{equation}
\frac{1}{T_1}(\theta)
\propto T\sum_{\bf k}
|A_{hf}^z|^2\frac{\chi^{\prime\prime}_{zz}({\bf k},\omega)}{\omega}\cos^2\theta.
\end{equation}

In assumption that the fluctuations of the hyperfine e-m field on the Co sites are determined by the fluctuations of magnetization of the subsystem of localized moments as we did  discussing the neutron scattering relaxation rate we can use the  formula 
\begin{equation}
\frac{\chi^{\prime\prime}_{zz}({\bf q},\omega)}{\omega}=\frac{A}{\omega^2+\Gamma_k^2}, ~~~~
\Gamma_{k}=2A(a+\gamma_{ij}k_ik_j),
\end{equation}
where
\begin{equation} 
a=\alpha_z+\beta_{yz}M_y^2+6\beta_zM_z^2= 2\beta_z(3M_z^2-M_{z0}^2).
\end{equation}
To obtain the last equality, as in the Eq.(\ref{az}), we have used the equilibrium condition
$$
2\alpha_z+2\beta_{yz}M^2_y+4\beta_zM_z^2=\frac{H_z}
{M_z}.$$
Here 
\begin{equation}
M_z=M_z(H_y,H_z)=M_z(H\cos\theta,H\sin\theta),~~~M_{z0}=M_z(H,0)
\end{equation}
are the equilibrium components of magnetization in the field
 \begin{equation}
 {\bf H}=H_y\hat y+H_z\hat z=H\cos\theta~\hat y+H\sin\theta~\hat z.
 \end{equation}
 At arbitrary temperatures below the Curie temperature one can use experimental values for field dependent
magnetization $M_z(H_y,H_z)$.

 The NMR measurements are done at the  frequencies $\omega<<\Gamma_{\bf q}$, hence,
 the relaxation rate is determined as
 \begin{equation}
\frac{1}{T_1}(\theta)
\propto T
|A_{hf}|^2A\cos^2\theta\int\frac{d^3k}{(2\pi)^3\Gamma_{\bf k}^{2}}.
\label{T_1}
\end{equation}
For the simplicity one can calculate  the converging  integral in spherical approximation  
$$
\int\frac{d^3k}{(2\pi)^3\Gamma_{\bf k}^{2}}\approx \int_0^\infty\frac{4\pi k^2dk}{(2\pi)^3(2A)^2(a+\gamma k^2)^2}=\frac{1}{32\pi A^2\sqrt{a\gamma^3}}.
$$  
Thus, keeping only the field dependent part in Eq.(\ref{T_1}) we obtain 
 \begin{equation}
\frac{1}{T_1}(\theta)
 \propto\frac{cos^2\theta}{\sqrt{a}}=\frac{1-\frac{H^2_z}{H^2}}{\sqrt{2\beta_z(3M_z^2-M_{z0}^2)}}.
\label{T_11}
\end{equation}

 The  measurements of  the NMR relaxation rate  dependence from the magnetic  field orientation was performed in relatively small fields $H<3.5$ Tesla \cite{Hattory2012}.
 This case, $M_{z0}=M_z(H,0)$ is almost independent from the field $H=H_y$ and one can take it approximately equal to the  spontaneous magnetization $M_{z0}(0,0)$. On the other hand, $M_z=M_z(H_y,H_z)\approx M_z(0,H_z)$  quickly  grows with increase of $H_z$. For instance, in field $H_z= 1$ Tesla the magnetization $M_z(0,H_z)$ is twice large than it is at $H_z=0$ \cite{Huy2008}. Hence, according to Eq.(\ref{T_11}), taking $H=3.5$ Tesla and  $H_z= 1$ Tesla, we obtain 
 \begin{equation}
\frac{1}{ T_1}(H_z=1T)\approx 0.4\frac{1}{ T_1}(H_z=0).
 \end{equation}
 In comparison with this estimation the relaxation rate  $\frac{1}{ T_1}(H_z)$  measured at $T=1.7$ Kelvin \cite{Hattory2012}  experiences  faster falling with growing of the $H_z$ component of magnetic field   ($H_c$ in  Fig.15). This is not astonishing in view of 
  roughness of our approximations made at the Eq.(\ref{T_11}) derivation. 
  
Thus, 
the NMR  relaxation rate  dependence from the  magnetic field along the easy magnetization axis originates from the corresponding field dependence of the longitudinal component of magnetic susceptibility.
 
\subsection{ Upper critical field anisotropy}
 
The anomalous upper critical field  anisotropy in UCoGe \cite{Aoki2009,Aoki14} also  finds the natural explanation  in terms of strong field dependence of the longitudinal susceptibility determining the constant of pairing interaction.
We have already demonstrated this dependence in the case of magnetic field parallel to the
direction of the spontaneous magnetization that is along the $c$ axis (see chapter IVB).  If  the magnetic field is directed along the $b$ crystallographic axis
the critical temperature in neglect of orbital effects  is determined by  Eqs. (\ref{e11}) and (\ref{e21}). In a field directed in the $bc$ plane ${\bf H}=H_y\hat y+H_z\hat z$ the equations (\ref{e11}) and (\ref{e21}) keep their form but  the Green  functions  now are
\begin{equation}
G^{{\uparrow,\downarrow}}=\frac{1}{i\omega_n-\xi^{{\uparrow,\downarrow}}_{{\bf k}}\pm\mu_B\sqrt{(h+H_z)^2+H_y^2}
}
\label{nG}
\end{equation}
and the angle is given by
$$
\tan\varphi=\frac{H_y}{h+H_z}.
$$
The susceptibilities are
\begin{equation}
\chi^u_{zz}({\bf k},{\bf k}')\cong
\frac{\gamma_{ij}k_F^2}{
4\left [\beta_z(3M_z^2-M_{z0}^2)+\gamma k_F^2\right]^2}\hat k_i \hat k_j^\prime,
\label{Chi11}
\end{equation}
\begin{equation}
\chi^u_{yy}({\bf k},{\bf k}')\cong
\frac{\gamma_{ij}k_F^2}{
\left [\alpha_y+\beta_{yz}M_z^2+2\gamma k_F^2\right]^2}\hat k_i \hat k_j^\prime,
\label{Chi12}
\end{equation}
where
$$M_z=M_z(H_y,H_z)=M_z(H\cos\theta,H\sin\theta),~~~M_{z0}=M_z(H,0)$$
are the equilibrium components of magnetization in the field $
 {\bf H}=H_y\hat y+H_z\hat z$.
 
 As usual one can neglect the $\chi^u_{yy}$ in comparison with $ \chi^u_{zz}$.
 Then, as in  the chapter IVB, assuming that the largest critical temperature corresponds to the $(\hat k_x\eta_x^\uparrow,~\hat k_x\eta_x^\downarrow)$ superconducting state,   we have in the single band approximation
\begin{equation}
\ln \frac{\varepsilon}{T_{sc}}=\frac{1}{g_{1x}^\uparrow}\propto\frac{\left [\beta_z(3M_z^2-M_{z0}^2)+\gamma^z k_F^2\right]^2}{\cos^2\varphi
}.
\label{H1}
\end{equation}
 There are no experimental data  about the low temperature behavior of $M_z=M_z(H_y,H_z)$  as function of both its arguments. All the measurements have been performed  in the field directed along the crystallographic axies $a,b,c$ \cite{Huy2008, Knafo2012}.
 However, looking at the data \cite{Knafo2012} in the strong field - low temperature region, where the phenomenon of the strong upper critical field anisotropy has been revealed, one can expect  that an increase $H_z$ at fixed $H_y$  strongly increases $M_z$. 
 Also a decrease of $H_y$    causing the increase of Curie temperature $T_c(H_y)$ increases  $M_z$ as well.  Thus, a magnetic field declination  from the $b$ or $a$ direction toward $c$ axis leads to the  increase of  $M_z=M_z(H_y,H_z)$, hence, 
 to the sharp
  drop in the constant of pairing interaction described by Eq.(\ref{H1}). This explains the low-temperature - high field upper critical field anisotropy observed in UCoGe  \cite{Aoki2009,Aoki14} (see Fig.16).

 \section{First order phase transition to ferromagnet state in UG${\bf e}_{2}$}

The pressure-temperature phase diagrams of several weak ferromagnets exhibit similarity. The  transition from the paramagnetic to the ferromagnetic states at ambient pressure occurs  by means of the second-order phase transition. The phase transition temperature decreases with pressure increase such that it reaches the zero value at some pressure $P_0$.  In  a  pressure interval  below  $P_0$ the ordered ferromagnetic moment disappears discontinuously. Thus at high pressures and low temperatures  the ferromagnetic and the paramagnetic states are divided by the   first-order type transition whereas at higher temperatures and  lower pressures this transition is of the second-order. Such type of behavior is typical for 
MnSi \cite{Pfleiderer1997,Stishov2007,Otero2009,Uemura2007},  UGe$_2$ 
\cite{Pfleiderer2002,Taufour2010} (see Fig.17), ZrZn$_2$ \cite{Uhlartz2004}. The same behavior has been established  in the ferromagnetic compounds Co(Si$_{1-x}$Se$_x$)$_2$ 
 \cite{Goto1997}
and (Sr$_{1-x}$Ca$_x$)RuO$_3$ \cite{Uemura2007} where the role of governing parameter plays 
the concentration of Se and Ca correspondingly.

 \subsection{Phase transition to ferromagnetic state in Fermi liquid theory}
 
 The phase transition from paramagnetic to itinerant ferromagnetic state is usually considered in frame of Stoner theory where it is the transition of the second order \cite{Hertz}.
 Some time ago, Belitz, Kirkpatrick, and Vojta (BKV) have argued that the phase transition in
clean itinerant ferromagnets is  of first-order at low temperatures, due to the correlation effects that lead
to a logarithmic term in the free energy density expansion in powers of dimensionless magnetization M
\cite{Belitz1999}:
\begin{equation}
E=E_0+\alpha M^2+\beta M^4+v M^4\ln |M|+. . .
\end{equation} 
 Indeed, at  positive coefficient $ v$ the effective fourth order term in this formula is negative at small M , and the transition to ferromagnet state is of the first-order. 
 
 The logarithm correction to the fourth order term has the long story. For the first time it was calculated  in 1970 by  S.Kanno \cite{Kanno1970} in the dilute Fermi gas model in the second order in respect to dimensionless gas parameter
$k_Fa$,  where $k_F$ is
the Fermi momentum related to the total density $$n = n^\uparrow + n^\downarrow=
\frac{k_F^3}{3\pi^2}$$ and $a > 0$ is the s-wave scattering length. 
 In general, to solve the phase transition problem at  $T=0$,
  one must calculate the Fermi-gas energy density
  $$E(x) = \frac{3}{5}
n\varepsilon_Ff (M)$$ 
as a function of the dimensionless spin polarization (magnetization)
  $$M =
\frac{n^\uparrow-n^\downarrow}{n^\uparrow + n^\downarrow}$$
   at given $k_Fa$.
Here $\varepsilon_F=k_F^2
/2m$. 

In the first order in $k_Fa$ the ferromagnetic phase transition is of the second-order and occurs \cite{Huang} at $k_Fa=\pi/2$.
 The second-order perturbation theory predicts a first-order phase transition \cite{Duine,Conduit}
at $k_Fa = 1.054$, consistent with the BKV argument.
 However, since the critical gas parameter is expected to
be of order $O(1)$, perturbative predictions may be unreliable. The nonperturbative effects are studied by He and Huang \cite{He}
by summing the particle-particle ladder diagrams to all orders in the gas
parameter. The theory predicts a second-order
phase transition, which indicates that ferromagnetic transition in  Fermi liquid occurs not according  
to BKV scenario. The predicted \cite{He} critical gas parameter $k_Fa = 0.858$ is in good agreement with the recent
quantum Monte Carlo result $k_Fa = 0.86$ for a nearly zero-range potential \cite{Pilati}. 
 
 So, the first order phase transition in UGe$_2$ cannot be explained in frame of isotropic Fermi liquid theory even if we will  forget that  this compound  presents the 
 strongly anisotropic ferromagnetic metal with magnetization mostly supported by the magnetic moments localized at uranium atoms.
 
 Finally, it should be noted that an isotropic  ferromagnetic Fermi liquid is unstable  in respect to the transversal inhomogeneous deviations of magnetization \cite{Mineev2005,MineevArXiv2012}.  So, the problem about the isotropic Fermi liquid phase transition into ferromagnetic state has only academic interest.
 
 \subsection{Magneto-elastic mechanism  of development of the first order type instability}
 
 The magneto-elastic mechanism  of development of the first order type instability  has been put forward  
in the paper \cite{Bean1961}
 where it was demonstrated that the change of transition character from the second to the first-order
  takes place at  large compressibility and at strong enough steepness  of the exchange interaction dependence from the  interatomic distance.
This can be considered in frame of the Landau  theory of the phase transitions. Namely,  in neglect the shear deformation the free energy density near  the phase transition to the  Ising type ferromagnetic state  has the following form
\begin{equation}
F=\alpha_0(T-T_c)M^2+\beta M^4+\frac{K}{2}\varepsilon^2-q\varepsilon M^2.
\label{B1}
\end{equation}
Here, $M$ is the magnetization density, $\varepsilon$ is the relative volume change, $K$ is the bulk modulus.
The coefficient $q$ is related to the Curie temperature pressure dependence as
\begin{equation}
q=\alpha_0\frac{d T_c}{d\varepsilon}=-\alpha_0K\frac{d T_c}{d P}.
\label{B2}
\end{equation}
At fixed pressure, that is when the specimen volume changes  are not accompanied by   pressure changes 
in environment media
$
\frac{\partial F}{\partial \varepsilon}=0$, the deformation is determined by square
of magnetization $\varepsilon=\frac{q}{K}M^2$ what yields
\begin{equation}
F=\alpha M^2+\left ( \beta-\frac{q^2}{2K}   \right )M^4.
\label{B3}
\end{equation}
Hence, at $
\frac{q^2}{2K} >\beta$ the phase transition changes its character from the second  to the first order.
This inequality can be rewritten through the measurable parameters as
\begin{equation}
\frac{K\Delta C}{T_c}\left (\frac{d T_c}{d P}\right)^2>1,
\label{B4}
\end{equation}
where we used  the formal expression $\Delta C=\frac{\alpha_0^2}{2\beta}T_c$ for the specific heat jump at phase transition of the second order.

The magneto-elastic interaction also produces
another general mechanism
 for instability of second-order phase transition 
 toward to the discontinuous  formation of ferromagnetic state from  the paramagnetic one.  
 For the first time it was pointed out  by O.K.Rice \cite{Rice1954}
who has demonstrated that at small enough distance from the volume dependent critical temperature $T_c(V)$, where the specific heat $C_{fl}(\tau) \sim \tau^{-\alpha},$ $\tau = \frac{T}{T_c(V )} -1$, tends to infinity due to the critical fluctuations,
the system bulk modulus $K= -V \frac{\partial P}{\partial V}=V\frac{\partial^2 FV}{\partial V^2}$, expressed through the free
energy density $F = F_0 + F_{fl}, ~~F_{fl}\sim -T_c\tau^{2-\alpha} $ starts to be negative
\begin{equation}	
K=K_0-A\frac{C_{fl}(\tau)V^2}{T_c}\left (\frac{\partial T_c}{\partial V}\right)^2 
=K_0-\left.AK_0^2\frac{C_{fl}(\tau)}{T_c}\left (\frac{\partial T_c}{\partial P}\right)^2 \right |_{\tau\to 0}< 0~,
\label{B5}
\end{equation}	
that	contradicts	to thermodynamic stability of the system. 
 In reality, at temperature decreasing  before  the temperature corresponding to $K=0$ is reached
 the system undergoes  the first-order transition, such that to jump over the instability region directly in the ferromagnetic state with finite magnetization and related to it striction deformation. This transition is similar to the jump over the region with $\partial P/\partial V >0$  on the van der Waals isotherm at the liquid-gas transition. 
 
 The condition of the first order instability (\ref{B5})  can be  written in similar to Eqn.(\ref{B4}) form
 \begin{equation}	
\frac{K_0C_{fl}(\tau)}{T_c}\left (\frac{\partial T_c}{\partial P}\right)^2 >1.
\label{B6}
\end{equation}	
Unlike to Eq. (\ref{B4}) this formula demonstrates that the  first-order instability is inevitable due to infinite increase of fluctuation specific heat.

The striction interaction can change the shape of the free energy singularity in respect to its form at fixed volume. More elaborate treatment \cite{LarkinPikin1969} taking into account this effect leads to the following condition of the first order instability 
\begin{equation}
\frac{1}{T_c}\frac{4\mu K}{3K+4\mu}f''(x)\left (\frac{\partial T_c}{\partial P}\right )^2>1.
\label{B7}
\end{equation}
Here the function $f(x)$ determines the fluctuation part of free energy $F=-T_cf\left (\frac{T-T_c}{T_c}\right)$, $\mu$ is the shear modulus. 

Usually, the left hand side in Eqn. (\ref{B6}) is quite small
and the transition of the first order occurs at temperature $T^\star$ close to the critical temperature  where fluctuation specific heat is large enough. It means that the  temperature difference $T^\star-T_c$ is much smaller than the critical temperature $T_c$. The latent  heat at this transition 
\begin{equation}
q\approx C_{fl}(T^\star-T_c)
\label{B8}
\end{equation}
proves to be extremely small. So, the first-order phase transition is practically indistinguishable from the second-order one and called {\bf weak first order phase transition or the phase transition of the first order closed to the second order}. 

According to Eqs.  (\ref{B4}), (\ref{B6})  the magneto-elastic mechanism  effectively leads to the first-order transition
when the critical temperature is strongly pressure dependent. This is the case in all mentioned above materials.  To check the criteria (\ref{B4}), (\ref{B6}) one must calculate the mean field jump and fluctuation part of the specific heat near  Curie temperature for each particular material. 
To be concrete, here, we will do these calculations for
UGe$_2$ \cite{Mineev2012} characterized by strong magnetic anisotropy and by
 the precipitous drop of the critical temperature at pressure increase near 14-15 kbar \cite{Saxena2000}. 
 
\subsection{Specific heat near the Curie temperature}

UGe$_2$ is the orthorhombic crystal with ferromagnetic order at ambient pressure found below $T_c=53~K$. 
Magnetic measurements reveal a very strong magnetocrystalline anisotropy \cite{Onuki1992} with ${\bf a}$ being the easy axis.  We shall denote it as $z$ direction.  As in the previous Chapter  we shall take into account only the easy axis order parameter fluctuations.
Above the Curie temperature they are determined by deviation of system free energy 
\begin{equation}
{\cal F}=\int d^3{\bf r}\left\{\alpha M^2+\beta M^4
+\gamma_{ij}\nabla_iM\nabla_jM
-\frac{1}{2}\frac{\partial^2M({\bf r})}{\partial z^2}\int\frac{M({\bf r}')d^3{\bf r}'}{|{\bf r}-{\bf r}'|}\right\}
\label{C1}
\end{equation}
from the equilibrium value. $\alpha=\alpha_0(T-T_c) $.
Here, the gradient terms are written taking into account the orthorhombic anisotropy
$$
\gamma_{ij} = \left(\begin{array}{ccc} \gamma_{xx} & 0 & 0\\
0 & \gamma_{yy} & 0 \\
0 & 0 & \gamma_{yy}
\end{array} \right),
$$
where the $x, y, z$ axes are pinned to the $b, c, a$ directions.
The last nonlocal term in Eq. (\ref{C1}) corresponds to magnetostatic energy \cite{StatPhysII} $-{\bf M}{\bf H}-H^2/8\pi$, where internal magnetic field ${\bf H}$ expressed in terms of magnetization density by means of the Maxwell equations $$rot{\bf H}=0,~~~~~div({\bf H}+4\pi{\bf M})=0.$$
We shall use the following estimations for the coefficients in the Landau free energy functional
\begin{eqnarray}
&\alpha_0&=\frac{1}{m^2n},
\label{8}\\
&\beta&=\frac{T_c}{2(m^2n)^2n},
\label{9}\\
\gamma_x\approx&\gamma_y&\approx\gamma_z\approx\frac{T_ca^2}{m^2n}.
\label{10}
\end{eqnarray}
Here, $m$ is the magnetic moment per uranium atom at zero temperature,  $m=1.4\mu_B$ at ambient pressure \cite{Kernavanois2001},
$n=a^{-3}$ is the density of uranium atoms, which can be approximately taken equal to cube of the inverse nearest-neighbor uranium atoms separation $a=3.85$ \AA \cite{Huxley2001}.

The  mean field magnetization and the jump of specific heat are
\begin{eqnarray}
&M^2&=-\frac{\alpha}{2\beta}=(mn)^2\frac{T_c-T}{T_c}\\
&\Delta C&=\frac{T_c\alpha_0^2}{2\beta}=n.
\label{12}
\end{eqnarray}
The experimentally found specific heat jump $\Delta C_{exp}\approx 10\frac{J}{mol K}\approx 1$ per uranium atom \cite{Huxley2001} is in the remarkable correspondence with Eq.(\ref{12}).

For calculation of the fluctuation specific heat we use the Fourier representation of the quadratic in the order parameter part of  Eq.(\ref{C1})
\begin{equation}
{\cal F}=\sum_{\bf k}\left (\alpha +\gamma_{ij}k_ik_j+2\pi k_z^2/k^2\right)
M_{\bf k}M_{-{\bf k}},
\end{equation}
where $M_{\bf k}=\int M({\bf r})e^{-i{\bf k}{\bf r}}d^3{\bf r}$. 
The last term in this expression corresponds to magnetostatic energy  \cite{StatPhysII,Mineev2012}.

The corresponding  thermodynamic potential and the specific heat found in the similar uniaxial segnetoelectric model are \cite{Levanyuk1965}
\begin{equation}
\Omega=-\frac{T}{2}\sum_{\bf k}\ln\frac{\pi T}{\alpha +\gamma_{ij}k_ik_j+2\pi k_z^2/k^2},
\end{equation}
\begin{equation}
C_{fl0}=\frac{T^2\alpha_0^2}{2(2\pi)^3}\int\frac{dk_xdk_ydk_z}{[\alpha+\gamma_{ij}k_ik_j+2\pi \hat k_z^2]^2}. 
\end{equation}
Proceeding to spherical coordinates  and performing integration over modulus $k$ we come to
\begin{equation}
C_{fl0}=\frac{T^2\alpha_0^2}{32\pi^2}\int_0^1d\zeta\int_0^{2\pi}\frac{d\varphi}{(\alpha+2\pi \zeta^2)^{1/2}(\gamma_{\perp}+\zeta^2(\gamma_z-\gamma_{\perp}))^{3/2}}. 
\label{C}
\end{equation}
Here, $\gamma_{\perp}(\varphi)=\gamma_{x}\cos^2\varphi+\gamma_{y}\sin^2\varphi$.
At critical temperature $\alpha=0$ and the integral diverges. 
Hence, performing integration over $\zeta$ with logarithmic accuracy we obtain
\begin{equation}
C_{fl0}=\frac{T_c^2\alpha_0^2}{32\pi\sqrt{2\pi}\gamma^{3/2}}\ln\frac{2\pi}{\alpha}\approx\frac{n}{32\pi}\sqrt{\frac{T_c}{2\pi m^2n}}\ln\frac{2\pi m^2n}{T-T_c},
\label{CC}
\end{equation}
where $$\frac{1}{\gamma^{3/2}}=\frac{1}{2\pi}\int_0^{2\pi}\frac{d\varphi}{\gamma_\perp^{3/2}(\varphi)}.$$

The used condition $\alpha \ll 2\pi$ at $T_c=10K$ is realized at
\begin{equation}
\frac{T-T_c}{T_c}<\frac{2\pi m^2n}{T_c}\approx 0.015.
\label{T}
\end{equation}
In view of roughness of the parameter estimation the region of logarithmic increase of specific heat
can be in fact broader.

The  calculation taking into account the interaction of fluctuations in the formally similar uniaxial segnetoelectric model has been performed by Larkin and Khmelnitskii \cite{Larkin1969}.
In our notations
the expression  for the fluctuation specific heat at const pressure obtained in this paper is
\begin{equation}
C_{fl}=\frac{3^{1/3}T_c^2\alpha_0^2}{16\pi\gamma_{LK}^{2/3}\gamma^{3/2}}
\left (\ln\frac{2\pi}{\alpha}\right )^{1/3}.
\label{Khm}
\end{equation}
Here $\gamma_{LK}=\frac{3T_c\beta}{\sqrt{32\pi}\gamma^{3/2}}$ is the effective constant of interaction.  Using the Eqs. (\ref{8})-(\ref{10}) one can rewrite Eq. (\ref{Khm}) as
\begin{equation}
C_{fl}\approx\frac{n}{10}\left (\frac{T_c}{2\pi m^2n} \right)^{1/6}\left (\ln\frac{2\pi m^2n}{T-T_c}\right )^{1/3}.
\label{Khme}
\end{equation}
The power of the logarithm $(\ln\frac{\alpha_x}{2\pi})^{1/3}$ is the quite slow function slightly exceeding unity, 
hence, in the temperature region given by inequality (\ref{T})
one may estimate the fluctuation specific heat as 
\begin{equation}
C_{fl}\approx\frac{n}{5}.
\label{Khmel}
\end{equation}
We see that the fluctuation specific heat is smaller than the mean field jump given by Eqn. (\ref{12}). Hence,  to check the first order phase transition instability in UGe$_2$ one must to proceed with the criterium (\ref{B4}).

\subsection{ First-order type transition in UGe$_2$}

The Curie temperature in UGe$_2$ falls monotonically with increasing pressure from 53 K at ambient pressure and drops precipitously above 15 Kbar.\cite{Saxena2000}  The average value of the critical temperature derivative can be estimated as 
\begin{equation}
 \frac{\partial T_c}{\partial P}\approx\frac{40~Kelvin}{14~kbar}=4\times 10^{-25}~cm^3
 \label{est}
\end{equation} 
 For the bulk modulus we have 
 \begin{equation}
 K= \rho c^2\approx 10^{11}erg/cm^3,
 \end{equation}
 where we have substituted  typical sound velocity  $c\approx10^5~cm/sec$ and used known \cite{Boulet1997} density value $\rho=10.26~g/cm^3$. Thus,  
 we have for the combination Eq.(\ref{B4})
 
\begin{equation}
\frac{Kn}{T_c}
\left (\frac{\partial T_c}{\partial P}\right )^2=0.2~.
\label{LPi}
\end{equation}
At $T_c\approx10K$  the pressure derivative of the critical temperature is much higher (and its square is even more higher) than its average value given by Eq. (\ref{est}). So, we come to conclusion 
that at critical temperature of the order  10 K  the criterium ({\ref{B4}) is fulfilled and the phase transition of the second order turns into
 the first order one.

\subsection{ Concluding remarks}
The magneto-elastic interaction provides development of the first-order instability at the phase transition to the ordered state in a ferromagnet. However,  actual temperature interval of this  instability development is negligibly small and the first-order transition looks almost indistinguishable from the second order one.
The particular feature of the anisotropic ferromagnet UGe$_2$ is the precipitous drop of the Curie temperature as the function of pressure near 14-15 kbar. Due to this property
at about these pressures the second order phase transition (or very weak transition of the first-order)  to the ferromagnet state turns into the real first-order type transition.  

At low temperatures according to the Nernst law and the Clausius-Clapeyron relation
\begin{equation}
\frac{d T_c}{d P}=\left.\frac{v_1-v_2}{s_1-s_2}\right|_{T\to0}\rightarrow\infty
\end{equation}
 the drop of  transition temperature with pressure  begins to be  infinitely fast. It means that weak first order transition has the tendency to be stronger and stronger  as temperature decreases.  Hence, the effect of magneto-elastic interaction or, more generally,  of the order parameter interaction with the elastic degrees of freedom 
at arbitrary type of ordering raises  the doubts upon the existence of quantum critical phenomena.

 \section{Superconducting order in ferromagnetic UI${\bf r}$}
 
 UIr has the monoclinic PbBi-type structure without inversion symmetry shown in Fig.18. The magnetic property at ambient pressure is Ising-like ferromagnetic with the Curie temperature $T_c=46K$. The magnetic susceptibility follows the Curie-Weiss law with an effective moment $\mu_{eff}=2.4\mu_B/U$ while the ordered moment is $0.5\mu_B/U$. The pressure-temperature phase diagram of UIr consists of a low pressure phase FM1, a high pressure magnetic phase FM2 and a superconducting phase as shown in Fig.19a.
 The discrete change of  the ordered moment indicates that the FM1-FM2 transition is of the first-order. The FM2-nonmagnetic transition is of the second-order \cite{Kobayashi2007}.
 
 For UIr, it is still not clear whether the superconductivity coexists with magnetic ordering or not \cite{Kobayashi2007}.  However, if it is the case, here we deal with unique 
 situation when superconducting state arises in material with broken space and time parity.  Hence, it is instructive to describe the symmetry and the order parameter of this type superconductivity.
  
  The group of symmetry of the normal nonmagnetic state  
 \begin{equation}
 G_N=(E,C_{2b})\times R\times U(1)
 \end{equation}
includes the point symmetry group $C_2=(E,C_{2b})$, where $C_{2b}$ is the rotation around the $b$-axis on angle $\pi$ (see Fig.19), the time inversion $R$ and the gauge group $U(1)$.
In the FM2 state the time inversion is broken and the symmetry group 
 \begin{equation}
 G_{FM}=(E,RC_{2b})\times U(1)
 \end{equation}
 includes now the combination of rotation on angle $\pi$ around the $b$-axis and the operation $R$ changing the direction of spontaneous magnetization, lying in the $(a,c)$ plane, to the opposite. Finally, in the superconducting state coexisting with FM2 state the gauge invariance is broken and the symmetry group is
  \begin{equation}
 G_{FM+SC}=(E,e^{2i\varphi}RC_{2b}).
 \end{equation}
 
 The space parity is broken, hence the magnetic pairing interaction inevitably  includes the  Dzyaloshinskii-Moriya-type propagators. \cite{Samokhin}
 As result the superconducting order parameter consists of sum of triplet and singlet parts
  \begin{equation}
 \hat\Delta=i({\bf d}\mbox{\boldmath$\sigma$})\sigma_y+id_0\sigma_y.
 \end{equation}
 The triplet part has the usual form
  \begin{equation}
 {\bf d}({\bf k},{\bf r})=\frac{1}{2}\left [  -(\hat x+i\hat y)\Delta^\uparrow({\bf k},{\bf r})+(\hat x-i\hat y)\Delta^\downarrow({\bf k},{\bf r})  \right ]+\Delta^0({\bf k},{\bf r})\hat z,
 \end{equation}
 however, unlike the orthorhombic crystals  we considered in Chapter II, the coordinate axis of nonunitary superconducting ordering does not coincide with monoclinic
crystallographic directions. Namery, here $\hat z$ is the unit vector aligned parallel to spontaneous magnetization lying in $(a,c)$ plane  (direction $[1,0,\bar 1]$),
$\hat x$ is the unit vector directed along $b$-axis, 
and $\hat y=\hat z\times\hat x$. 
 \begin{equation}
\Delta^\uparrow({\bf k},{\bf r})=k_x\eta_x^\uparrow({\bf r})+ik_y\eta_y^\uparrow({\bf r}),~~~~\Delta^\downarrow({\bf k},{\bf r})=k_x\eta_x^\downarrow({\bf r})+ik_y\eta_y^\downarrow({\bf r}),~~~~\Delta^0({\bf k},{\bf r})=k_z\eta_z^0({\bf r}),
 \end{equation} 
 where $k_x,k_y,k_z$ are projections of momentum on the axis $\hat x, \hat y, \hat z$ determined above.
 The singlet part of the order parameter is
  \begin{equation}
d_0({\bf k},{\bf r})=F\eta_0({\bf r}),
 \end{equation}
 where $F$ is some function of $k_x^2,k_y^2,k_z^2$.

 \section{Conclusion}

  The treatment  of properties of uranium compounds developed in the present paper is based on the symmetry of superconducting states with triplet pairing in the orthorhombic ferromagnets.
   The  phenomenological considerations  were supported by the microscopic calculations
performed  in frame of the  weak coupling superconductivity theory. We have operated with pairing interaction expressed through the frequency-independent magnetic susceptibility of anisotropic ferromagnetic media. 
This approach reproduces the structure of superconducting states found on pure symmetry ground and allows to explain qualitatively many experimental observations.

The topic, that  was out  our attention,  is  the ARPES experiments and the band structure calculations which are still in not complete agreement. We can recommend the two recent papers 
reporting the  ARPES studies in URhGe \cite{Fujimori2014}
and in UGe$_2$ and UCoGe \cite{Fujimori2015},  the comparison with the  band structure calculations  and the vast list of references to the previous studies.

\section*{Acknowledgements}

I am grateful to A. Huxley, J.-P. Brison, S. Raymond, K. Ishida, D. Aoki, M. Zhitomirsky and J. Flouquet for the discussions of physics of  uranium ferromagnets. I am also indebted to M. Sadovskii who invited me to give the lectures at Winter Theoretical School "Kourovka-2016" what has obviously helped  this paper emergence.

\begin{figure}[p]
\includegraphics
[height=.8\textheight]
{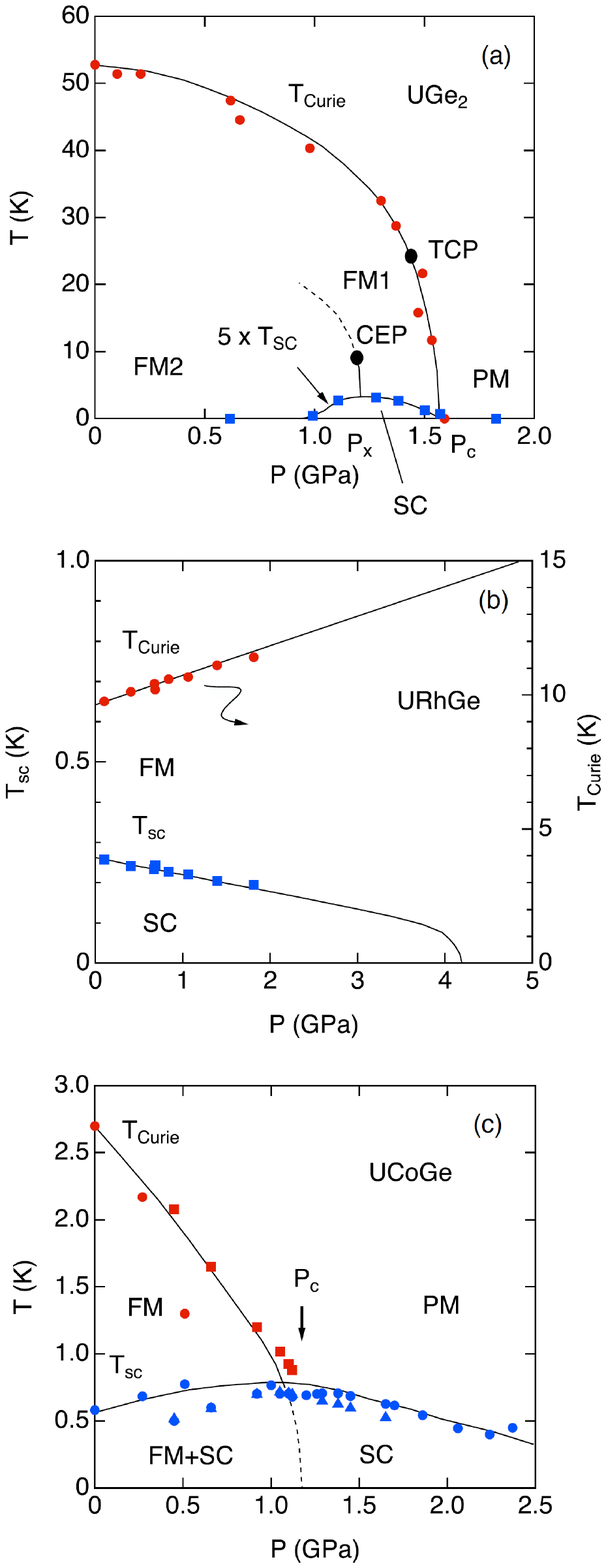}
 \caption{(Color online) Temperature-pressure phase diagram of UGe2,
URhGe, and UCoGe. Notations  FM, SC and PM have been used for ferromagnetic, superconducting and paramagnetic  phases correspondingly, TCP is the tricritical point, CEP is the critical end point.
\cite{Aoki14} }
\end{figure}

\begin{figure}[p]
\includegraphics 
[height=.8\textheight]
{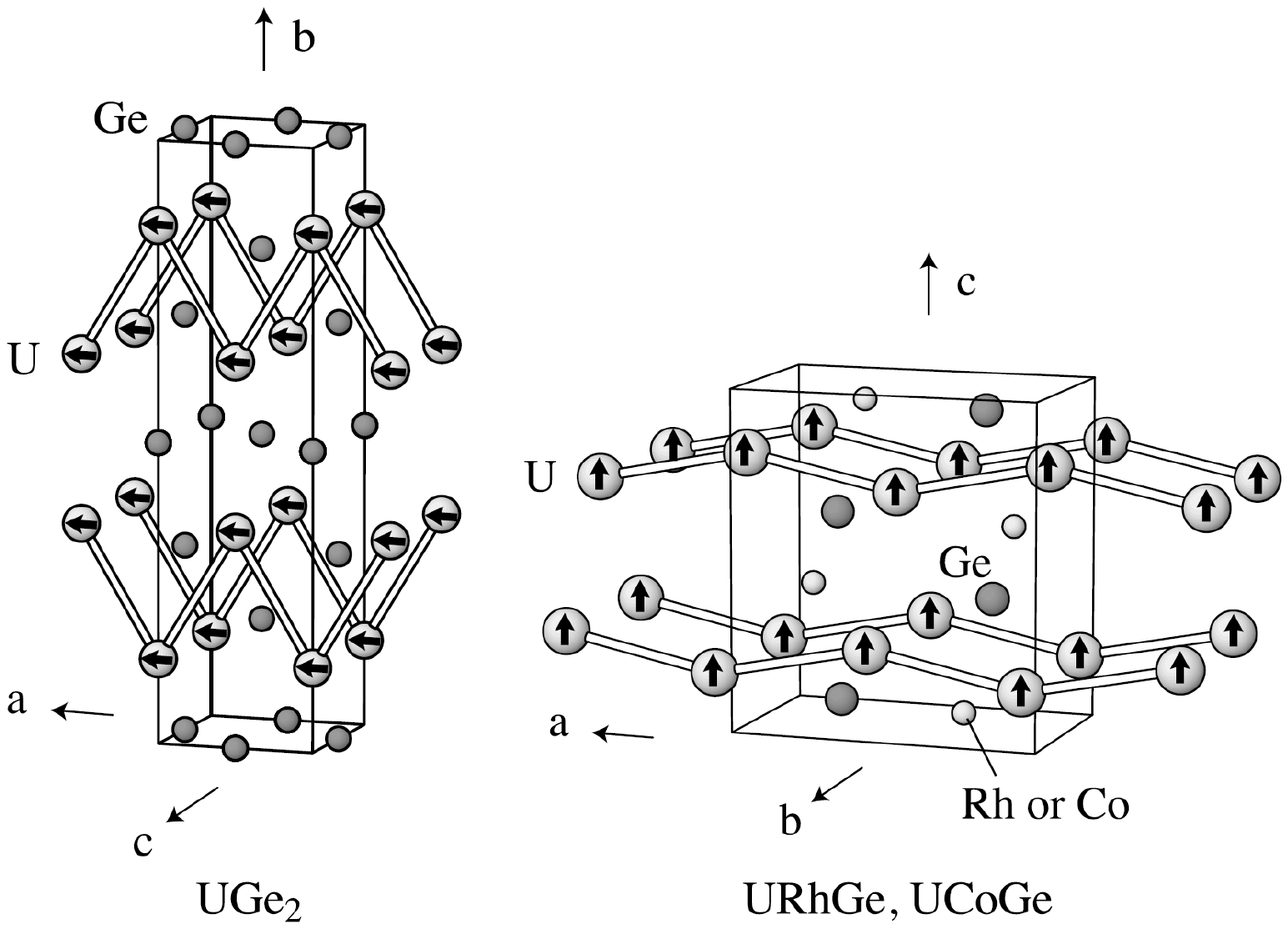}
 \caption{(Color online) 
 Crystal structures of UGe2, URhGe, UCoGe.
\cite{Aoki14} }
\end{figure} 
\begin{figure}[p]
\includegraphics 
[height=.8\textheight]
{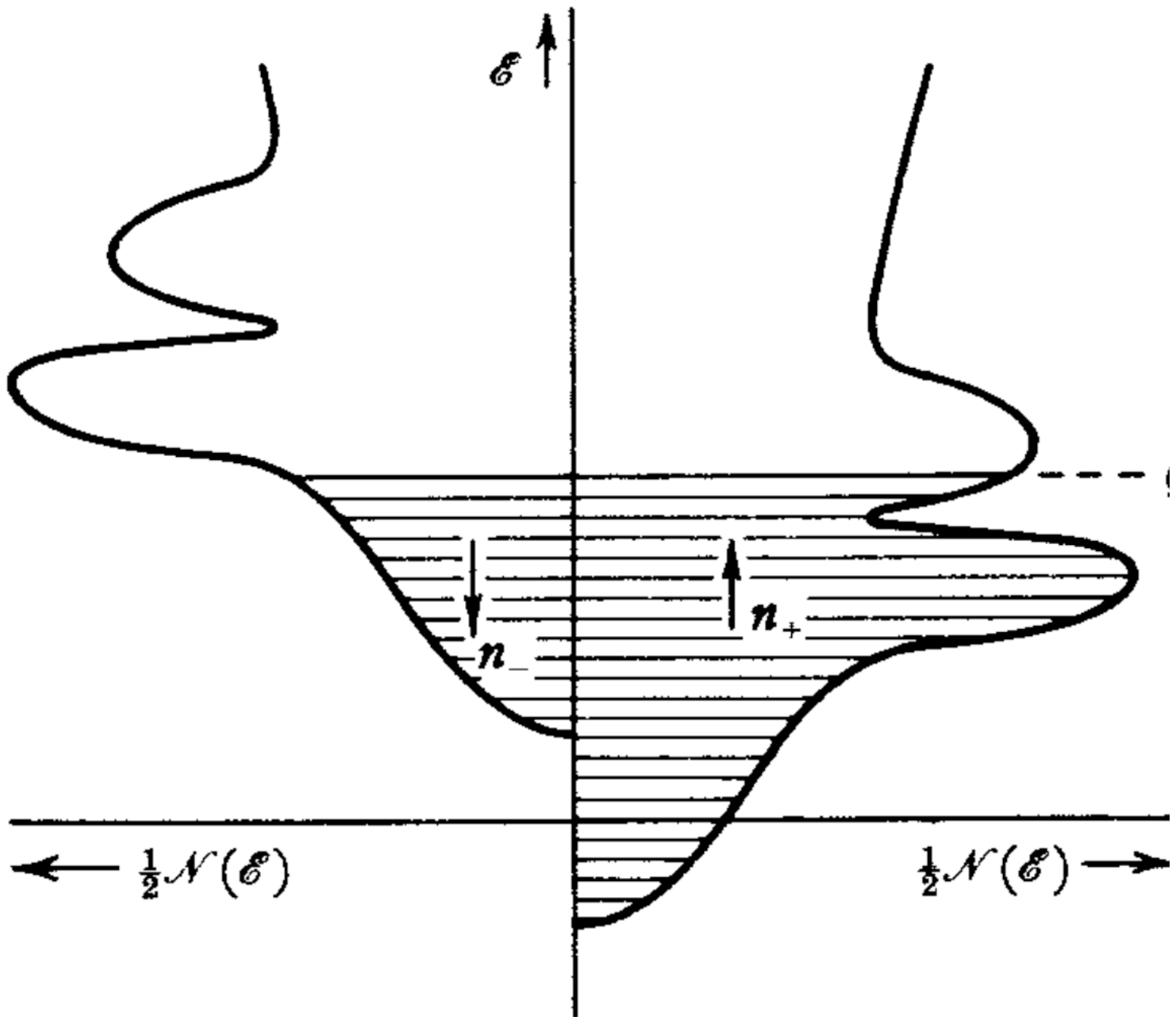}
 \caption{ Density of states for spin-up and  spin-down electron bands in 
a  ferromagnetic metal.}
\end{figure} 

\begin{figure}[p]
\includegraphics
[height=1.0\textheight]
{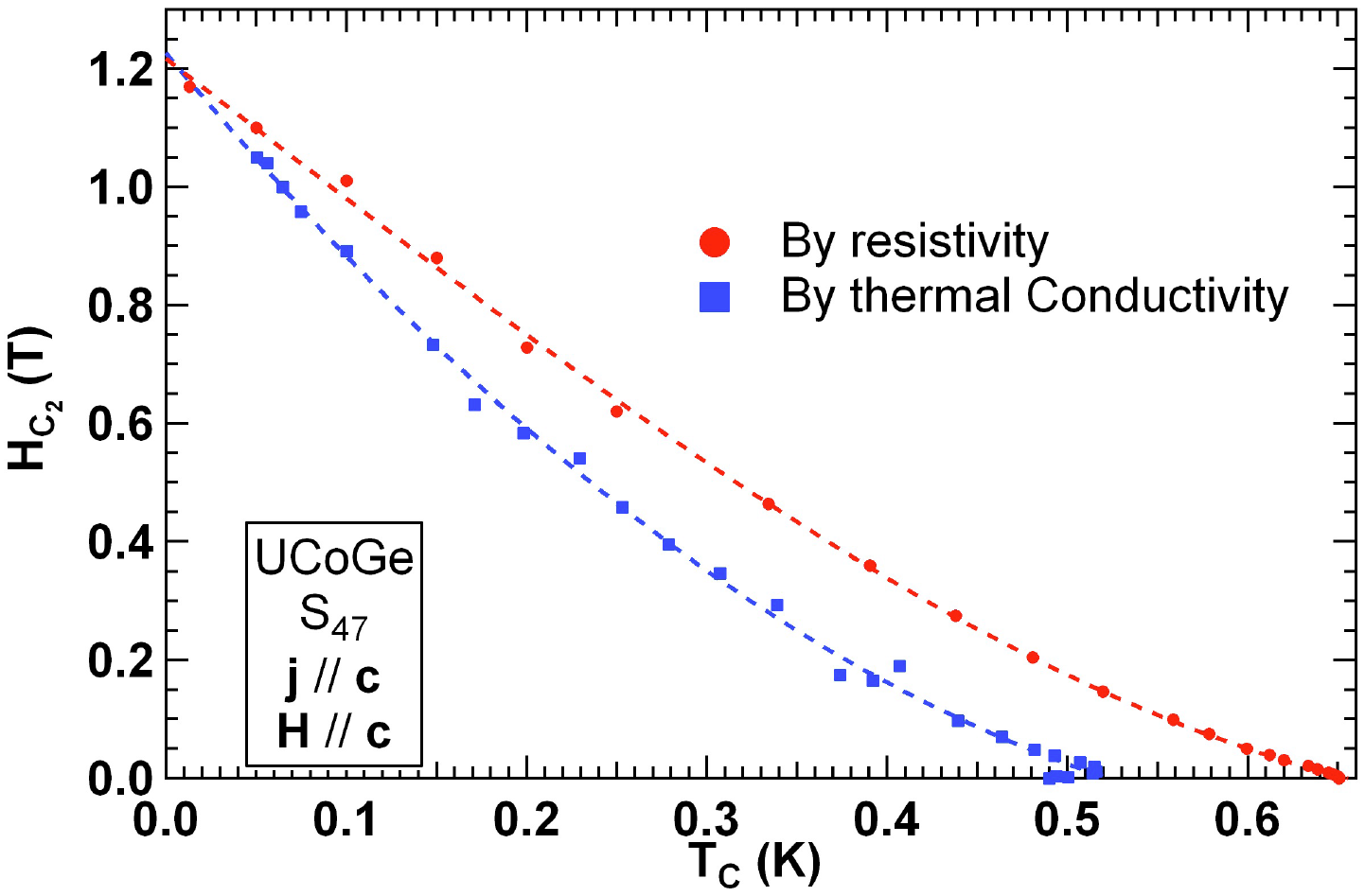}
 \caption{(Color online) 
The upper critical field 
$H_{c2}$ in UCoGe extracted from the resistivity and the thermal conductivity measurements.
(M.Taupin, unpublished (2016)). }
\end{figure} 

\begin{figure}[p]
\includegraphics
[height=1.0\textheight]
{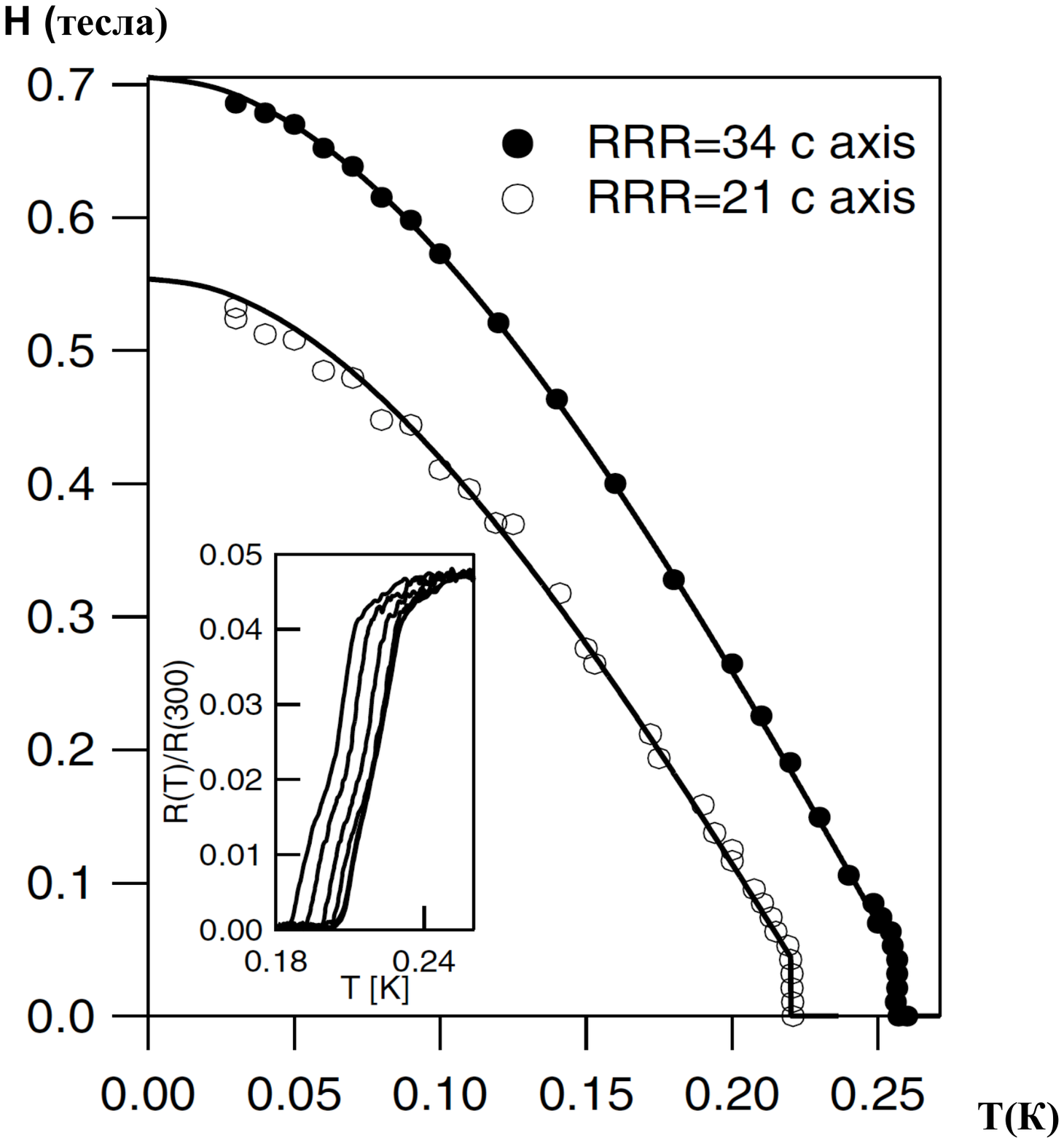}
 \caption{ 
The temperature dependence of the applied field at
which superconductivity is destroyed for two URhGe crystals with
$RRR=34$ and $RRR=21$ for fields applied parallel to the $c$
axis.\cite{Hardy2005}}
\end{figure} 

\begin{figure}[p]
\includegraphics
[height=1.0\textheight]
{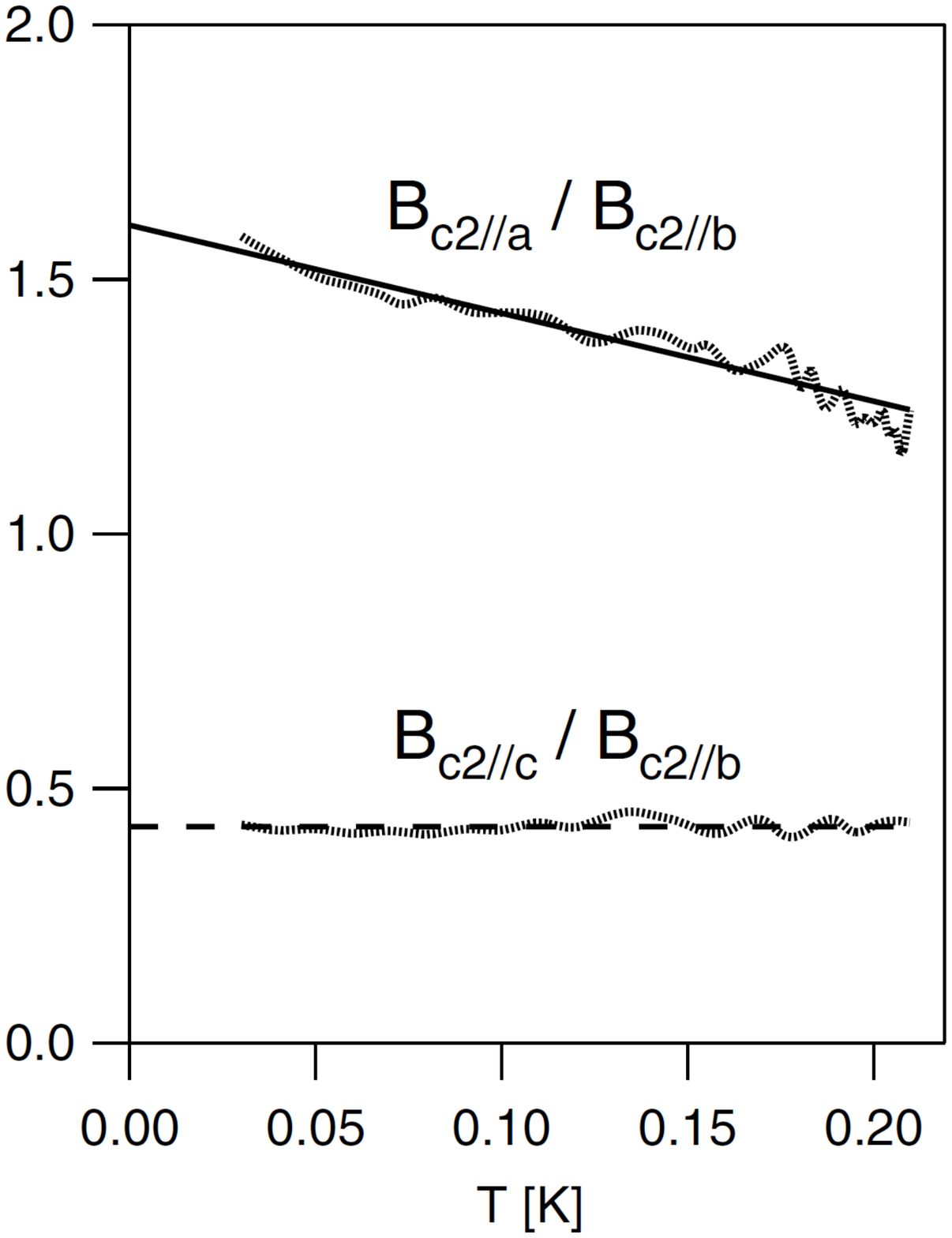}
 \caption{ 
The temperature dependence of the ratio of the upper critical fields parallel
to different axes in URhGe.\cite{Hardy2005}
}
\end{figure} 
\begin{figure}[p]
\includegraphics
[height=1.0\textheight]
{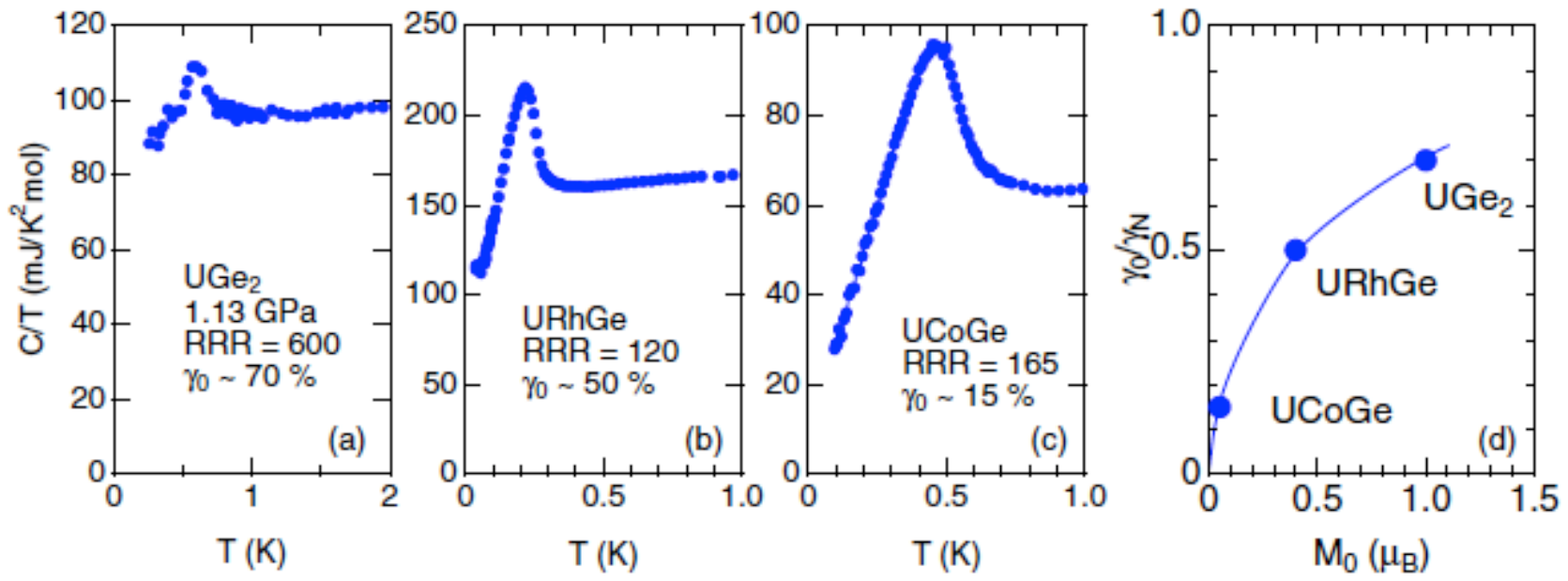}
 \caption{ 
(Color online) Specific heat  divided by temperature $C/T=\gamma$  in (a) UGe2, (b) URhGe, and (c) UCoGe. (d) Scaled residual $\gamma_0/\gamma_N$-value as a function of ordered
moment. \cite{Aoki14}
   }
\end{figure} 

\begin{figure}[p]
\includegraphics
[height=1.0\textheight]
{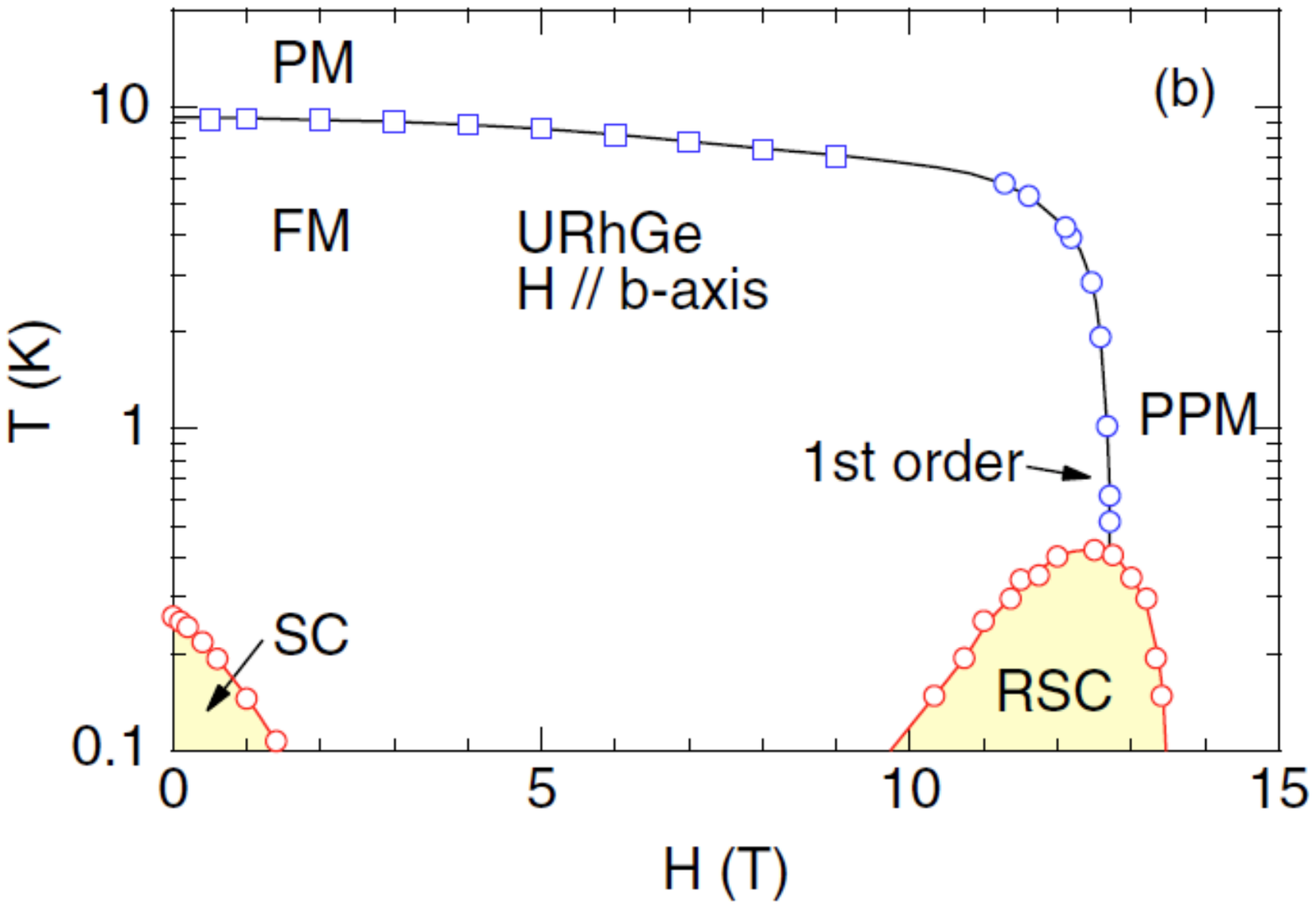}
 \caption{(Color online)
Temperature-field phase diagram for $H\parallel b$-axis
in URhGe. PM, FM and PPM denote paramagnetic, ferromagnetic and strongly polarized paramagnetic
states, 
repectively. SC and RSC
denote superconducting and reentrant
superconducting states, respectively.\cite{AokiKnebel2014}
}
\end{figure}

\begin{figure}[p]
\includegraphics
[height=0.8\textheight]
{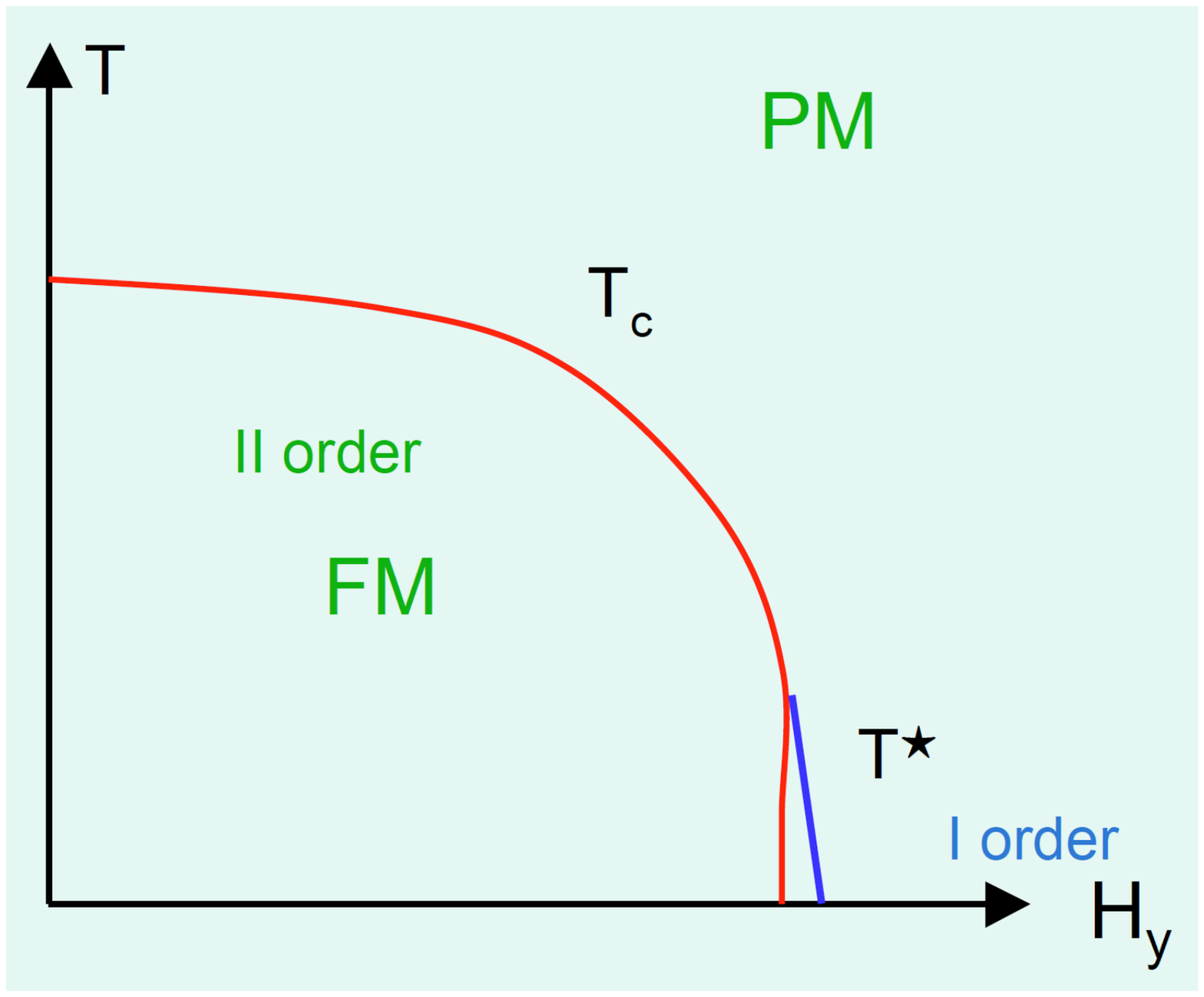}
 \caption{(Color online)
$(H,T)$ phase diagram of an uniaxial ferromagnet under magnetic field perpendicular to spontaneous.magnetization.
PM and  FM denote paramagnetic and ferromagnetic 
states, 
respectively. Red line is the  Curie temperature line. Blue line is the line of the first-order transition.
}
\end{figure}
 
\begin{figure}[p]
\includegraphics
[height=1.0\textheight]
{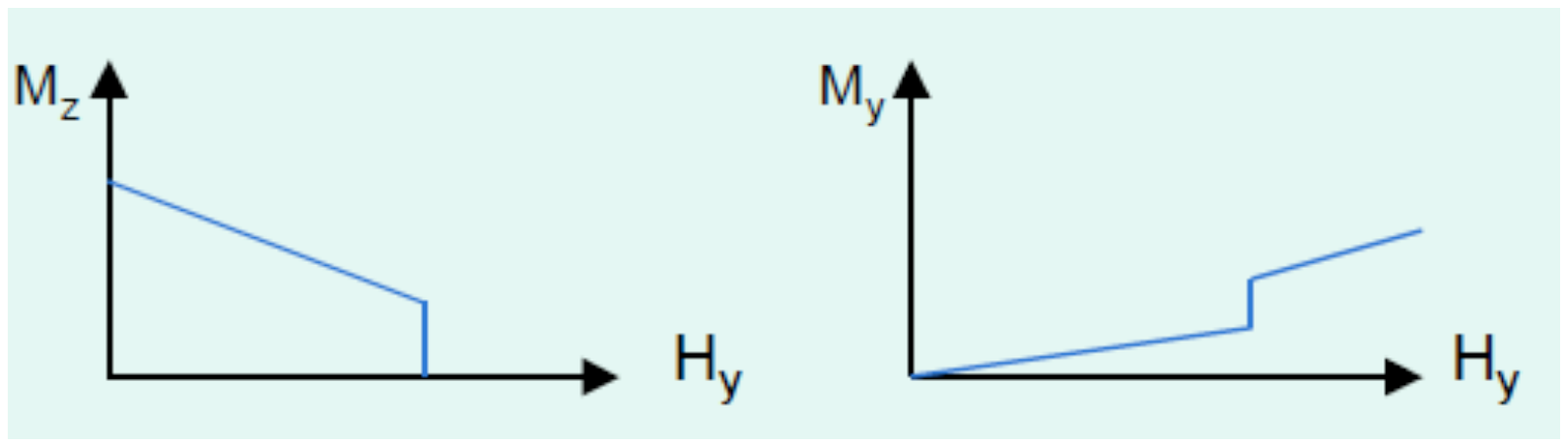}
 \caption{(Color online)
Magnetic field  dependences of $M_z(H_y)$ and $M_y(H_y)$  with the jumps  at the first-order transition.
}
\end{figure} 

\begin{figure}[p]
\includegraphics
[height=1.0\textheight]
{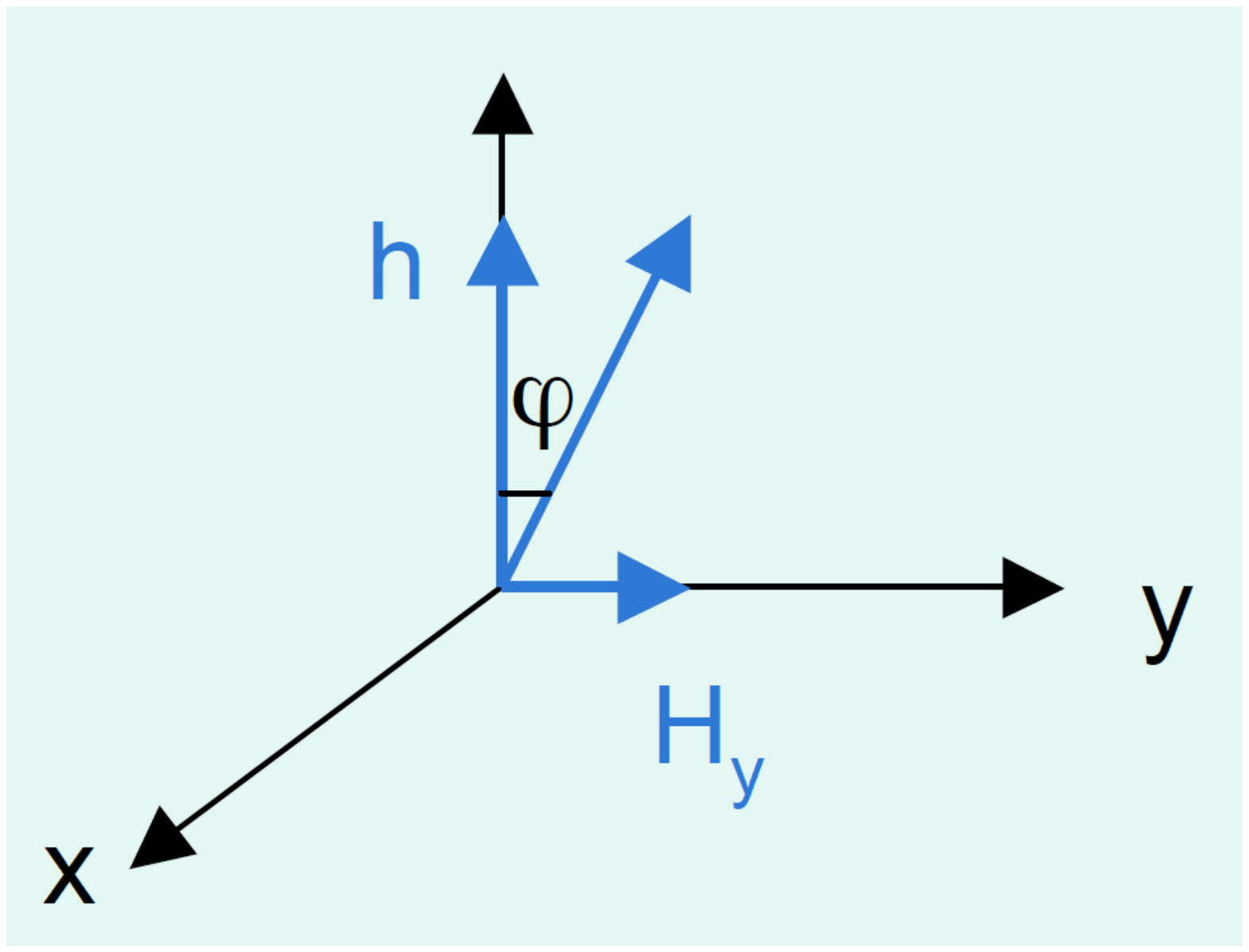}
 \caption{
(Color online) 
Magnetic field $H_y$ directed perpendicular to the exchange field ${\bf h}$.
}
\end{figure}

\begin{figure}[p]
\includegraphics
[height=1.0\textheight]
{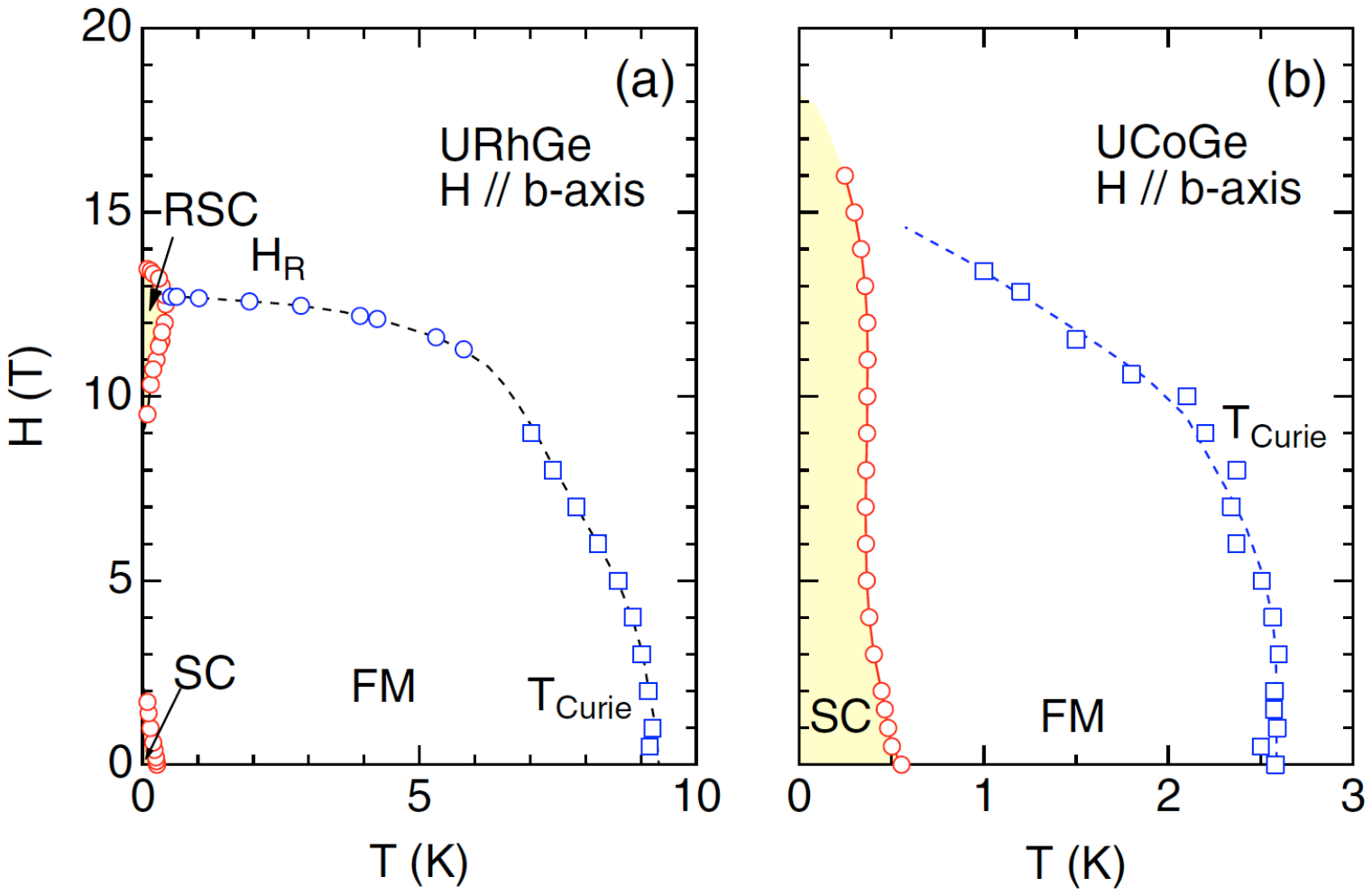}
 \caption{
(Color online) Field-temperature phase diagram of URhGe and
UCoGe for the field along the $b$-axis.\cite{Aoki14}
}
\end{figure}

\begin{figure}[p]
\includegraphics
[height=1.0\textheight]
{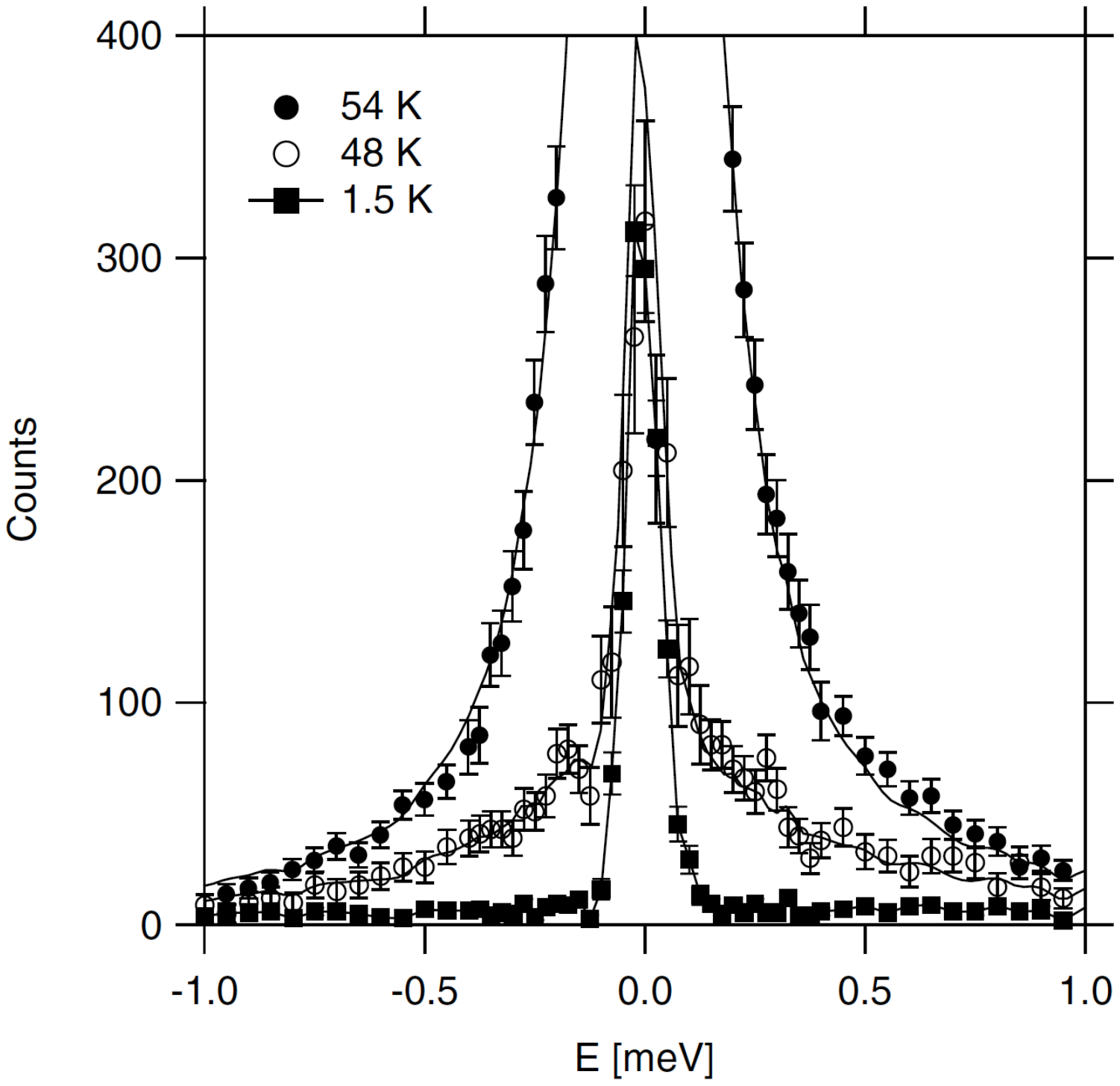}
 \caption{
The detected neutron scattering (normalized to a
fixed incident beam monitor count) is shown as a function of
energy transfer at ${\bf Q}=(0,0,1.04)$ reciprocal lattice units, just
above $T_c$, at 48 K and at 1.5 K \cite{Huxley2003}.
}
\end{figure} 
\begin{figure}[p]
\includegraphics
[height=1.0\textheight]
{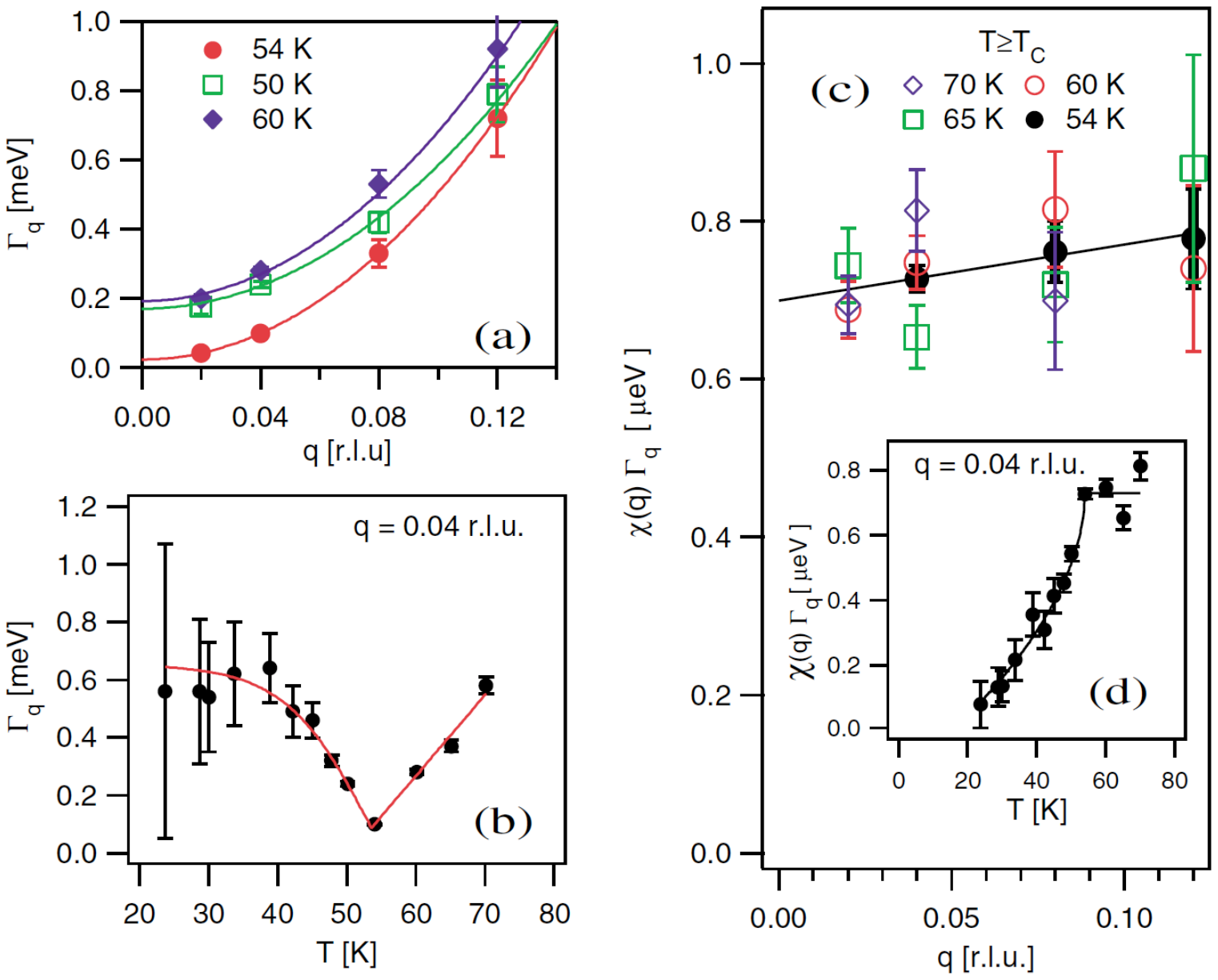}
 \caption{
(Color online). Panel (a) shows the $q$ dependence (${\bf q}\parallel 
{\bf c}$) of 
$\Gamma_{\bf q}$ at three different temperatures. Panel (b) shows the
temperature dependence of  $\Gamma_{\bf q}$ at ${\bf q}=(0,0,0.04)$. Panel (c)
shows the $q$ dependence of the product $\chi({\bf q})\Gamma_{\bf q}$
at different
temperatures above $T_c$. Panel (d) shows the temperatures dependence
of 
$\chi({\bf q})\Gamma_{\bf q}$
 at ${\bf q}=(0,0,0.04)$
above and below $T_c$. 
\cite{Huxley2003}
}
\end{figure} 
\begin{figure}[p]
\includegraphics
[height=1.0\textheight]
{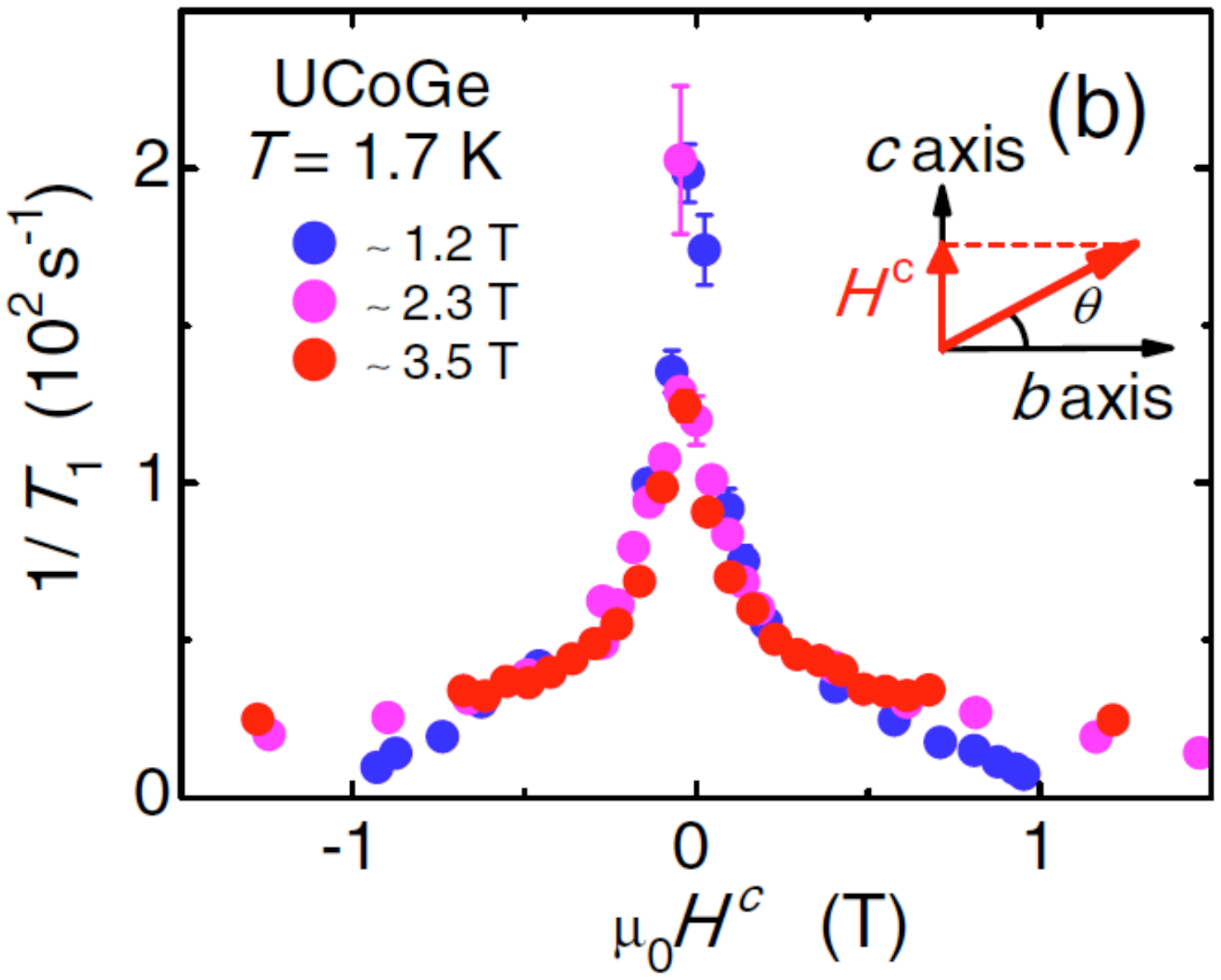}
 \caption{
(Color online). 
Plot of the $1/T_1$ against $H_c$ \cite{Hattory2012}.
}
\end{figure} 
\begin{figure}[p]
\includegraphics
[height=1.0\textheight]
{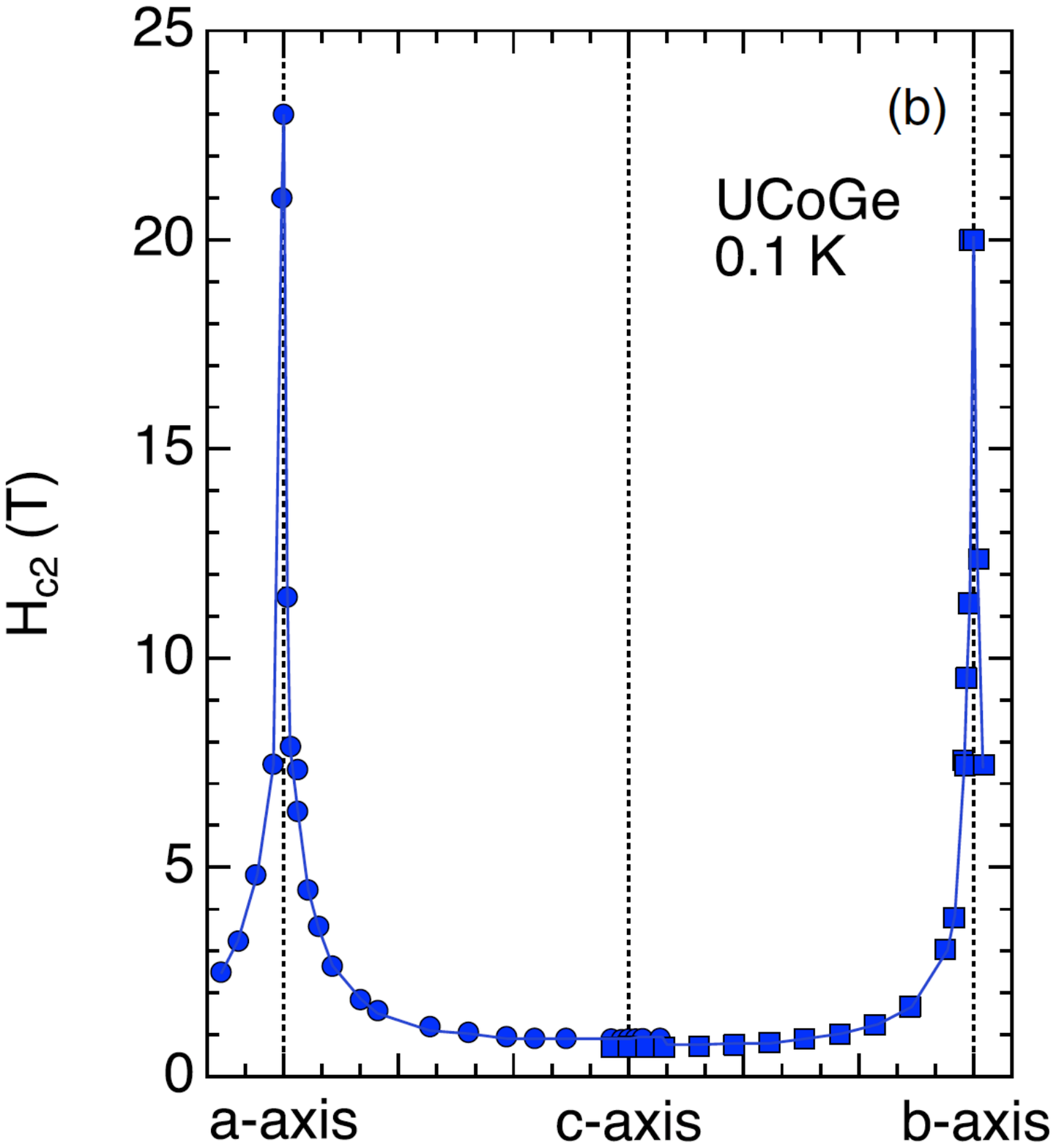}
 \caption{
(Color online).
Angular dependence of $H_{c2}$ at $0.1K$ in UCoGe. $T_{sc}(H=0)\approx 0.6 K$
\cite{Aoki14}.
}
\end{figure} 
\begin{figure}[p]
\includegraphics
[height=1.0\textheight]
{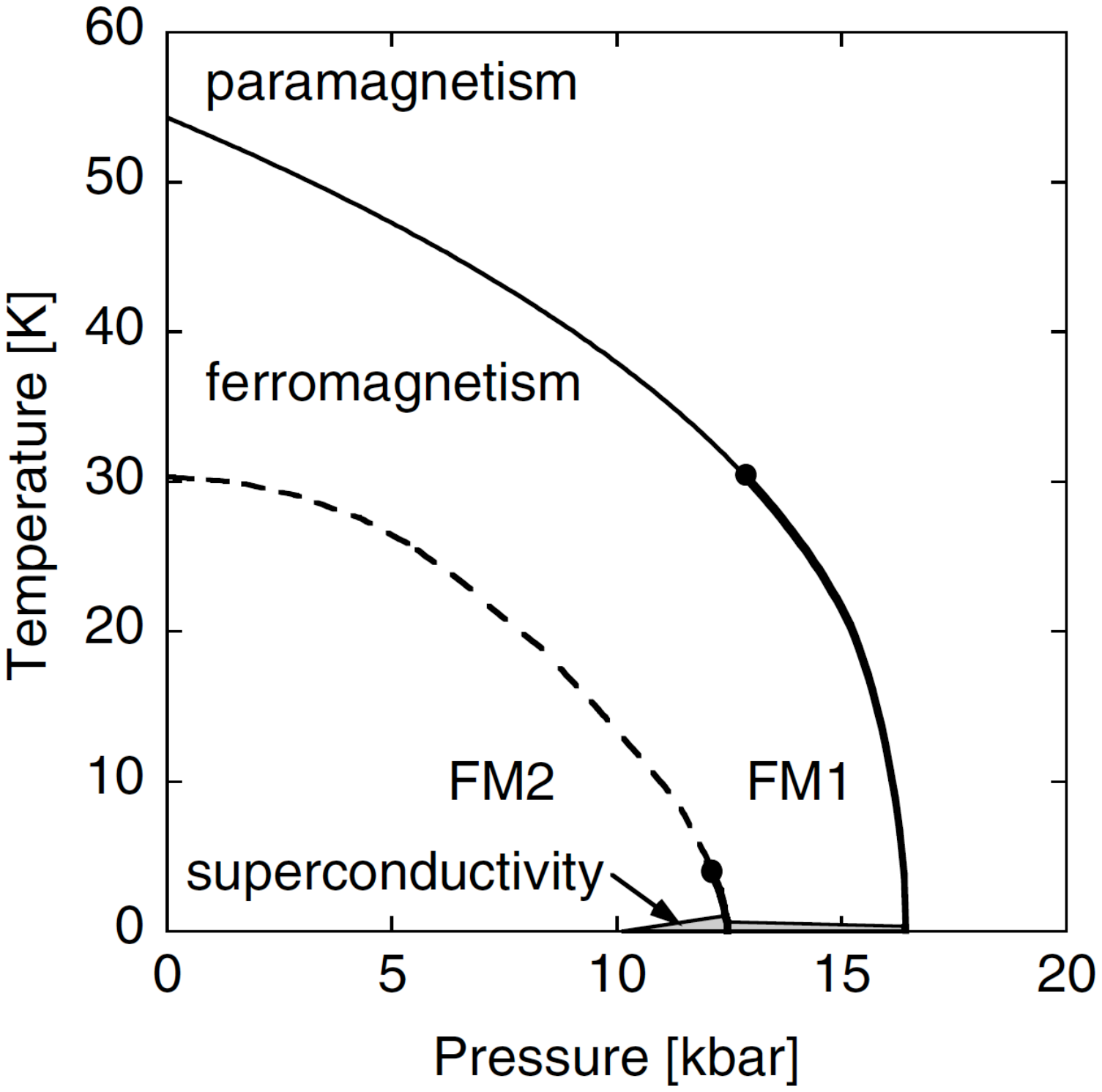}
 \caption{The schematic phase diagram of UGe2. Thick lines
denote first order transitions and fine lines second order transitions.
The dashed line is a crossover. Dots mark the positions
of critical (tricritical) points. The region where superconductivity
occurs is shaded.\cite{Huxley2003}
}
\end{figure}
\begin{figure}[p]
\includegraphics
[height=1.0\textheight]
{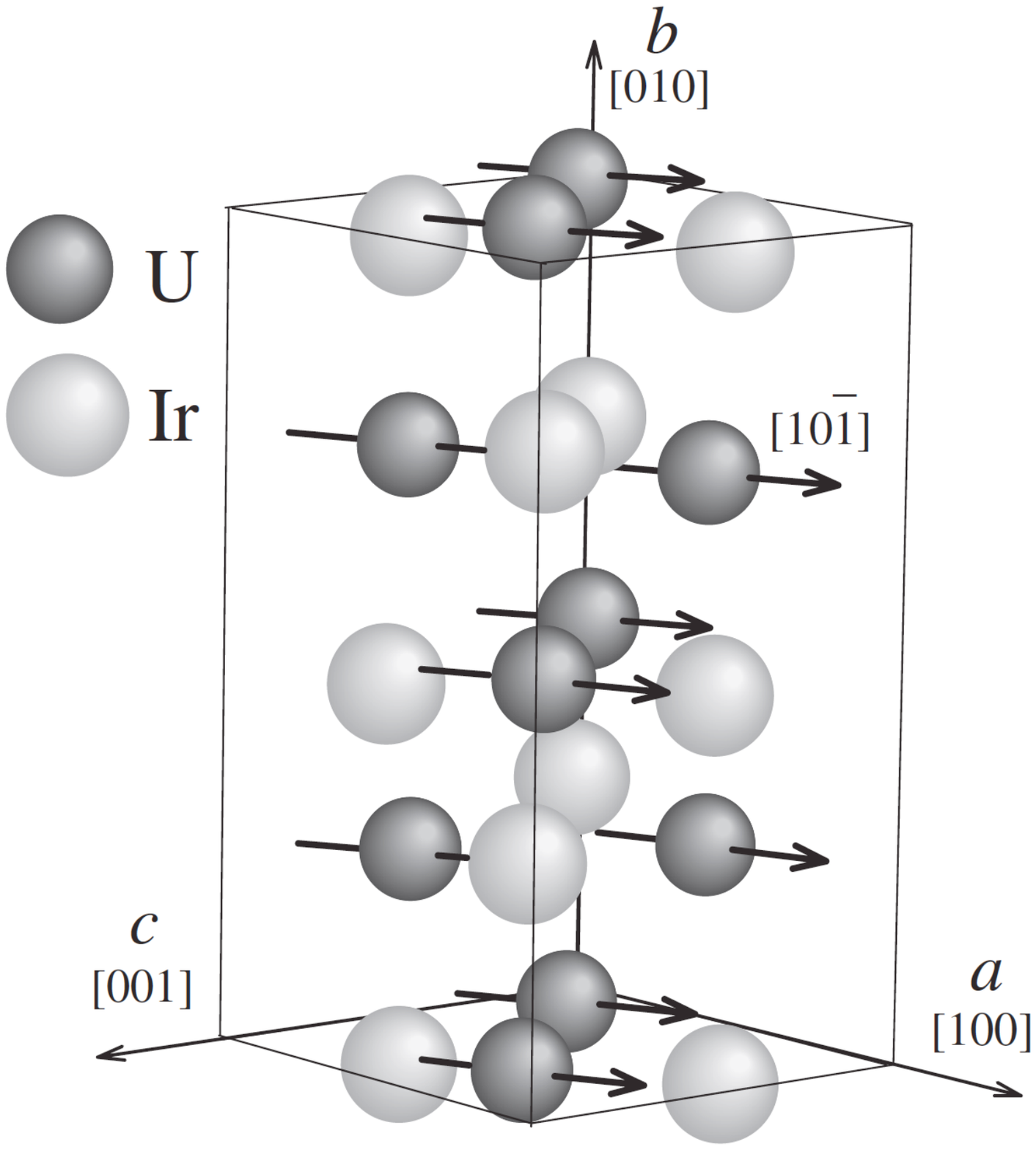}
\caption{The monoclinic structure of UIr. The arrows indicate the direction of spontaneous magnetization.
\cite{Galatanu2004}}
\end{figure} 
\begin{figure}[p]
\includegraphics
[height=1.0\textheight]
{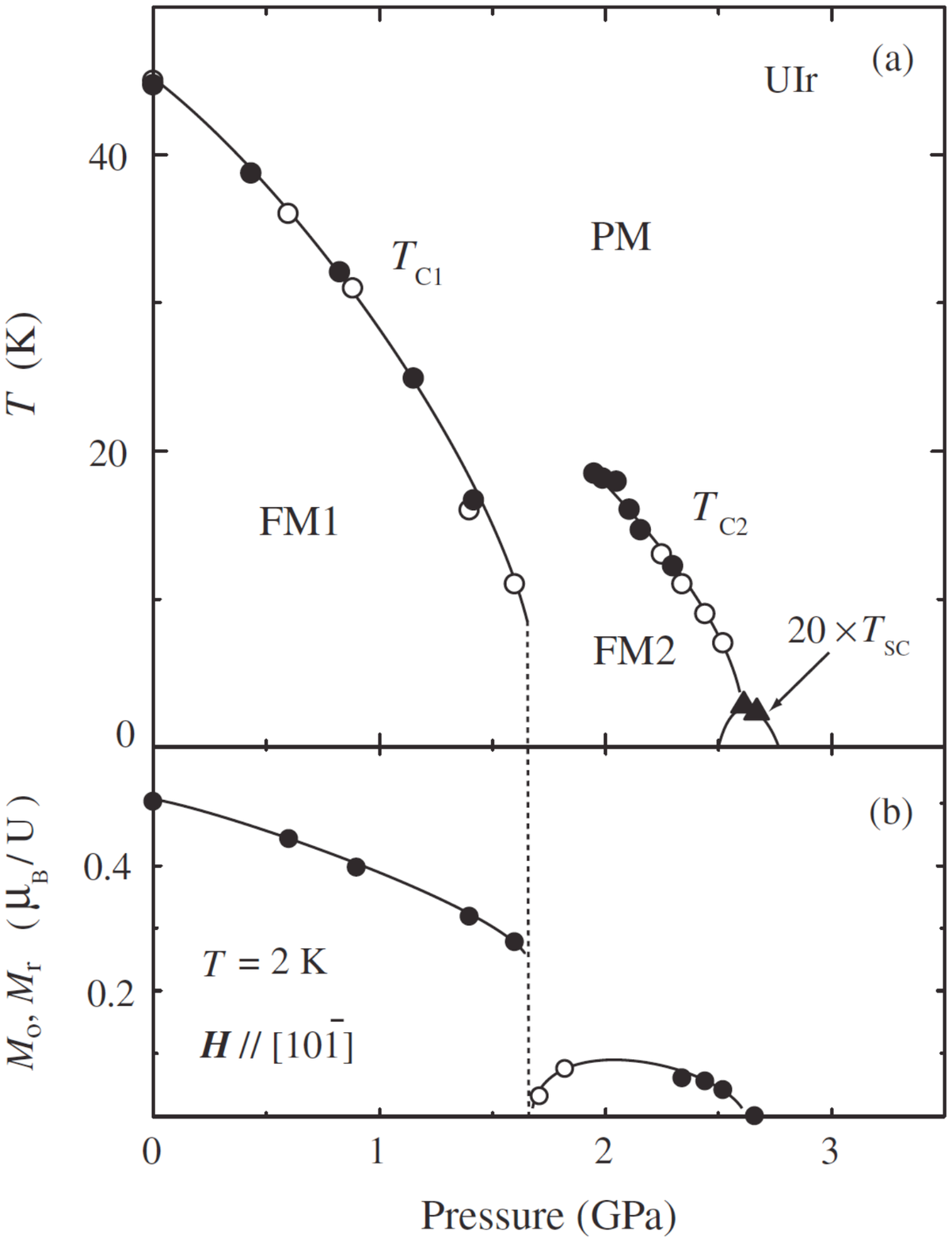}
\caption{
(a) PÐT phase diagram of UIr. Solid and open circles are  determined by the resistivity
and magnetization measurements, respectively. The superconductivity is observed near the critical
pressure where the FM2 phase disappears. (b) Pressure dependence of the ordered moment along
$[1, 0, \bar1]$ at 2 K. Solid circles are the ordered moments $M_0$ determined from the Arrott plot. Open
circles are the residual magnetizations $M_r $.
\cite{Kobayashi2007}}
\end{figure}


\begin{thebibliography}{220}
\bibitem{fn4}  Any superconductor  is diamagnet, however, here and what follows  we shall call by "paramagnetic" superconducting state
the state which does not possess  a spontaneous magnetic moment.
\bibitem{footnote}The case  of UCoGe at pressure about 1.1 GPa when the Curie temperature is almost equal to the temperature of superconducting transition demands the particular treatment.
\bibitem{residual} Small residual linewidth $\Gamma_0\sim0.02~mev$  has been registered also in ferromagnet MnP. \cite{Yamada}
\end{thebibliography}
\end{document}